\documentclass{aa}

\usepackage{graphicx}
\usepackage{txfonts}
\usepackage{siunitx}
\usepackage{textcmds}
\usepackage{subfigure}

\usepackage{url}

\begin{document}

   \title{Hydrostatic Mass Profiles of Galaxy Clusters in the eROSITA Survey}


   \author{Dominik Scheck\inst{1,2}
          \and
          Jeremy S. Sanders\inst{1}\fnmsep\thanks{\email{jsanders@mpe.mpg.de}}
          \and
          Veronica Biffi\inst{3,4}
          \and
          Klaus Dolag\inst{4,5}
          \and
          Esra Bulbul\inst{1}
          \and
          Ang Liu\inst{1}
          }

   \institute{Max-Planck-Institut für extraterrestrische Physik, Gießenbachstraße 1,            D-85748 Garching, Germany 
        \and
            Technical University of Munich, Germany; Department of Physics
        \and
            INAF — Osservatorio Astronomico di Trieste, via Tiepolo 11, I-34143
            Trieste, Italy
        \and
            Universitäts-Sternwarte, Fakultät für Physik, Ludwig-Maximilians-Universität München, Scheinerstr.1, 81679 München, Germany
        \and
            Max-Planck-Institut für Astrophysik, Karl-Schwarzschild-Straße 1, 85741 Garching, Germany
            }

   \date{Accepted by A\&A, 15 pages, 18 figures}


\abstract
{To assume hydrostatic equilibrium between the intracluster medium and the gravitational potential of galaxy clusters is an extensively used method to investigate their total masses.}
{We want to test hydrostatic masses obtained with an observational code in the context of the Spectrum-Roentgen-Gamma / eROSITA survey.}
{We use the hydrostatic modeling code MBProj2 to fit surface-brightness profiles to simulated clusters with idealized properties as well as to a sample of 93 clusters taken from the Magneticum Pathfinder simulations. We investigate the latter under the assumption of idealized observational conditions and also for realistic eROSITA data quality. The comparison of the fitted cumulative total mass profiles and the true mass profiles provided by the simulations allows to gain knowledge both about the validity of hydrostatic equilibrium in each cluster and the reliability of our approach. Furthermore, we use the true profiles for gas density and pressure to compute hydrostatic mass profiles based on theory for every cluster.}
{For an idealized cluster that was simulated to fulfill perfect hydrostatic equilibrium, we find that the cumulative total mass at the true $r_{\mathrm{500}}$ and $r_{\mathrm{200}}$ can be reproduced with deviations of less than 7\%. For the clusters from the Magneticum Pathfinder simulations under idealized observational conditions, the median values of the fitted cumulative total masses at the true $r_{\mathrm{500}}$ and $r_{\mathrm{200}}$ are in agreement with our expectations, taking into account the hydrostatic mass bias. Nevertheless, we find a tendency towards a too high steepness of the cumulative total mass profiles in the outskirts. For realistic eROSITA data quality, this steepness problem intensifies for clusters with high redshifts and thus leads to too high cumulative total masses at $r_{\mathrm{200}}$.
For the hydrostatic masses based on the true profiles known from the simulations, we find a good agreement with our expectations concerning the hydrostatic mass.}
{}


\maketitle


\section{Introduction}
\label{section_Introduction}

The study of clusters of galaxies often requires accurate and precise mass determinations, including when investigating the physical processes taking place within them, their evolution over time and when using them as cosmological probes. The eROSITA X-ray survey \citep{Merloni_2012, Predehl_2021} is expected to detect around 100\,000 galaxy clusters \citep{Liu_2022, Bulbul_2022}, with the primary aim of using them to study cosmology. The number of clusters as a function of mass is a sensitive probe of cosmology \citep[e.g.][]{Pillepich12}.

There are different ways in which the masses of clusters can be obtained within a survey. One approach is to measure the masses of a subset of clusters using weak lensing in order to calibrate X-ray observable to mass scaling relations \citep[e.g.][]{Chiu21}. The observable can then be used to infer the cluster mass in the wider sample. Masses of cluster samples have also been measured by assuming hydrostatic equilibrium within their intracluster medium \citep[ICM; e.g.][]{Pointecouteau05,Vikhlinin06,Martino14,Ettori19,Pratt_2019}. Different methods of mass measurement are subject to different biases and it is therefore important to make comparisons between them to ensure consistency. For example, although weak lensing might be considered the most pure mass measurement, in the sense it is only sensitive to mass, it can be affected by various biases, including projection of other structure, noise and miscentering \citep[e.g.][]{Grandis21}.

Hydrostatic mass measurements assume that the hot atmosphere of the cluster is a static atmosphere in hydrostatic equilibrium \citep[see e.g.][]{Boehringer_2009}. The gravitational acceleration at a location in a cluster can be calculated from the pressure gradient and gas density under this assumption. This in turns allows the total mass to be inferred. However, real galaxy clusters are not in complete hydrostatic equilibrium. Motions of the ICM (e.g. turbulence) can give rise to a hydrostatic mass bias, $b$, which is the fractional difference the hydrostatic mass is below the true mass of a cluster. Simulations \citep[e.g.][]{Lau_2009,Lau13,Biffi16} predict values of $b \sim 0.1-0.2$ (also compare to \citealt{Hoekstra_2015} and \citealt{von_der_Linden_2014}).

There are also various observational effects which could potentially bias obtained hydrostatic masses. These include the choice of center (the ICM may be offset from the dark matter potential), the assumption of a particular geometry (spherical geometry is often assumed), temperature biasing due to multiple components \citep{Mazzotta_2004,Vikhlinin_2006_Predicting}, clumping within the cluster (the X-ray emissivity is proportional to the density-squared of the ICM), how contaminating substructures are masked, and instrument calibration uncertainties \citep{Schellenberger15}.

To use hydrostatic masses measurements from a survey such as eROSITA, it is important to understand the total contribution of these potential biases. In addition, many of the clusters observed by eROSITA will only be detected with a few hundred X-ray counts and so it will be useful to test the methods in this high statistical noise regime. One way in which this can be done is to make realistic simulated observations of clusters and use a method to obtain hydrostatic masses which is designed to work on real data.

To determine hydrostatic masses of clusters there are various different approaches depending on what level the data are analyzed \citep[see e.g.][]{Ghirardini18}. The method we use here is full forward modeling using the MBProj2 code\footnote{\url{https://github.com/jeremysanders/mbproj2}} \citep{Sanders_2018}. MBProj2 predicts the surface brightness profiles of a cluster in multiple X-ray bands given a model of the gas density and dark matter mass, in the case of a hydrostatic model, or gas density and temperature, without assuming hydrostatic equilibrium.
As the number of energy bands increases, the method is equivalent to spectral fitting, and reasonable numbers of bands give results consistent with spectra \citep{Sanders_2014}.
Forward modeling has the advantage of being able to easily include observational effects within the model. In addition as MBProj2 computes a likelihood for the observed profiles given the model assuming Poisson statistics, it can work directly on data with relatively few X-ray counts.

\section{Synthetic datasets}

We use two sets of cluster models in our analysis. The first is a sample of model galaxy clusters taken from Magneticum Pathfinder simulations we use to investigate clusters with realistic properties. The second consists of simple models of clusters with idealized properties to verify our approach. For all models, we know the theoretical radial profiles for the different quantities. We refer to these profiles as the \qq{true} profiles of the quantities for the individual clusters.

\subsection{Generating synthetic observations}
\label{section_SIXTE}
To make the synthetic eROSITA data we wish to analyze, we take our cluster models and process them through the software package SIXTE (Simulation of X-ray Telescopes)\footnote{\url{http://www.sternwarte.uni-erlangen.de/research/sixte/}}, which is the official end-to-end simulator of eROSITA \citep{Dauser_2019}. This package simulates the instrumental effects appropriate for X-ray observations \citep[see][]{Clerc_2018}. For example, it is able to simulate the effects of the point spread function (PSF) specifically for the instrument. It furthermore considers the correct on-axis ancillary response file (ARF) and redistribution matrix file (RMF) of the detector \citep[see][]{George_2007}. The version we use is SIXTE-2.6.2.10.

We consider two different eROSITA-like configurations when making our synthetic observations.
The first is an idealized setup, designed to look for systematic errors, where we modify the instrumental configuration to have no background (instrumental or X-ray), a long \SI{16}{ks} flat exposure time across the field, no PSF (although the eROSITA pixel size is included), no vignetting, and fixing the effective area curve to match the survey eROSITA average.
The second is a standard eROSITA setup, using realistic exposure times and spacecraft attitude, background, PSF, vignetting and effective area curves.

To further process the outputted synthetic data for analysis, we make use of the eROSITA Science Analysis Software System (eSASS) \citep{Brunner_2018, Brunner_2021, Predehl_2021}. We use the tool called \textit{evtool}\footnote{\url{https://erosita.mpe.mpg.de/edr/DataAnalysis/evtool_doc.html}} to create images of the clusters in different energy bands (for images of a cluster see Sect. \ref{section_Clusters_from_Magneticum_Pathfinder_simulations_in_realistic_observations}).
In these images, each pixel has a width of 4 arcsec.
For realistic observational conditions, we use \textit{expmap}\footnote{\url{https://erosita.mpe.mpg.de/edr/DataAnalysis/expmap_doc.html}} to create exposure maps.

\subsection{Magneticum Pathfinder simulations}
\label{section_Magneticum_Pathfinder_simulations}

The Magneticum Pathfinder simulations\footnote{\url{http://www.magneticum.org}} \citep[see][]{Biffi_2021} are cosmological hydrodynamical simulations of a large cosmic volume. 
A lightcone is constructed from the so-called Box2 at high resolution (\qq{hr}),
which comprises a comoving volume of (352\hspace{1.9pt}$h^{\mathrm{-1}}$cMpc)$^{\mathrm{3}}$, resolved with $2 \times 1584^{3}$ particles.
From this lightcone, we extract a sample of 93 galaxy clusters with masses $M_{\mathrm{500}} > \SI{1e14}{M_{\odot}/\textit{h}}$,
spanning 5 snapshots (hereafter, \qq{snap}) in redshift. The snaps have the numbers 140, 136, 132, 128 and 124. All clusters in one snap have the same redshift (Table \ref{tab:snaps}).
\begin{table}[t]
    \centering
    \caption{The distribution of clusters among five snapshots in the Magneticum Pathfinder simulations.}
    \begin{tabular}{|c|c|c|c|c|c|}\hline
        snap & 140 & 136 & 132 & 128 & 124 \\\hline
        $\sum$ clusters & 4 & 6 & 17 & 27 & 39 \\\hline
        $\approx zz$ & 0.0326 & 0.0663 & 0.1011 & 0.1371 & 0.1742 \\\hline
        bin width & 10 & 5 & 3 & 3 & 2 \\\hline
    \end{tabular}
    \tablefoot{This table shows the properties of clusters from a lightcone extracted from five redshift snapshots (\qq{snaps}) in the Magneticum Pathfinder simulations. The four rows show the snap number, the number of clusters in the corresponding snap, the redshift $zz$ of all clusters in the corresponding snap and the bin width for all the clusters in the snap in pixels. The bin width refers to the width of one bin in the radial surface-brightness profiles of a cluster (see Sect. \ref{section_Fitting_process}).}
    \label{tab:snaps}
\end{table}
The simulations take many essential physical processes into account (e.g. cooling, star formation, winds \citep{Springel_2003, Springel_2003_2, Springel_2003_3}, chemical enrichment \citep{Tornatore_2004, Tornatore_2007} and many others; see the webpage for more detail). A standard $\Lambda$CDM model and a cosmology based on the seven-year Wilkinson Microwave Anisotropy Probe \citep[WMAP7][]{Komatsu_2011} is assumed, with a total matter density $\Omega_{\mathrm{M}} = 0.272$, dark energy with $\Omega_{\mathrm{\Lambda}} = 0.728$, a baryon density $\Omega_{\mathrm{b}} = 0.0451$, a Hubble parameter of $h = 0.704$ and an overall normalization of the power spectrum of $\sigma_{\mathrm{8}} = 0.809$.

To calculate the projected emission spectra of the ICM in the X-ray band, the program PHOX is used \citep{Biffi_2012, Biffi_2013, Biffi_2018}, which is based on XSPEC \citep{Arnaud_1996}. This emission is computed given the temperature, density and chemical abundances of the gas elements in the simulated clusters.
Absorption with an equivalent Hydrogen column density of $N_\mathrm{H}=\SI{0.01e22}{cm^{-2}}$ \citep{Morrison_1983} is included.
The output photon lists are used as input to SIXTE to generate the observed counts.
For these clusters we generate synthetic observations using both the idealized (Sect. \ref{section_Clusters_from_Magneticum_Pathfinder_simulations_in_idealized_observations}) and realistic eROSITA configurations (Sect. \ref{section_Clusters_from_Magneticum_Pathfinder_simulations_in_realistic_observations}).

\subsection{Idealized cluster models}
\label{section_Ideal_cluster_simulations}

We model two different kinds of spherically-symmetric clusters with idealized properties:
\begin{itemize}
    \item Isothermal clusters with six different temperatures:\\
    0.5, 1.0, 2.0, 4.0, 8.0 and \SI{12.0}{keV}.
    \item A cluster which perfectly fulfills hydrostatic equilibrium.
\end{itemize}
All these clusters models are generated assuming a standard $\Lambda$CDM model with the following parameters: $\Omega_{\mathrm{M}} = 0.3$, $\Omega_{\mathrm{\Lambda}} = 0.7$ and $H_{\mathrm{0}} = 70 \hspace{0.08cm} \text{(km/s)/Mpc}$. We assume a redshift of $z = 0.1$. All models assume an electron-density profile of
\begin{align}
    n_{\mathrm{e}}^{2} = n_{\mathrm{0}}^{2} \frac{(r/r_{\mathrm{c}})^{-\alpha}}{\left(1+r^{2}/r_{\mathrm{c}}^{2}\right)^{3 \beta - \alpha /2}},
    \label{ideal_simulations_model_ne}
\end{align}
where the parameters are $n_{\mathrm{0}} = \SI{0.005}{cm^{-3}}$, $r_{\mathrm{c}} = \SI{300}{kpc}$, $\alpha = 0.2$ and $\beta = 2/3$ (see Sects. \ref{section_Modified_beta_model} and \citealt{Vikhlinin06} for details of the functional form).

The input to SIXTE is a table of spectra, simulated using XSPEC, emitted from a set of 3D shells in each model cluster out to 3 Mpc.
For these idealized clusters we only simulate using the idealized eROSITA configuration.
The spectra were simulated assuming a metallicity of $0.3$~Z$_{\odot}$ \citep{Anders_1989} and absorption associated with an equivalent Hydrogen column density $N_\mathrm{H}=\SI{0.02e22}{cm^{-2}}$ \citep{Balucinska-Church_1992}.
In the isothermal case, for each cluster the shells have the same spectra but different normalizations.
In each cluster, these shells and their spectra are simulated using an associated input image giving the distribution of projected emission from that shell.

For the ideal hydrostatic cluster, we assume an NFW dark matter profile \citep{Navarro_1996}, where $r_{\mathrm{200, DM}}$, the average overdensity of the dark matter is 200 times the critical density of the universe at the corresponding redshift, is $\SI{1.2}{Mpc}$, and the concentration, $c$, is $5$.
The temperature for the shells is calculated assuming hydrostatic equilibrium given the total mass profile.
In detail, we work from the outermost shell, computing the contributions to the increasing pressure given the mass profile and gas density profile.
The temperature profile is computed from the final pressure and density profiles assuming an ideal gas.

\section{Data analysis}
\label{section_Data_analysis}

\subsection{Estimating the position of the center of the cluster}
\label{section_Estimating_the_position_of_the_center_of_the_cluster}

We fit models to the radial binned count profiles of the simulated galaxy clusters using the code MBProj2 (Sect. \ref{section_Fitting_process}). These profiles are extracted from images of the clusters (see Sect. \ref{section_Clusters_from_Magneticum_Pathfinder_simulations_in_realistic_observations}), given positions of their centers. We use a three-step process to find the cluster centers from idealized cluster images, created between energies of $0.3$ and \SI{9.0}{keV}, assuming a constant flat exposure time of \SI{16000}{s}, and without including masking, background and a realistic PSF. The procedure is as follows
\begin{itemize}
    \item Given a starting point (initially the center of the image in the first iteration), we construct a circle with a radius of 1000 pixels. By computing the weighted mean of the encircled counts, we determine an updated center position of the circle. We repeat this procedure until the position of the center of the circle does not change any more.
    \item We reduce the radius of the circle to 50 pixels, and use the latest value of the former step as the initial center of the circle. We then repeat the mechanism of the former step.
    \item Finally we reduce further the radius of the circle to 15 pixels, and again repeat the above centroid finding. We consider the final result to be the center of the cluster.
\end{itemize}
Using three steps instead of one makes the final result more independent from the starting point. Although the true minima of the potential wells of the individual clusters are known from the simulations, we use this estimate to be closer to real observational analyses. The determined center positions for the majority of the clusters are in agreement with an estimation by eye.

There is one cluster in snap 128 for which our approach leads to an obviously aberrant center position. For this cluster, we replace the three steps of our mechanism by one step with a circle with a radius of 10 pixels.

\subsection{Masking}
\label{section_Masking}

Bright substructures in the outskirts of the clusters prevent the MBProj2 fitting procedure from fitting realistic forms of the profiles (see Sect. \ref{section_Fitting_process} for more information).
Therefore we apply a masking procedure to the clusters from the Magneticum Pathfinder simulations. We manually detect and mask obvious substructures in the outer parts of 80 of the 93 clusters of our sample. The masking is based on looking at the same idealized images used when estimating the position of the centers (Sect. \ref{section_Estimating_the_position_of_the_center_of_the_cluster}).
If for a cluster there are several bright regions near the probable center and one cannot determine which one is the actual center, neither of these regions are masked.
For idealized isothermal and hydrostatic clusters, no masking is applied.

\subsection{Fitting process}
\label{section_Fitting_process}

The fitting code MBProj2 is a forward modeling code that is able to fit surface-brightness profiles in multiple energy bands \citep{Sanders_2018, Sanders_2014}. The code fits the number of counts in radial annular bins, which are obtained from images (see the images in Sect. \ref{section_Clusters_from_Magneticum_Pathfinder_simulations_in_realistic_observations}). To do this, it is necessary to assume a certain center position (see Sect. \ref{section_Estimating_the_position_of_the_center_of_the_cluster}) and a maximum radius. In our analysis, this maximum radius is set to a value so that the true $r_{\mathrm{200}}$ lies within the outermost three fitted bins \citep[compare to][]{Ettori_2013}. However, for three out of the four clusters in snap 140, where the image does not extend to $r_{200}$, we take the maximum possible radius that ensures the whole fitted region to be covered by the image.
The bin width depends on the snap of the cluster and can be seen in Table \ref{tab:snaps}. It is chosen to ensure that all clusters have a comparable number of bins.

Given an input model (either hydrostatic or non-hydrostatic), the code, given exposure times, bin radius and area, predicts the number of counts in the bin in different energy bands. Comparing calculated and observed numbers of counts allows the model to be fitted to the data, by maximizing the Poisson likelihood.

For our analysis, we fitted a hydrostatic model to the clusters, which requires assuming a profile for the electron density in the gas, a dark matter profile, a metallicity profile and an outer pressure. First, the code calculates the pressure profile inwards by summing up the corresponding contributions given the assumption of hydrostatic equilibrium. The gas density is characterized by an electron density profile $n_{\mathrm{e}} (r)$ that is modeled by either a radially-binned or a parametric model. Given a parameterized model for the dark matter mass profile, the gravitational acceleration is calculated. With the ideal gas law, gas density and pressure are used to calculate the temperature. Using the metallicity and the absorbing column density $N_{\mathrm{H}}$, the code is able to determine the X-ray emissivity in various energy bands. To match the input simulations, our analysis of the clusters from the Magneticum Pathfinder simulations (Sect. \ref{section_Magneticum_Pathfinder_simulations}) uses XSPEC version 12.11.0 \citep{Arnaud_1996}, the plasma emission model APEC version 2.0.1 \citep{Smith_2001} and the photoelectric absorption model WABS \citep{Morrison_1983}. The analysis of the clusters with idealized properties (Sect. \ref{section_Ideal_cluster_simulations}) uses APEC version 2.0.2 and PHABS \citep{Balucinska-Church_1992}. The three-dimensional emissivities are projected onto the sky to compute the model count profiles.

Following the initial fit an MCMC (Markov Chain Monte Carlo) analysis is used to examine the uncertainties of the profiles \citep{Goodman_2010, Foreman-Mackey_2013, Bryan_1998}.
For this, we use a chain length of 15\,000 steps; for the burn-in period, we use 5000 steps. The number of walkers we use is calculated as four times the number of all parameters in our hydrostatic model. For a detailed explanation of the functionality of MBProj2 we refer to \cite{Sanders_2018} and the webpage of MBProj2\footnote{\url{https://mbproj2.readthedocs.io/en/latest/README.html}}.

For our analysis, we use eight energy bands between edges of 0.3, 0.6, 1.1, 1.6, 2.2, 3.5, 5.0, 7.0 and \SI{9.0}{keV}, unless stated otherwise. These edges are chosen to span the energy range of eROSITA, but also align to energies where the eROSITA effective area changes \citep{Predehl_2021}.

\subsection{Modeling}

In order to fit a hydrostatic model with MBProj2, models describing the gas electron density profile (also simply called density profile), dark matter mass profile, metallicity and outer pressure have to be chosen. We use a constant flat metallicity value of \SI{0.3}{Z_{\odot}} \citep{Anders_1989}. The outer pressure is parameterized logarithmically with a default value of \SI{e-14}{erg/cm^{3}}, a minimum possible value of \SI{e-16}{erg/cm^{3}} and a maximum possible value of \SI{e-8}{erg/cm^{3}}. The different models we use for density and dark matter profiles are listed in the following subsections. To avoid problems in the outermost fitted bins, which can appear because the cluster model does not extent beyond the outer radius of the data, we exclude the outer four bins in the fitting process, independently of the chosen models. This does not hold for the isothermal clusters discussed in Sect. \ref{section_Ideal_clusters}.

\subsubsection{Modified $\beta$-model}
\label{section_Modified_beta_model}

A popular parameterization of the gas density profile of a cluster is a modified $\beta$-model \citep{Vikhlinin06}. A simple $\beta$-model \citep{Cavaliere_1978} is modified to be a more realistic description of real clusters \citep[compare to][]{Pointecouteau_2004, Vikhlinin_1999}, and includes an inner core and break in slope at larger radius. The model gives the number density of electrons $n_{\mathrm{e}}$ as
\begin{align}
    n_{\mathrm{e}}^{2} = n_{\mathrm{0}}^{2} \frac{(r/r_{\mathrm{c}})^{-\alpha}}{\left(1+r^{2}/r_{\mathrm{c}}^{2}\right)^{3 \beta - \alpha /2}} \frac{1}{\left(1+r^{\gamma}/r_{\mathrm{s}}^{\gamma}\right)^{\epsilon/\gamma}} + \frac{n_{\mathrm{02}}^{2}}{\left(1+r^{2}/r_{\mathrm{c2}}^{2}\right)^{3 \beta_{2}}}
    \label{Vikhlinin_model_formula}.
\end{align}
The second summand in Eq. \ref{Vikhlinin_model_formula} models a second $\beta$-component, and can optionally be included in MBProj2. For the fit, we assume a frozen value of $\gamma=3$. We use flat, wide priors for the remaining parameters. The minimum allowed value of $\epsilon$ is $-5$. We refer to this model as a modified $\beta$-model (abbreviated mod. $\beta$).

\subsubsection{Interpolation model}

A way to more freely model the gas density is by a simple interpolation model. We take several radial control points, with the density at these radii parameters in the fit, and use interpolation to find the density in between. The interpolation is performed in logarithmic space and includes spline smoothing \citep{Schoenberg_1964}. After testing this model, we found it advantageous to use ten radial control points $r_{\mathrm{i}}$ with a square root radial scaling:
\begin{align}
    r_{\mathrm{i}} = \frac{\sqrt{10}-\sqrt{9-i}}{\sqrt{10}}\cdot r_{\mathrm{max, interp}},
\end{align}
where $i$ is an integer number from $0$ to $9$ and $r_{\mathrm{9}} = r_{\mathrm{max, interp}}$ is the maximum radius of the interpolation model. This radius is located in the outermost bin of the profile that is not excluded to mitigate the edge effect. The radii $r_{\mathrm{i}}$ are fixed during the fitting process. As it is important to have plausible starting values for the interpolation model before starting the fitting process, as the fit can become trapped in local minima, these were found from a modified $\beta$ model with a single $\beta$ component initially fitted to the data assuming isothermality.

\subsubsection{Binned model}

The third model we use for the gas density is a binned model.
In this model, the density profile is assumed to be constant within each radial bin, where the radial bins are at the same annuli as the input profiles (see Table \ref{tab:snaps} and Sect. \ref{section_Fitting_process}).
In order to ensure plausible starting values, the density is first estimated in MBProj2 assuming the cluster to be isothermal.

\subsubsection{NFW model}
\label{section_NFW_model}

One way to model the dark matter content of a cluster is the NFW model \citep{Navarro_1996}. It models the dark matter mass density as
\begin{align}
    \rho (r) = \frac{\delta_{\mathrm{c}} \rho_{\mathrm{c}}}{(r/r_{\mathrm{s}})(1+r/r_{\mathrm{s}})^2},
\end{align}
where $\rho_{\mathrm{c}} = (3 H^{2}(z))/(8 \pi G)$ is the critical density of the universe at redshift $z$, $r_{\mathrm{s}} = r_{\mathrm{200, DM}}/c$ is the scale radius, $c$ is the concentration and $\delta_{\mathrm{c}}$ is the characteristic overdensity of the halo, defined by
\begin{align}
    \delta_{\mathrm{c}} = \frac{200}{3} \frac{c^{3}}{\log(1+c) - c/(1+c)}.
\end{align}
Here, $H(z)$ is the Hubble constant at redshift $z$ and $G$ is the gravitational constant.
Note that the radius $r_{\mathrm{200,DM}}$ is not equivalent to $r_{\mathrm{200}}$, as the former refers to the dark matter, while the latter refers to the total mass. For our analysis, we constrain that $-2 \leq \log_{\mathrm{10}}(c) \leq 2$ and $-1 \leq \log_{\mathrm{10}}\left(r_{\mathrm{200, DM}}/\text{Mpc}\right) \leq 1$.

\subsubsection{GNFW model}

The generalized NFW model, also called GNFW model, extends the NFW model with an inner slope parameter $\alpha$:
\begin{align}
    \rho (r) = \frac{\rho_{\mathrm{0}}}{(r/r_{\mathrm{s}})^{\alpha} (1+r/r_{\mathrm{s}})^{3-\alpha}},
\end{align}
with the central density this time denoted as $\rho_{\mathrm{0}}$ \citep{Zhao_1996, Wyithe_2001, Schmidt_2007}. We use a flat prior of $0 \leq \alpha \leq 2.5$; with the remaining parameter priors chosen to be the same as the NFW model (see Sect. \ref{section_NFW_model}).

\subsubsection{Einasto model}

The third model we use to describe the dark matter mass density is the Einasto model with the functional form
\begin{align}
    \rho (r) = \rho_{\mathrm{s}} \exp{\left \lbrace -d_{\mathrm{n}} \left[\left(\frac{r}{r_{\mathrm{s}}}\right)^{1/n}-1\right]\right \rbrace}
    \label{Einasto_model_formula}
\end{align}
\citep{Retana-Montenegro_2012}. The positive number $n$, the Einasto index, defines the steepness of the profile. A sphere with the radius $r_{\mathrm{s}}$ encloses half of the integrated dark matter mass. The corresponding dark matter mass density at $r = r_{\mathrm{s}}$ is $\rho_{\mathrm{s}}$. $d_{\mathrm{n}}$ is a numerical constant that ensures $r_{\mathrm{s}}$ to be the true half-mass radius. We use the following flat priors: $12 \leq \log_{\mathrm{10}}\left(M_{\mathrm{tot, DM}}/\text{M}_{\odot}\right) \leq 16$, $0 \leq n \leq 20$ and $-1.3 \leq \log_{\mathrm{10}}\left(r_{\mathrm{s}}/\text{Mpc}\right) \leq 0.7$.

\subsection{Comparing profiles from simulations and observations}
\label{section_Comparing_profiles_from_simulations_and_observations}

The simulations provide the true profiles for the gas mass density (in g/cm$^{3}$), the temperature and the cumulative total mass.
We assume a hydrogen mass fraction of roughly 0.76 \citep[compare to e.g.][]{Anders_1989} and a proton mass of approximately \SI{1.0073}{u} \citep{Tanabashi_2018}. Furthermore, we assume that the gas entirely consists of hydrogen and helium and calculate the electron number density as $n_{\mathrm{e}} = n_{\mathrm{H}} + 2n_{\mathrm{He}}$. The electron pressure is calculated as electron number density times temperature.

The ICM can have more than one local temperature phase, which can cause difficulties while comparing spectral temperatures $T_{\mathrm{spec}}$ to the true temperature profiles provided by numerical simulations, as the weighting of the temperatures in each phase is important \citep{Boehringer_2009}. For the comparison of the temperature profiles for the clusters from the Magneticum Pathfinder simulations, we use the so-called spectroscopic-like temperature $T_{\mathrm{sl}}$, which was suggested by \cite{Mazzotta_2004} for observations with \textit{Chandra} and XMM-\textit{Newton}. For these instruments, the discrepancies between $T_{\mathrm{spec}}$ and $T_{\mathrm{sl}}$ is less than a few percent for temperatures above \SI{3}{keV}. However, larger deviations for eROSITA data could be conceivable as $T_{\mathrm{spec}}$ is instrument dependent.

When comparing the fitted temperature profile to the corresponding true profile it should be remembered that the positions used for the cluster centers may be different. While the fit assumes the center to be at the position determined as explained in Sect. \ref{section_Estimating_the_position_of_the_center_of_the_cluster}, the true profiles assume the center to be at the true minimum of the potential well of a cluster. An offset between these positions is referred to as \qq{shift of the center} and further discussed in Sect. \ref{section_Influence_of_the_shift_of_the_center}.


\subsection{Hydrostatic mass profiles based on theory}
\label{section_Hydrostatic_mass_profiles_based_on_theory}

From the simulations, we know the true profiles for electron gas density and the mass-weighted electron pressure in each cluster. These profiles and the assumption of hydrostatic equilibrium allow us to compute cumulative total mass profiles for the clusters. We use these profiles to evaluate the validity of hydrostatic equilibrium independently of a spectroscopic analysis and our fit with MBProj2. 
For our calculation, we fit a modified $\beta$-model with two $\beta$-components to the mass-weighted electron pressure profile. This fit is necessary because the pressure gradient is noisy in the original profiles. The fit is performed independently of MBProj2.
For the calculation of the hydrostatic mass profiles, we take into account the pressure components of hydrogen and helium nuclei as well as free electrons.

\section{Results}

\subsection{Ideal clusters}
\label{section_Ideal_clusters}

We test the validity of our approach with the idealized clusters (Sect. \ref{section_Ideal_cluster_simulations}), simulated with the idealized eROSITA configuration of flat exposure, no background and no PSF (Sect. \ref{section_SIXTE}).

First, we use the six simulated isothermal clusters to verify the temperature determination of MBProj2. The fit does not assume hydrostatic equilibrium, but fits the temperature and profiles simultaneously. The density is fit by a modified $\beta$-model with one $\beta$-component. We assume a flat profile for the temperature, while the minimum allowed value is \SI{0.1}{keV}, the maximum value is \SI{60}{keV} and the default value is \SI{4.0}{keV}. Deviating from Sect. \ref{section_Fitting_process}, we use only six energy bands for the cluster with \SI{0.5}{keV} because the two bands with highest energies do not include any counts. We find the deviations of fitted temperature and true temperature to be about 1\% of the true value or below for all fitted isothermal clusters except the one with a temperature of \SI{12.0}{keV}; for this cluster, the deviation is about 3.8\%.

\begin{figure}
    \centering
    \subfigure[different gas density models]{\includegraphics[width=0.24\textwidth]{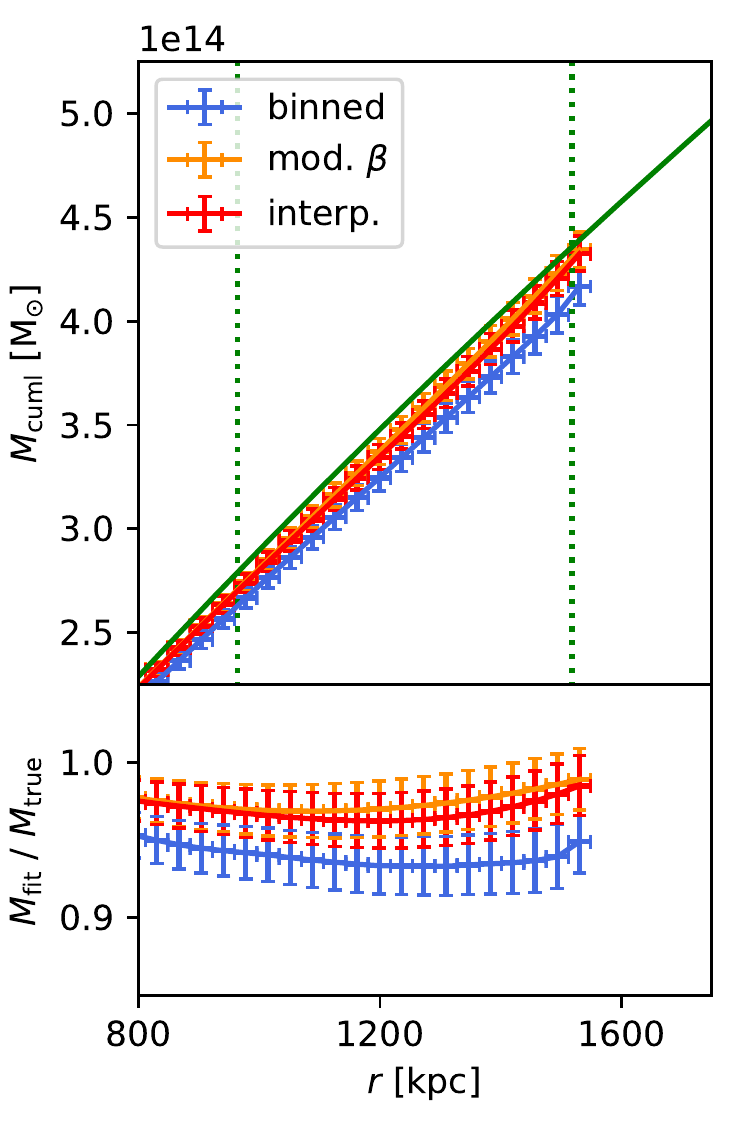}}
    \subfigure[different dark matter mass models]{\includegraphics[width=0.24\textwidth]{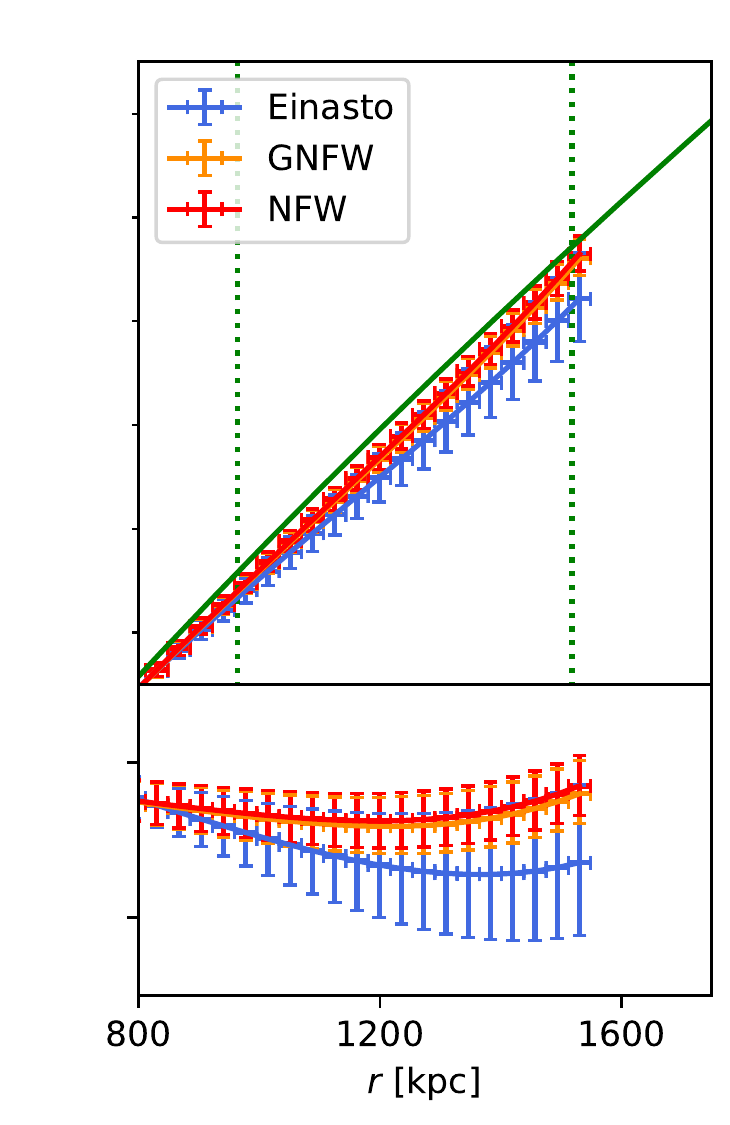}}
    \subfigure[different binnings]{\includegraphics[width=0.24\textwidth]{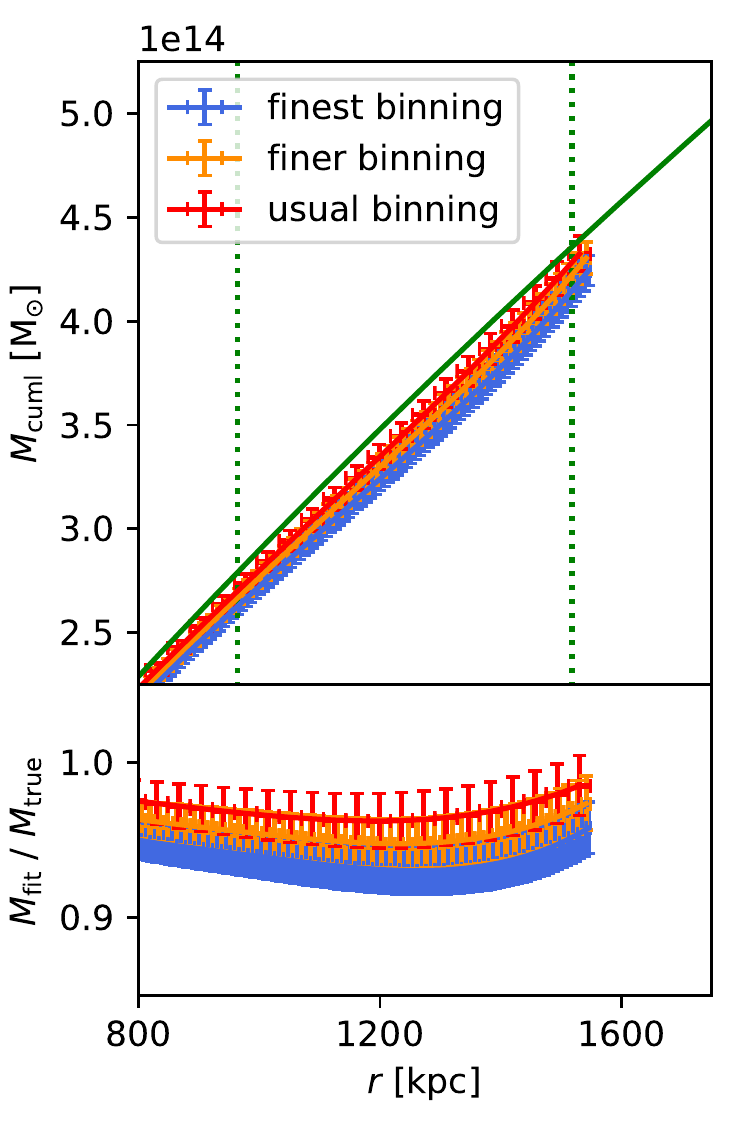}}
    \subfigure[different maximum radii of the fit]{\includegraphics[width=0.24\textwidth]{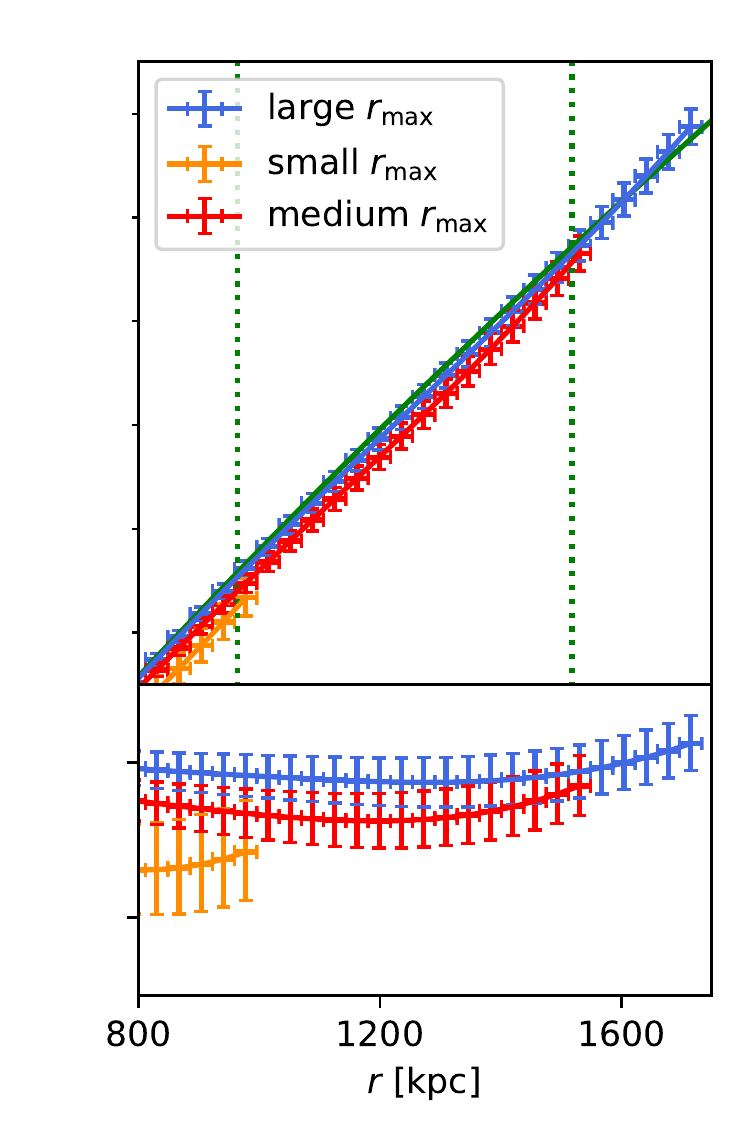}}
    \caption{Ideal clusters: upper panels: cumulative total mass (calculated as the sum of gas mass and dark matter mass) with different models, binnings and maximum radii for the cluster fulfilling perfect hydrostatic equilibrium. The green continuous line shows the true profile according to the simulations. The green vertical dotted lines show $r_{\mathrm{500}}$ and $r_{\mathrm{200}}$, computed based on the true profiles and the characteristic overdensity of the universe at the corresponding redshift. The value of $r_{\mathrm{200}}$ is remarkably larger than stated in Sect. \ref{section_Ideal_cluster_simulations}, as the gas mass accounts for a considerable part of the total mass for the cluster fulfilling perfect hydrostatic equilibrium. Lower panels: ratio of fitted cumulative total mass and corresponding true mass.}
    \label{fig:perfect_hydro}
\end{figure}

Second, we test the validity of our method on an idealized cluster that fulfills perfect hydrostatic equilibrium. We investigate the cumulative total mass (calculated as sum of gas mass and dark matter mass; compare to \cite{Chiu_2016}) testing different density models (binned model, modified $\beta$-model, interpolation model), dark matter mass models (Einasto model, NFW model, GNFW model), binnings and different choices of the maximum radius of the fit (Fig. \ref{fig:perfect_hydro}). For all tested models and choices, we find deviations at $r_{\mathrm{500}}$ and $r_{\mathrm{200}}$ that are smaller than 7\% of the true value. We can thus conclude that all the tested models are appropriate to describe the quantities of the cluster fulfilling perfect hydrostatic equilibrium in the outer parts of the cluster and that the tested variations of other factors do not have a decisive influence on the total cumulative mass. Because of the rather symmetric structure of this idealized cluster, the determined position of the center (see Sect. \ref{section_Estimating_the_position_of_the_center_of_the_cluster}) is only about \SI{8.4}{kpc} away from the true minimum of the gravitational potential according to the simulations. This allows a meaningful comparison of fitted profiles and true profile.

\subsection{Clusters from Magneticum Pathfinder simulations in idealized observations}
\label{section_Clusters_from_Magneticum_Pathfinder_simulations_in_idealized_observations}

In this section, we analyze the clusters from the Magneticum Pathfinder simulations under the idealized observational conditions. We use the input galactic absorption density of $N_\mathrm{H} = \SI{0.01e22}{cm^{-2}}$. Deviating from Sect. \ref{section_Fitting_process}, we only use seven energy bands for five clusters in snap 124 because the band with highest energy does not include any counts.

We are particular interested in comparing the fitted and true mass profiles. Therefore we define the parameter
\begin{align}
    \zeta = \frac{\text{cumulative hydrostatic mass}}{\text{cumulative true mass}},
    \label{equation_zeta}
\end{align}
which we explore at certain radii. The values at the true $r_{\mathrm{500}}$ and $r_{\mathrm{200}}$ are called $\zeta_{\mathrm{500}}$ and $\zeta_{\mathrm{200}}$, respectively. We also write $\zeta_{\mathrm{x}}$ for the ratio $\zeta$ at the corresponding $r_{\mathrm{x}}$.
The fitted cumulative total masses at the exact values of the true $r_{\mathrm{500}}$ and $r_{\mathrm{200}}$ are calculated via linear interpolation between the values at the centers of the neighboring bins. The corresponding uncertainties are determined via linear interpolation between the lower (upper) limits of the uncertainties of the mass values at the neighboring bins.
As mentioned in Sect. \ref{section_Fitting_process}, for three of four clusters in snap 140 the fitted region cannot extend to the true $r_{\mathrm{200}}$. These clusters have the numbers 323, 89 and 83. The maximum radius of the fit lies somewhere between the true $r_{\mathrm{500}}$ and $r_{\mathrm{200}}$ in these cases. We therefore use a linear extrapolation of the fitted cumulative total mass profiles in logarithmic space to compute $\zeta_{\mathrm{200}}$.

\subsubsection{Different models and other factors}

\begin{figure*}
    \centering
    \includegraphics[width=0.14565\textwidth]{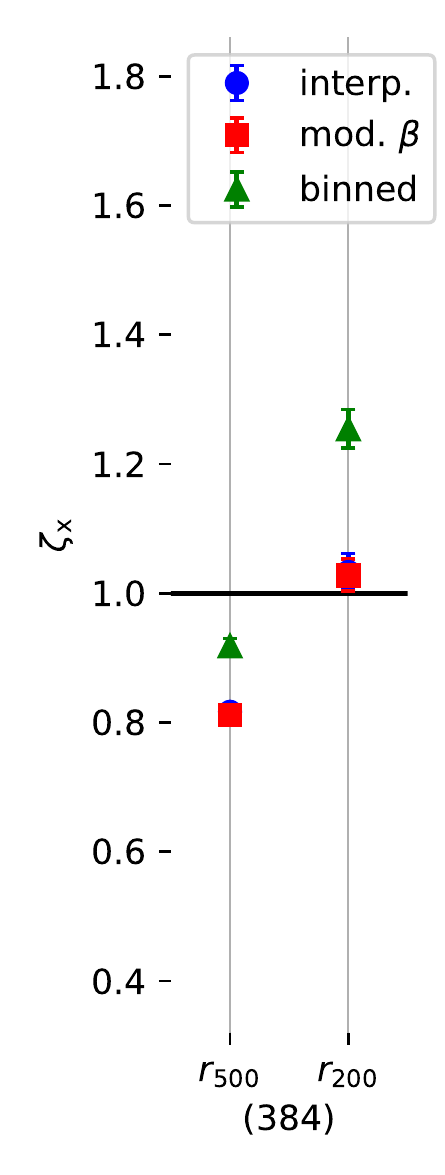}
    \includegraphics[width=0.102\textwidth]{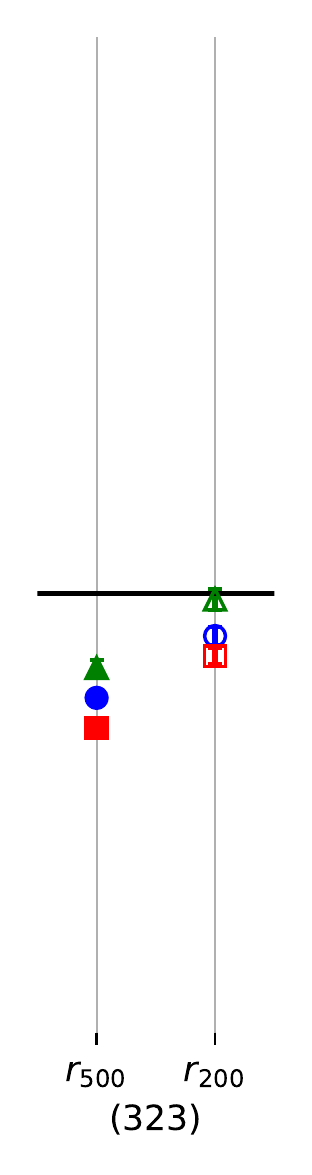}
    \includegraphics[width=0.102\textwidth]{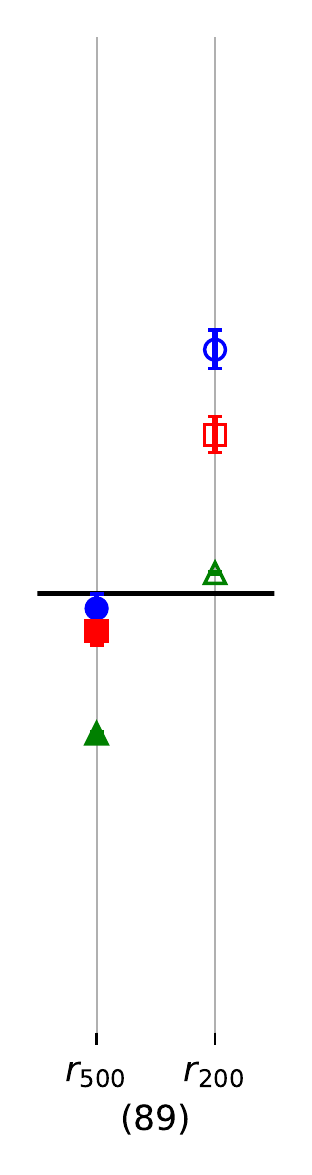}
    \includegraphics[width=0.102\textwidth]{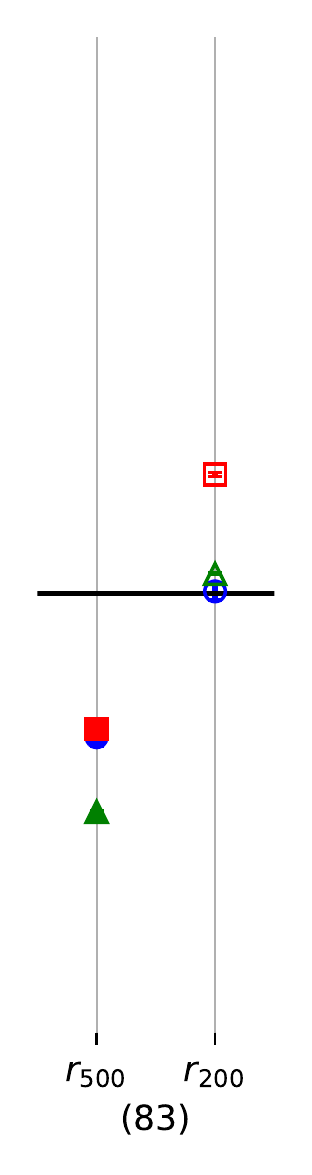}
    \includegraphics[width=0.14565\textwidth]{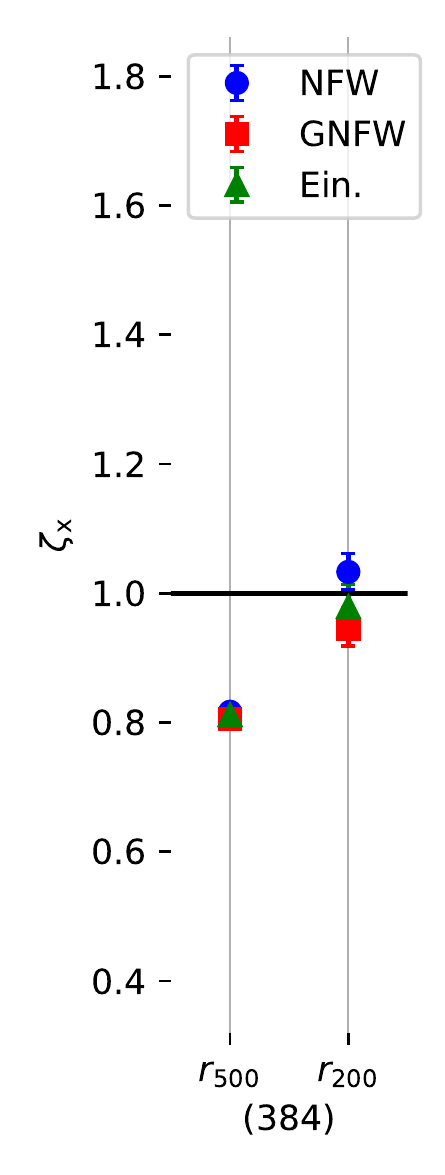}
    \includegraphics[width=0.102\textwidth]{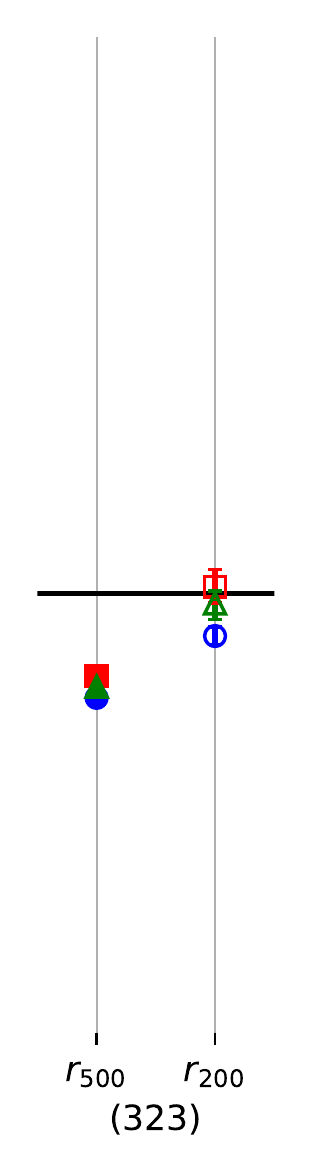}
    \includegraphics[width=0.102\textwidth]{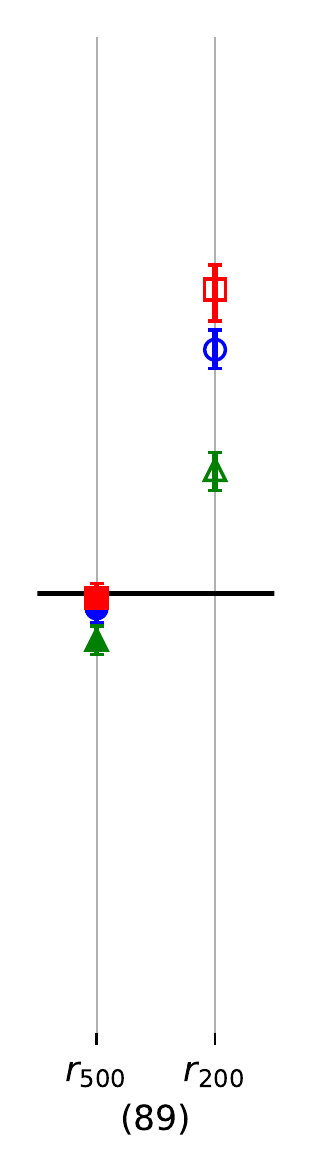}
    \includegraphics[width=0.102\textwidth]{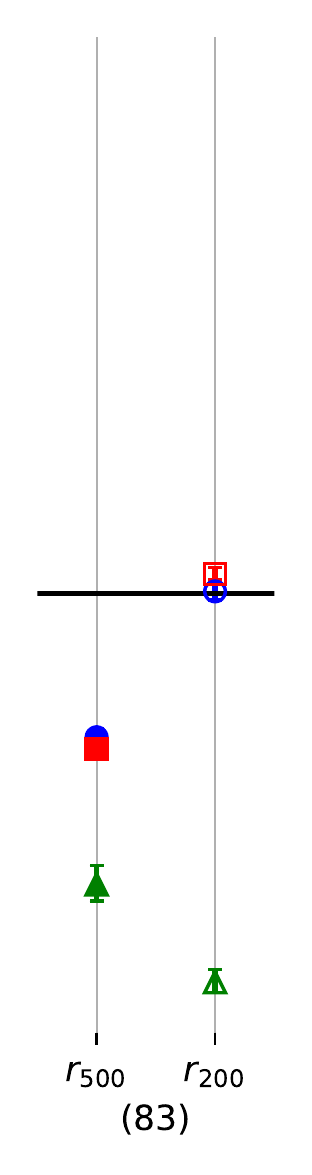}
    
    \includegraphics[width=0.14565\textwidth]{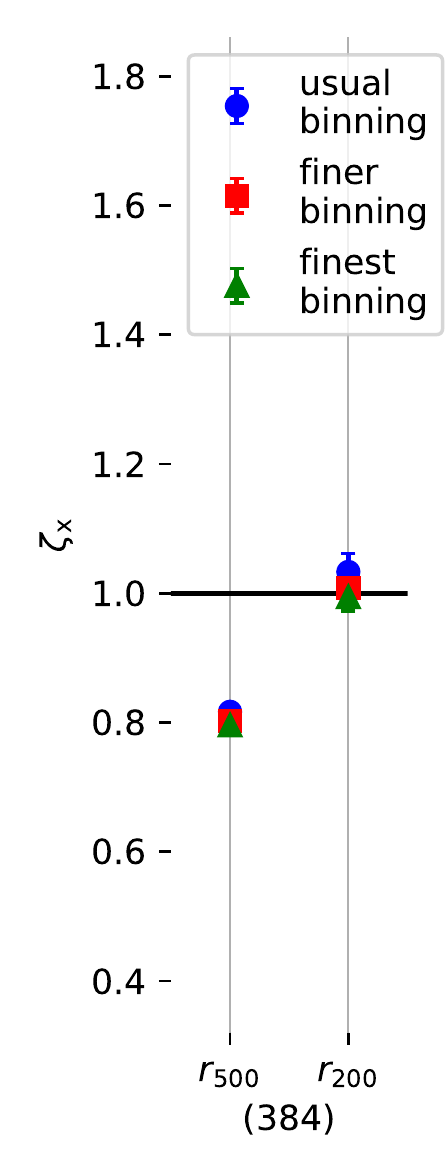}
    \includegraphics[width=0.102\textwidth]{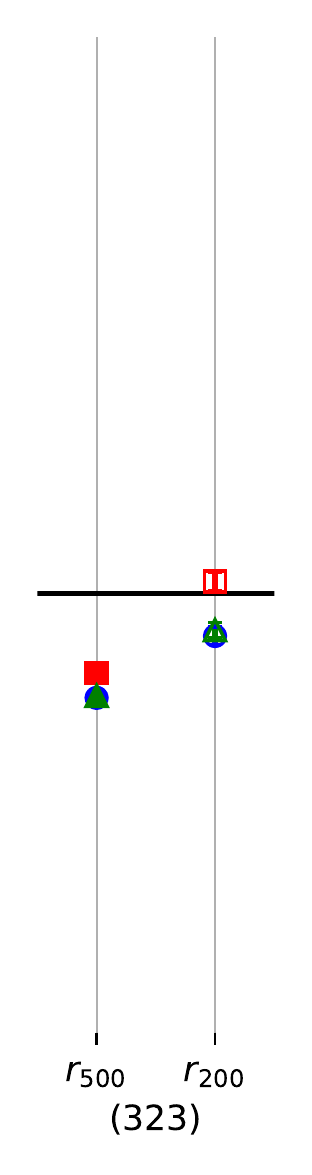}
    \includegraphics[width=0.102\textwidth]{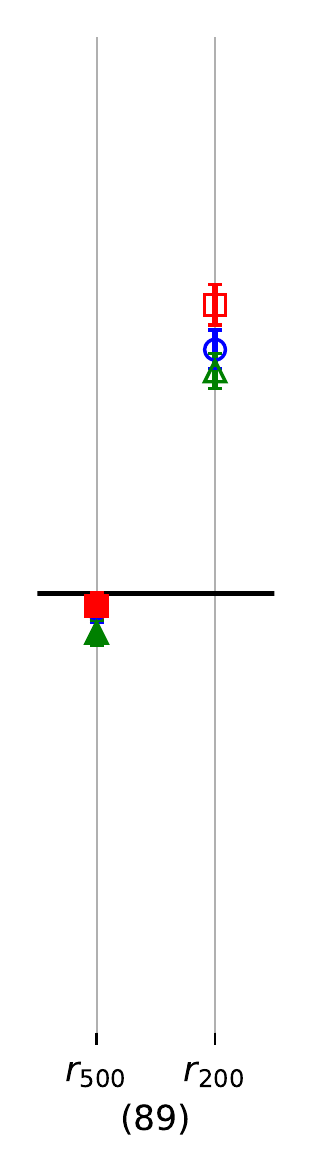}
    \includegraphics[width=0.102\textwidth]{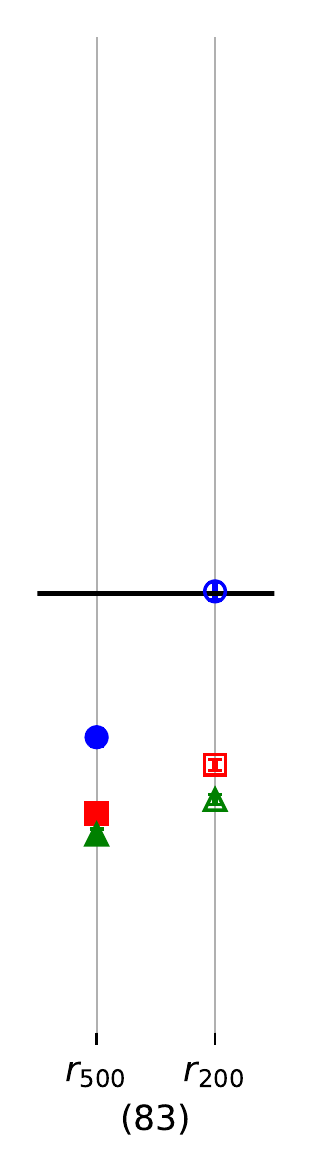}
    \includegraphics[width=0.14565\textwidth]{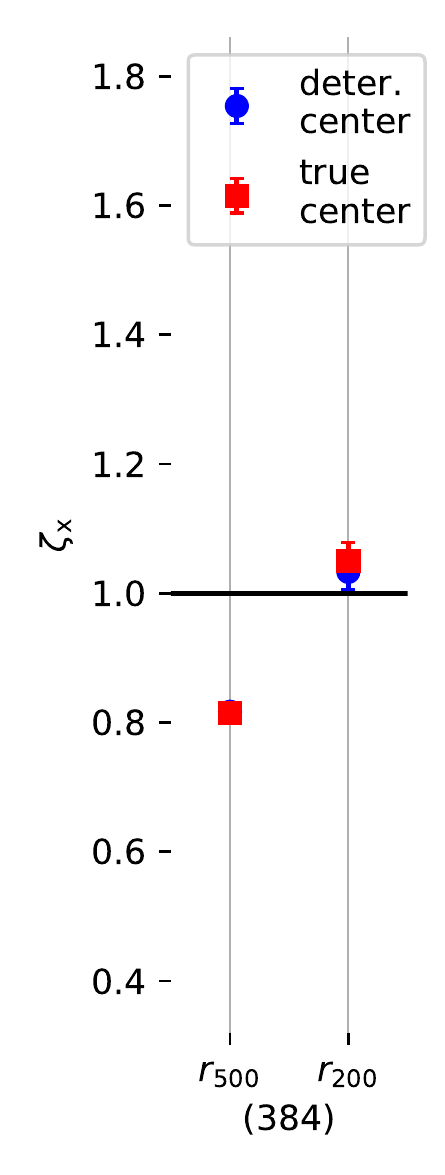}
    \includegraphics[width=0.102\textwidth]{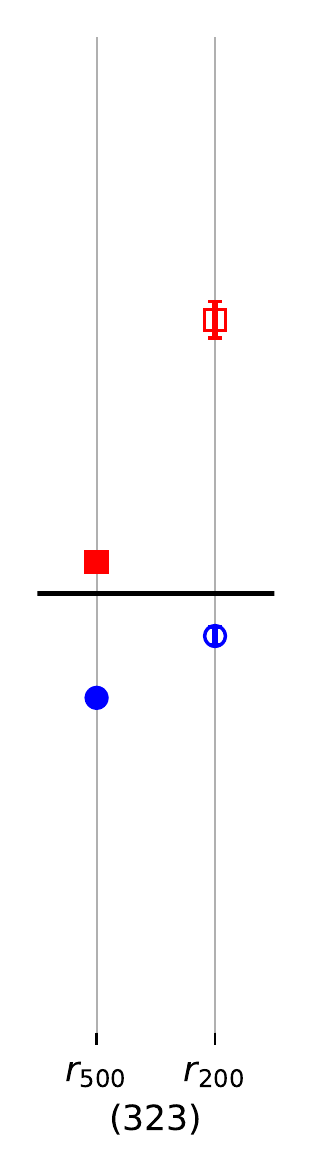}
    \includegraphics[width=0.102\textwidth]{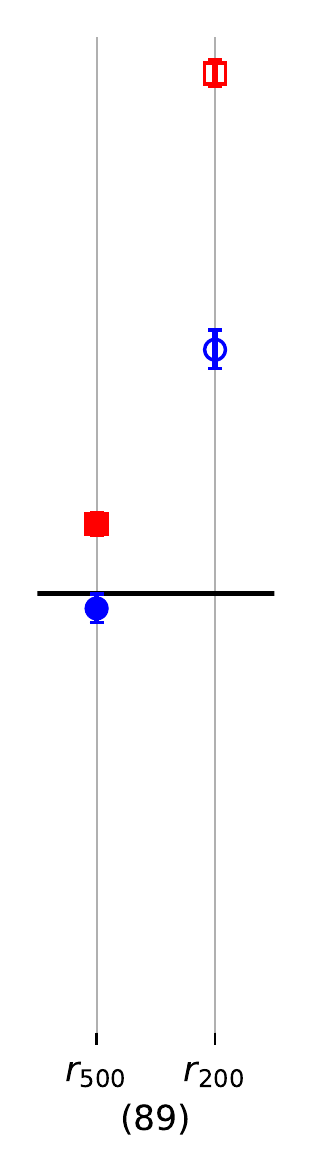}
    \includegraphics[width=0.102\textwidth]{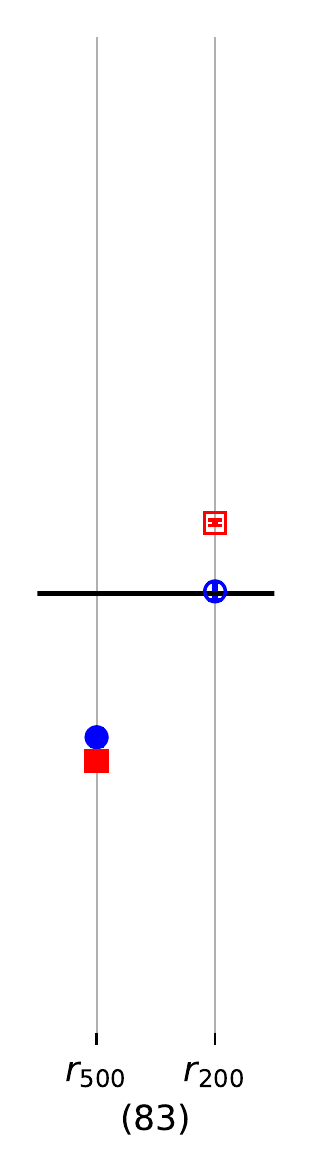}
    
    \caption{Idealized observations: $\zeta_{\mathrm{x}}$ at the corresponding true $r_{\mathrm{x}}$ varying models and other factors (different gas electron density models: interpolation model, modified $\beta$-model with one $\beta$-component, binned model (upper left); different dark matter mass models: NFW model, GNFW model, Einasto model (upper right); different binnings (lower left); different center positions for the fit: determined center according to Sect. \ref{section_Estimating_the_position_of_the_center_of_the_cluster}, true position of the minimum of the gravitational potential according to the simulations (lower right)). If not stated otherwise, we assume an interpolation model for the gas electron density and an NFW model for the dark matter mass. The data points at the true $r_{\mathrm{200}}$ that are based on an extrapolation of the fitted profile are shown as empty symbols. More information can be found in the text.}
    \label{fig:models_and_other_factors}
\end{figure*}

Similar to the approach for the idealized clusters (see Sect. \ref{section_Ideal_clusters}), we analyze the effects of different models and other factors also for the clusters in snap 140 from the Magneticum Pathfinder simulations. These are the clusters with the lowest redshift. Fig. \ref{fig:models_and_other_factors} shows the values of $\zeta$ at the true $r_{\mathrm{500}}$ and $r_{\mathrm{200}}$. Those values at $r_{\mathrm{200}}$ that are based on an extrapolation of the profiles are shown as empty symbols. At $r_{\mathrm{500}}$, no extrapolation is necessary, as the fitted profiles reach this radius. Table \ref{tab:models_and_other_factors} lists the individual values.
From Fig. \ref{fig:models_and_other_factors} one can see that the differences between the results using the interpolation model and the modified $\beta$-model are rather small at $r_{\mathrm{500}}$, whereas the results using the binned model diverge most from the others. The figure also shows that the results with the NFW and GNFW model do not differ much at $r_{\mathrm{500}}$ or $r_{\mathrm{200}}$. The small discrepancies might be due to the similar structure of the models.
Furthermore, Fig. \ref{fig:models_and_other_factors} shows the results with three different binnings. In the usual binning, one bin has a width of 10 pixels in radial direction for snap 140 (the width of one pixel is 4 arcsec). In the finer binning, one bin has a width of 5 pixels, and in the finest binning, one bin has a width of 2 pixels. The total number of bins is then determined by the maximum radius of the fit; in the usual binning, it lies between 44 and 47 for the four clusters in snap 140. It can be seen that finer binnings do not guarantee better results.
To investigate the influence of the position of the chosen center of the cluster, we repeat the fit taking the true minimum of the gravitational potential according to the simulations as the assumed center of the cluster (Fig. \ref{fig:models_and_other_factors}). The results assuming the true center position do not show systematically lower discrepancies from the ideal value one than the one assuming the determined center according to Sect. \ref{section_Estimating_the_position_of_the_center_of_the_cluster}. This indicates that choosing the region with the highest X-ray luminosity to be the center of the cluster fits well in the fitting process with MBProj2, probably as this choice of the center leads to a smoother decrease in the X-ray surface brightness with increasing radius.

For the future analysis of the clusters from the Magneticum Pathfinder simulations in idealized observations, we use the interpolation model for the gas electron density. This model is most suitable to fit the high data quality the mentioned idealizations entail, as we deduce from the residual distribution of the fitted surface brightness profiles (see Sect. \ref{section_Fitting_process}).

Furthermore, we use the NFW model for the dark matter mass, because it mostly provides reliable results for a relatively low number of free parameters.

\begin{table}
    \centering
    \caption{Comparison of results for idealized observations varying models and other factors.}
    \begin{tabular}{|c|c|c|}\hline
         384 & \hspace{0.42cm} interp. \hspace{0.42cm} & 0.817$^{+0.012}_{-0.012}$\hspace{0.35cm}1.033$^{+0.029}_{-0.027}$\\
         & mod. $\beta$ & 0.812$^{+0.011}_{-0.011}$\hspace{0.35cm}1.028$^{+0.025}_{-0.024}$\\
         & binned & 0.918$^{+0.012}_{-0.011}$\hspace{0.35cm}1.254$^{+0.030}_{-0.029}$\\\hline
         
         323 & interp. & 0.838$^{+0.011}_{-0.011}$\hspace{0.35cm}0.934$^{+0.014}_{-0.013}$\\
         & mod. $\beta$ & 0.792$^{+0.010}_{-0.009}$\hspace{0.35cm}0.902$^{+0.013}_{-0.012}$\\
         & binned & 0.884$^{+0.012}_{-0.013}$\hspace{0.35cm}0.990$^{+0.016}_{-0.016}$\\\hline
         
         89 & interp. & 0.976$^{+0.022}_{-0.021}$\hspace{0.35cm}1.377$^{+0.031}_{-0.029}$\\
         & mod. $\beta$ & 0.942$^{+0.022}_{-0.022}$\hspace{0.35cm}1.245$^{+0.028}_{-0.027}$\\
         & binned & 0.783$^{+0.002}_{-0.003}$\hspace{0.35cm}1.031$^{+0.002}_{-0.002}$\\\hline
         
         83 & interp. & 0.777$^{+0.014}_{-0.013}$\hspace{0.35cm}1.003$^{+0.014}_{-0.014}$\\
         & mod. $\beta$ & 0.790$^{+0.003}_{-0.003}$\hspace{0.35cm}1.184$^{+0.003}_{-0.003}$\\
         & binned & 0.662$^{+0.002}_{-0.001}$\hspace{0.35cm}1.030$^{+0.001}_{-0.001}$\\\hline
    \end{tabular}
    
    \begin{tabular}{|c|c|c|}\hline
         384 & \hspace{0.4835cm} NFW \hspace{0.4835cm} & 0.817$^{+0.012}_{-0.012}$\hspace{0.35cm}1.033$^{+0.029}_{-0.027}$\\
         & GNFW & 0.806$^{+0.013}_{-0.012}$\hspace{0.35cm}0.945$^{+0.032}_{-0.026}$\\
         & Ein. & 0.811$^{+0.012}_{-0.012}$\hspace{0.35cm}0.979$^{+0.034}_{-0.039}$\\\hline
         
         323 & NFW & 0.838$^{+0.011}_{-0.011}$\hspace{0.35cm}0.934$^{+0.014}_{-0.013}$\\
         & GNFW & 0.871$^{+0.015}_{-0.015}$\hspace{0.35cm}1.010$^{+0.027}_{-0.025}$\\
         & Ein. & 0.855$^{+0.012}_{-0.013}$\hspace{0.35cm}0.984$^{+0.019}_{-0.025}$\\\hline
         
         89 & NFW & 0.976$^{+0.022}_{-0.021}$\hspace{0.35cm}1.377$^{+0.031}_{-0.029}$\\
         & GNFW & 0.992$^{+0.023}_{-0.025}$\hspace{0.35cm}1.470$^{+0.038}_{-0.049}$\\
         & Ein. & 0.928$^{+0.021}_{-0.023}$\hspace{0.35cm}1.191$^{+0.027}_{-0.032}$\\\hline
         
         83 & NFW & 0.777$^{+0.014}_{-0.013}$\hspace{0.35cm}1.003$^{+0.014}_{-0.014}$\\
         & GNFW & 0.758$^{+0.010}_{-0.010}$\hspace{0.35cm}1.030$^{+0.010}_{-0.010}$\\
         & Ein. & 0.549$^{+0.029}_{-0.026}$\hspace{0.35cm}0.398$^{+0.020}_{-0.016}$\\\hline
    \end{tabular}
    
    \begin{tabular}{|c|c|c|}\hline
         384 & usual binning & 0.817$^{+0.012}_{-0.012}$\hspace{0.35cm}1.033$^{+0.029}_{-0.027}$\\
         & finer binning & 0.803$^{+0.011}_{-0.011}$\hspace{0.35cm}1.008$^{+0.023}_{-0.023}$\\
         & finest binning & 0.796$^{+0.011}_{-0.010}$\hspace{0.35cm}0.994$^{+0.023}_{-0.022}$\\\hline
         
         323 & usual binning & 0.838$^{+0.011}_{-0.011}$\hspace{0.35cm}0.934$^{+0.014}_{-0.013}$\\
         & finer binning & 0.876$^{+0.010}_{-0.010}$\hspace{0.35cm}1.018$^{+0.014}_{-0.014}$\\
         & finest binning & 0.841$^{+0.009}_{-0.009}$\hspace{0.35cm}0.943$^{+0.011}_{-0.011}$\\\hline
         
         89 & usual binning & 0.976$^{+0.022}_{-0.021}$\hspace{0.35cm}1.377$^{+0.031}_{-0.029}$\\
         & finer binning & 0.980$^{+0.021}_{-0.021}$\hspace{0.35cm}1.446$^{+0.031}_{-0.031}$\\
         & finest binning & 0.939$^{+0.019}_{-0.019}$\hspace{0.35cm}1.343$^{+0.027}_{-0.026}$\\\hline
         
         83 & usual binning & 0.777$^{+0.014}_{-0.013}$\hspace{0.35cm}1.003$^{+0.014}_{-0.014}$\\
         & finer binning & 0.659$^{+0.010}_{-0.009}$\hspace{0.35cm}0.734$^{+0.009}_{-0.008}$\\
         & finest binning & 0.627$^{+0.009}_{-0.009}$\hspace{0.35cm}0.681$^{+0.008}_{-0.008}$\\\hline
    \end{tabular}
    
    \begin{tabular}{|c|c|c|}\hline
         384 & \hspace{0.015cm} deter. center \hspace{0.015cm} & 0.817$^{+0.012}_{-0.012}$\hspace{0.35cm}1.033$^{+0.029}_{-0.027}$\\
         & true center & 0.815$^{+0.012}_{-0.012}$\hspace{0.35cm}1.050$^{+0.029}_{-0.028}$\\\hline
         
         323 & deter. center & 0.838$^{+0.011}_{-0.011}$\hspace{0.35cm}0.934$^{+0.014}_{-0.013}$\\
         & true center & 1.049$^{+0.016}_{-0.016}$\hspace{0.35cm}1.423$^{+0.028}_{-0.028}$\\\hline
         
         89 & deter. center & 0.976$^{+0.022}_{-0.021}$\hspace{0.35cm}1.377$^{+0.031}_{-0.029}$\\
         & true center & 1.107$^{+0.017}_{-0.016}$\hspace{0.35cm}1.804$^{+0.021}_{-0.020}$\\\hline
         
         83 & deter. center & 0.777$^{+0.014}_{-0.013}$\hspace{0.35cm}1.003$^{+0.014}_{-0.014}$\\
         & true center & 0.740$^{+0.005}_{-0.005}$\hspace{0.35cm}1.109$^{+0.004}_{-0.004}$\\\hline
    \end{tabular}
    
    \tablefoot{The table shows $\zeta_{\mathrm{x}}$ at $r_{\mathrm{500}}$ (left numbers) and $r_{\mathrm{200}}$ (right numbers) for the four different clusters in snap 140 varying gas electron density models, dark matter mass models, binnings and center positions assumed for the fit. The numbers correspond to Fig. \ref{fig:models_and_other_factors}; more information can be found in the caption there.}
    \label{tab:models_and_other_factors}
\end{table}

\subsubsection{Fraction $\zeta$}

\begin{figure}
    \centering
    \includegraphics[width=0.433\textwidth]{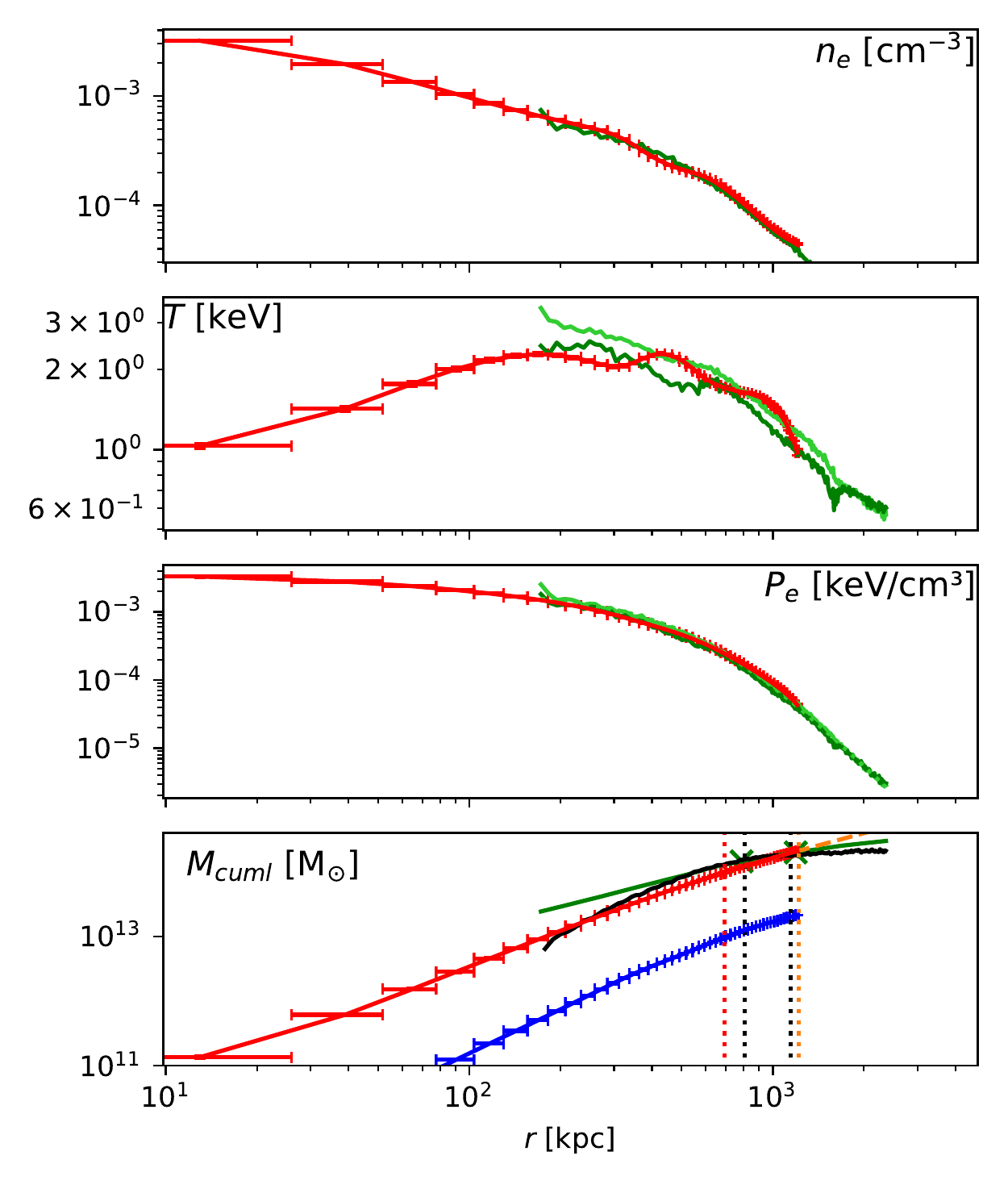}
    \caption{Idealized observations: gas electron density, temperature, electron pressure and cumulative mass as a function of the radius in physical kpc for cluster 384 in snap 140. In the first three panels, the red lines show the results of the fit with MBProj2. In the fourth panel, the red line shows the cumulative total mass and the blue line shows the cumulative gas mass according to the fit with MBProj2. The dashed orange line shows the linear extrapolation of the cumulative total mass profile. The green lines show the true profiles according to the simulations (for temperature and pressure: the dark green line represents the spectroscopic-like estimate of the profile (see Sect. \ref{section_Comparing_profiles_from_simulations_and_observations}); the light green line represents the mass-weighted profile). In the fourth panel, the green crosses show the true values of the cumulative total mass at the true $r_{\mathrm{500}}$ and $r_{\mathrm{200}}$. The black mass profile shows the hydrostatic mass according to theory (see Sect. \ref{section_Hydrostatic_mass_profiles_based_on_theory}); the green dotted line in the third panel shows the corresponding fit to the pressure profile.
    The vertical dotted lines in the fourth panel show the calculated values for $r_{\mathrm{500, calculated}}$ and $r_{\mathrm{200, calculated}}$ (red for the fit with MBProj2, orange for the extrapolation of the fit with MBProj2 and black for the hydrostatic profile according to theory).}
    \label{fig:combi_plot_140_384}
\end{figure}

\begin{figure}
    \centering
    \includegraphics[width=0.248\textwidth]{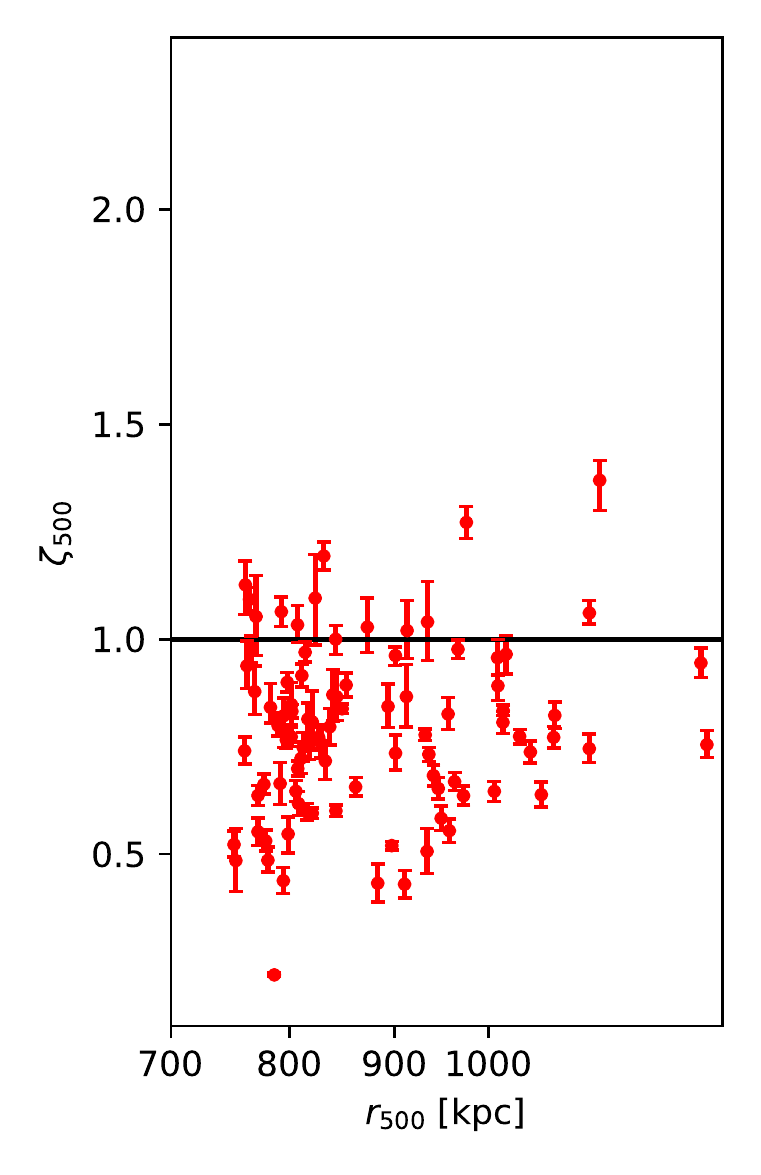}
    \includegraphics[width=0.21535\textwidth]{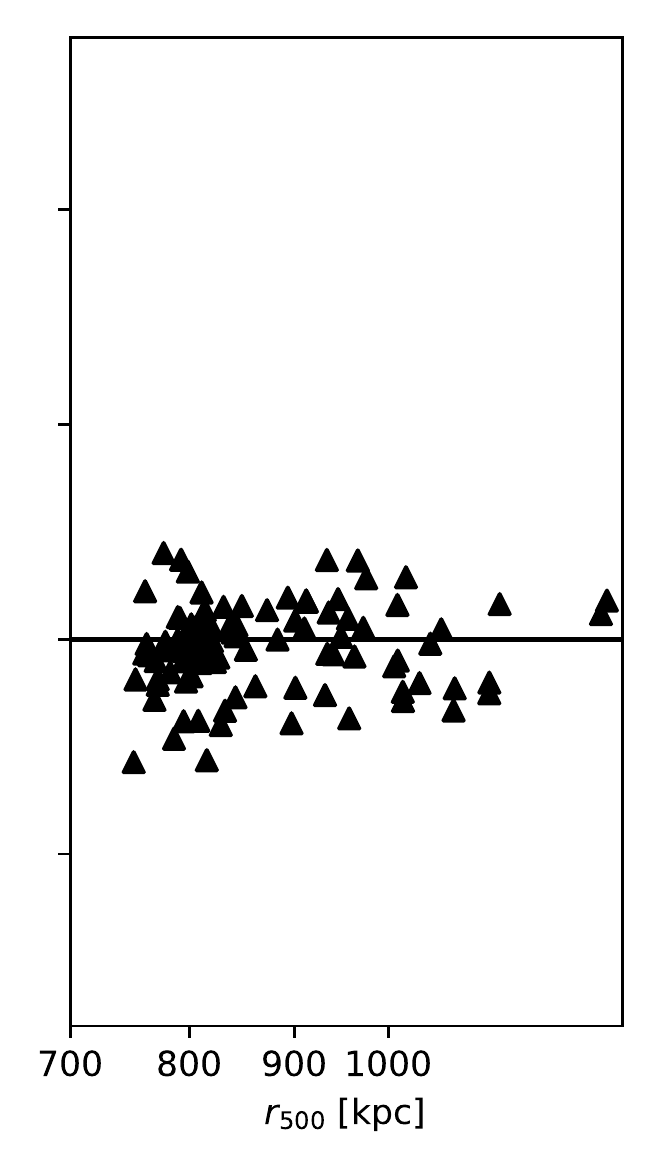}
    \includegraphics[width=0.248\textwidth]{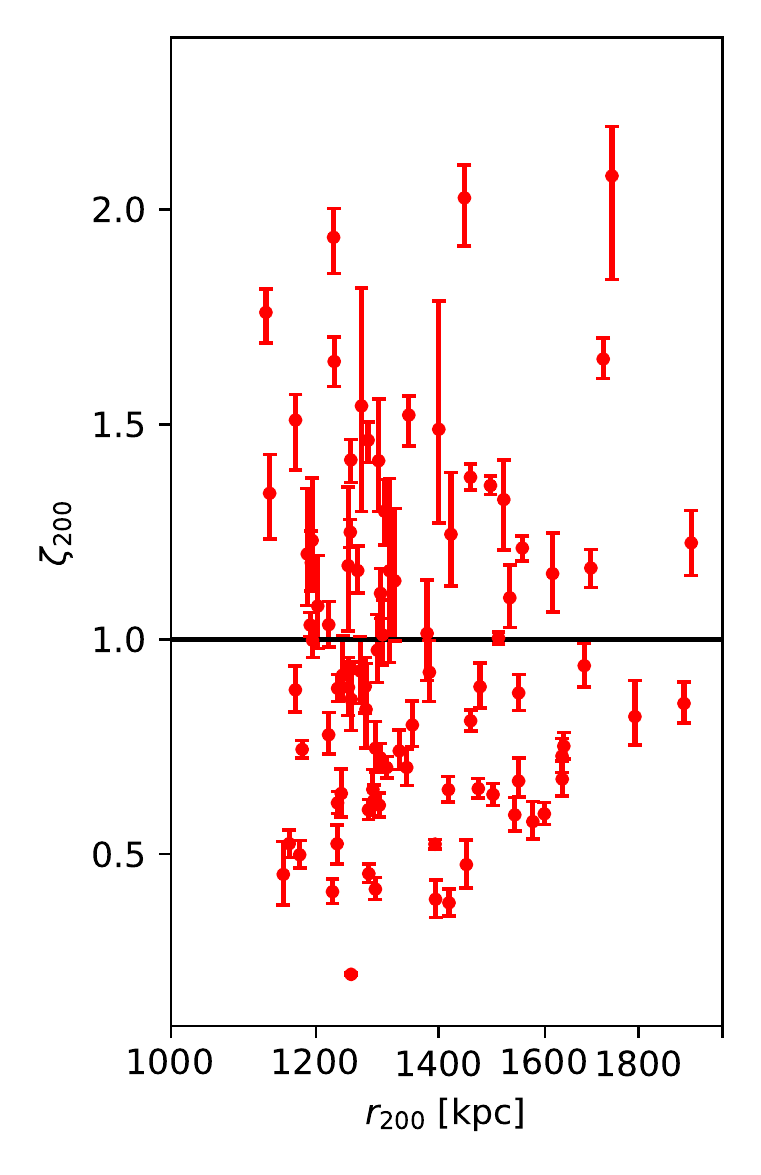}
    \includegraphics[width=0.21535\textwidth]{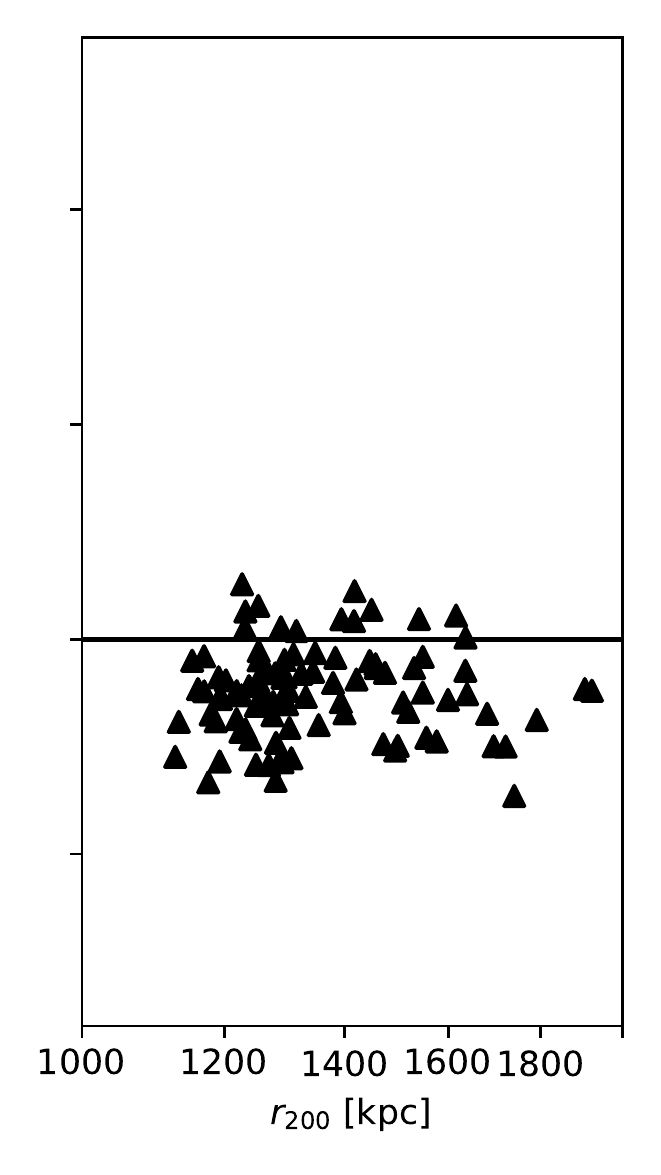}
    \caption{Idealized observations: $\zeta$ at the true $r_{\mathrm{500}}$ (upper row)/ $r_{\mathrm{200}}$ (lower row) for all 93 clusters from the Magneticum Pathfinder simulations. The two left panels show the results for the fits with MBProj2; the right panels show the results for the hydrostatic mass according to theory.}
    \label{fig:zeta}
\end{figure}

The results of the fits can be seen on the example of cluster 384 (snap 140) in Fig. \ref{fig:combi_plot_140_384}. We compare the results of the fits with MBProj2 to the true profiles according to the simulations by investigating the fraction $\zeta$ (Eq. \ref{equation_zeta}) for all 93 clusters from the Magneticum Pathfinder simulations: the left panels of Fig. \ref{fig:zeta} show the distribution of $\zeta_{\mathrm{x}}$ as a function of the true $r_{\mathrm{x}}$. The right panels show the corresponding results for the hydrostatic mass profiles according to theory (see Sect. \ref{section_Hydrostatic_mass_profiles_based_on_theory}). In all four panels the individual values scatter around the ideal value of one, while the scattering amplitude varies between the different cases. Fig. \ref{fig:zeta_bars} shows the corresponding medians and 1$\sigma$ intervals of the values in Fig. \ref{fig:zeta}. The calculation of the 1$\sigma$ intervals does not consider the uncertainties of the values for the individual clusters.
All median values shown in Fig. \ref{fig:zeta_bars} are between 0.77 and 0.99, which in principle fits our expectations of hydrostatic mass bias.
The 1$\sigma$ intervals for the results of the fits with MBProj2 are significantly larger than for the hydrostatic masses according to theory. This indicates that not only the validity of hydrostatic equilibrium is crucial for the mass determination, but also the spectral analysis and the fitting process lead to additional uncertainties.
One possible factor is the fact that our fit works with the two-dimensional projection of the cluster, whereas the true profiles, and therefore also the hydrostatic mass profiles according to theory, represent the three-dimensional quantities of the clusters. Another factor could be the shift of the center (see Sect. \ref{section_Comparing_profiles_from_simulations_and_observations}), which is further discussed in Sect. \ref{section_Influence_of_the_shift_of_the_center}.

\subsubsection{Steepness of the mass profiles}
\label{section_Steepness_of_the_mass_profiles}

\begin{figure}
    \centering
    \includegraphics[width=0.485\textwidth]{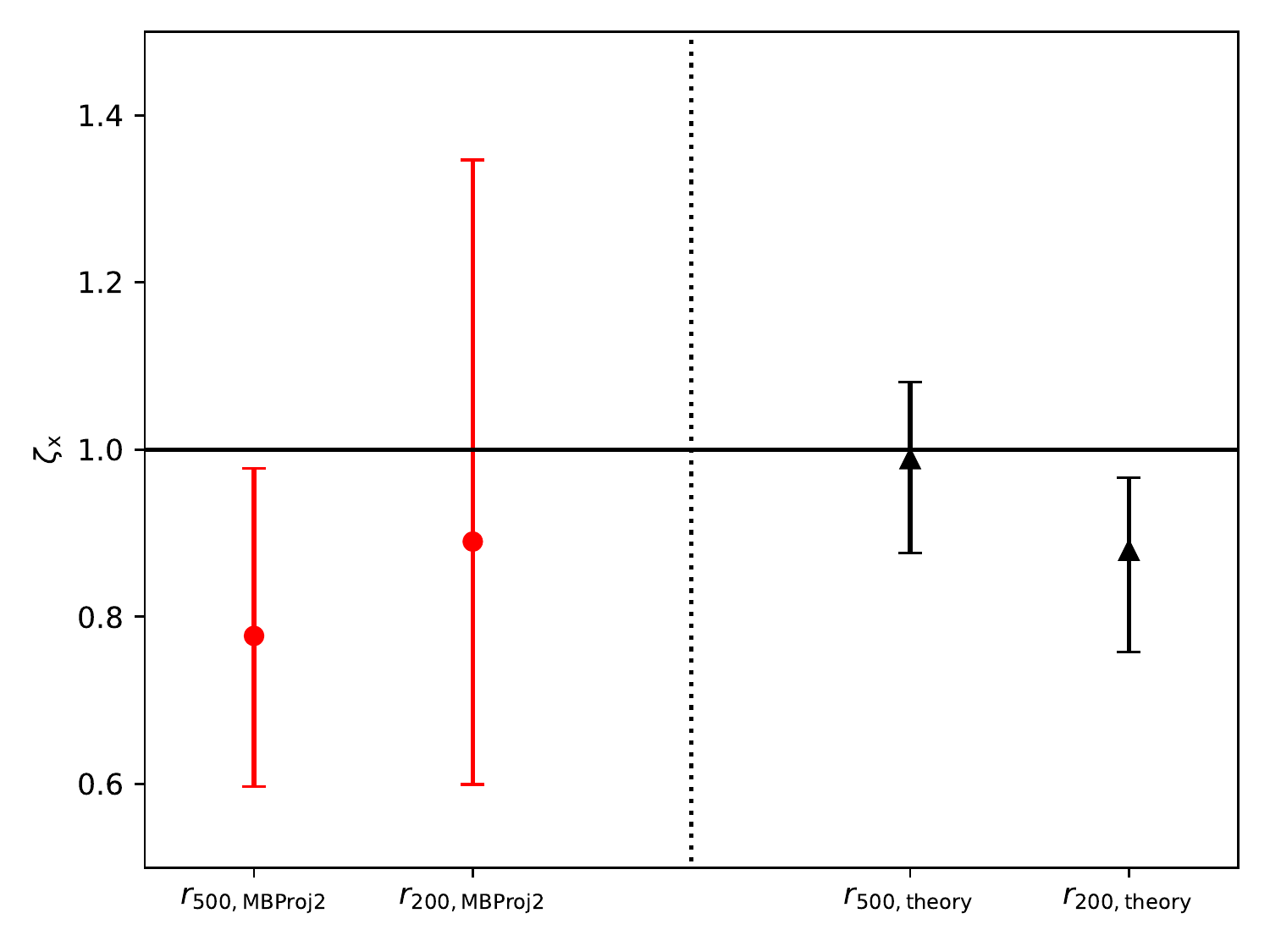}
    \caption{Idealized observations: medians and 1$\sigma$ intervals of $\zeta_{\mathrm{x}}$ for the profiles fitted with MBProj2 (left side) and for the hydrostatic mass according to theory (right side).}
    \label{fig:zeta_bars}
\end{figure}

\begin{figure}
    \centering
    \includegraphics[width=0.252\textwidth]{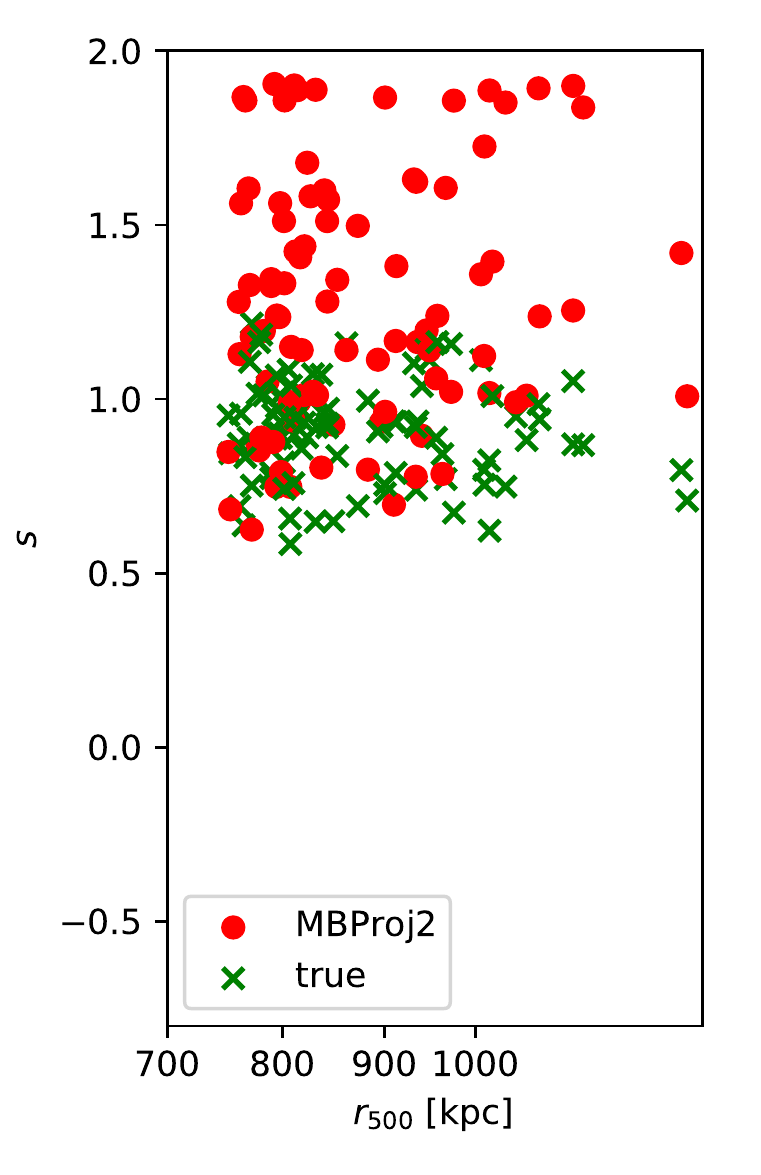}
    \includegraphics[width=0.2163\textwidth]{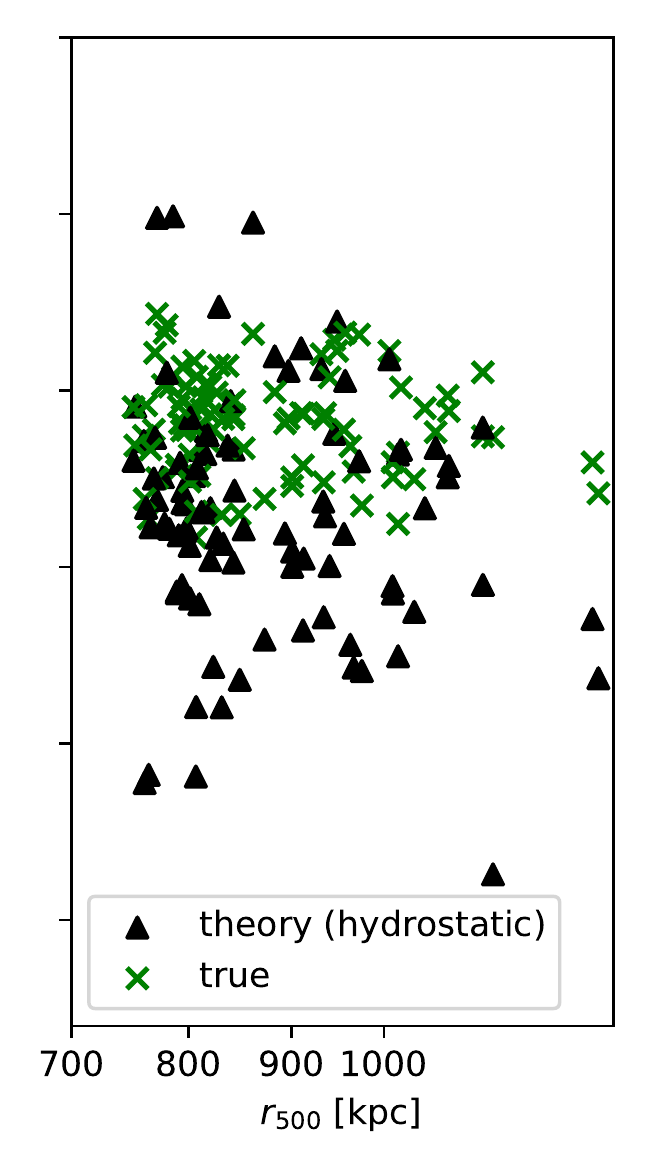}
    \caption{Idealized observations: steepness parameter $s$ for the 93 clusters from the Magneticum Pathfinder simulations as a function of the true $r_{\mathrm{500}}$. The left panel shows the results for the fits with MBProj2; the right panel shows the results for the hydrostatic mass according to theory. Both panels show the true values for comparison. We do not plot any uncertainties for $s$, as they are hard to determine because the uncertainties of the fitted values of $M_{\mathrm{500}}$ and $M_{\mathrm{200}}$ are correlated.}
    \label{fig:s_500}
\end{figure}

\begin{figure}
    \centering
    \includegraphics[width=0.485\textwidth]{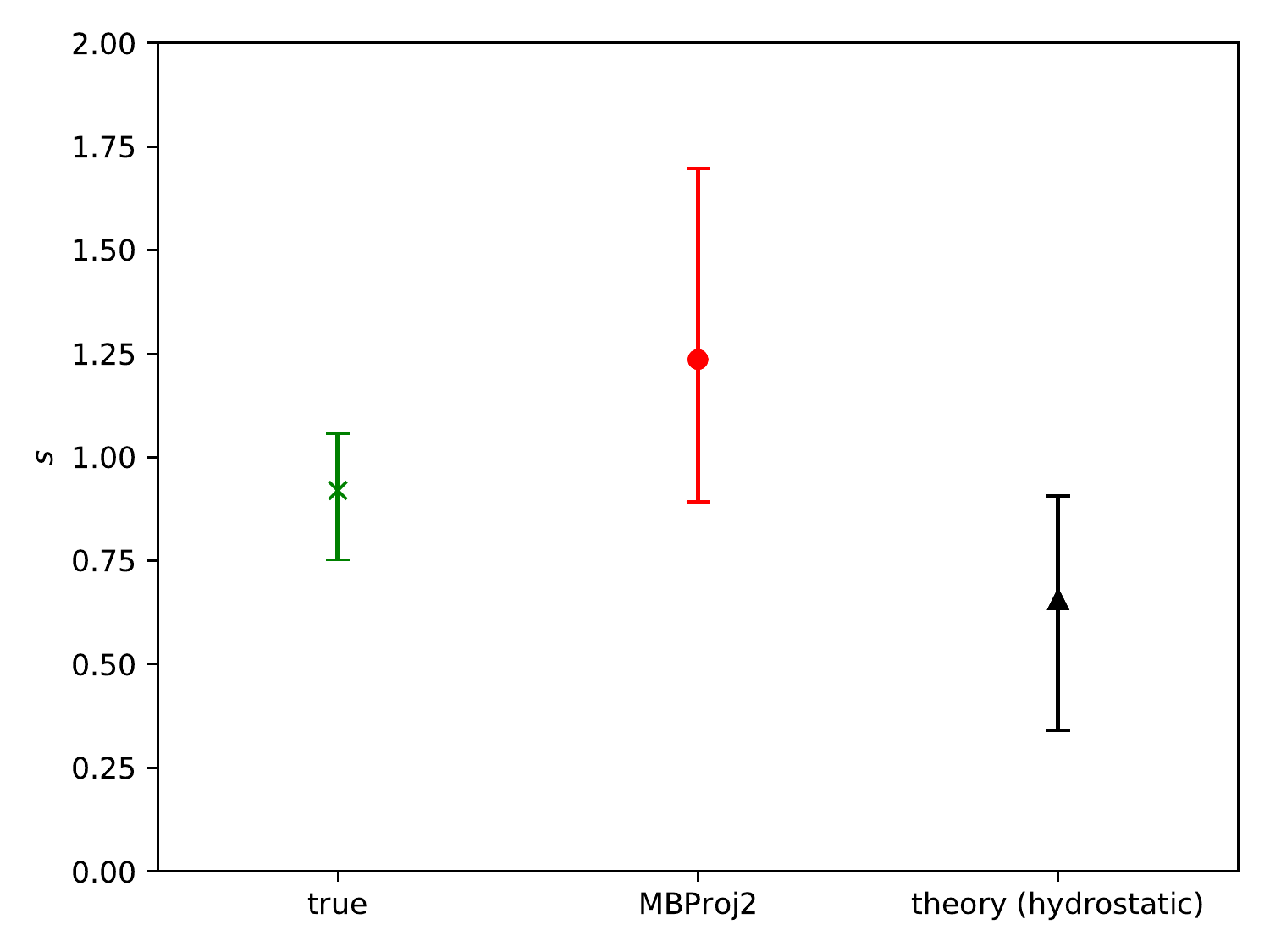}
    \caption{Idealized observations: medians and 1$\sigma$ intervals of $s$ for the true cumulative mass profiles according to the simulations, for the profiles fitted with MBProj2 and for the hydrostatic mass profiles according to theory.}
    \label{fig:s_bars}
\end{figure}

It is apparent that the median value of $\zeta_{\mathrm{200}}$ is higher than the median value of $\zeta_{\mathrm{500}}$ for the results of the fit with MBProj2, whereas it is the other way round for the hydrostatic masses according to theory (Fig. \ref{fig:zeta_bars}). To further analyze this effect, we consider the \qq{steepness} of the cumulative total mass profiles separately and define
\begin{align}
    s = \frac{\log{(M_{\mathrm{200}}/\text{M}_{\odot})}-\log{(M_{\mathrm{500}}/\text{M}_{\odot})}}{\log{(r_{\mathrm{200}}/\text{kpc})}-\log{(r_{\mathrm{500}}/\text{kpc})}}.
    \label{equation_s}
\end{align}
In Eq. \ref{equation_s}, we insert the true values of $r_{\mathrm{500}}$ and $r_{\mathrm{200}}$ and either the results of the fits for the cumulative total mass at these radii, or the corresponding results of the hydrostatic mass profiles according to theory, respectively (Fig. \ref{fig:s_500}). For comparison, we show $s$ inserting true values for masses and radii. Furthermore, we compute the corresponding median values and 1$\sigma$ intervals (Fig. \ref{fig:s_bars}).
The median $s$ is too high for the fits with MBProj2, while the hydrostatic profiles according to theory exhibit a rather too low steepness. The reason for the smaller values for the hydrostatic profiles according to theory could be the fact that they display features of the hydrostatic mass bias \citep[compare to][]{Lau_2009}.
As the cluster fulfilling perfect hydrostatic equilibrium (see Sect. \ref{section_Ideal_clusters}) does not have a systematically too high steepness, we conclude that the too high $s$ for many clusters according to the fit with MBProj2 is a consequence of the fit to the asymmetric and \qq{clumpy} structure of the clusters from the Magneticum Pathfinder simulations \citep[compare to results from observations, e.g.][]{Simionescu_2017}. However, we do not find any unique characteristics of a cluster that certainly lead to too high values of $s$. Furthermore, we are not able to solve the problem of too high $s$ by introducing priors to the model. Moreover, it seems to be unlikely that the high $s$ for the fits with MBProj2 is caused by the hydrostatic mass bias, as the hydrostatic mass profiles according to theory show the opposite behavior.

\subsubsection{Calculation of $r_{\mathrm{x}}$ and $M_{\mathrm{x}}$ based on the mass profiles}
\label{section_Calculation_of_rx_and_Mx_based_on_the_mass_profiles}

\begin{figure}
    \centering
    \includegraphics[width=0.2641\textwidth]{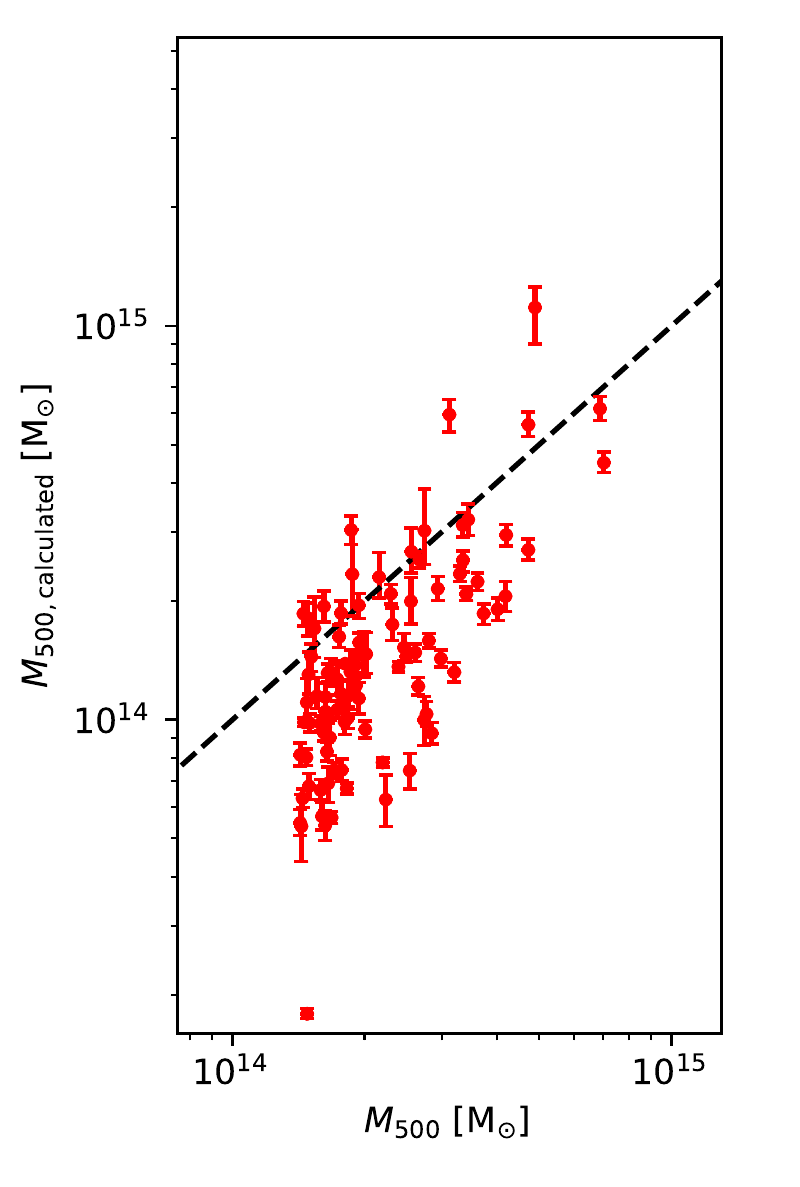}
    \includegraphics[width=0.2135\textwidth]{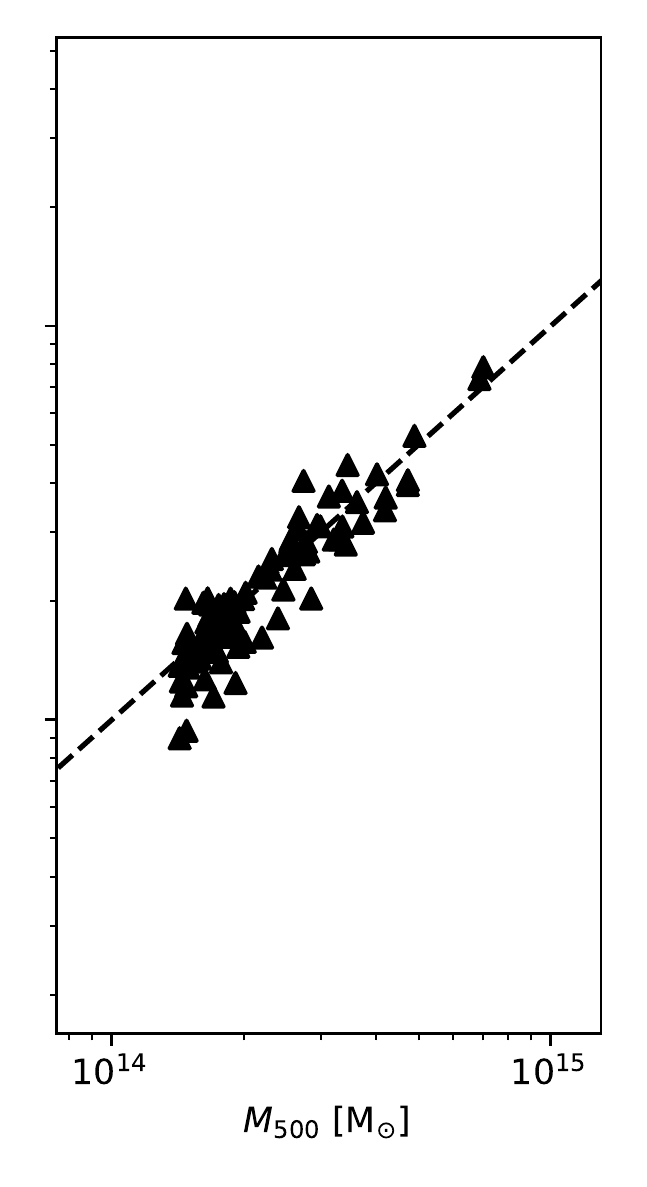}
    \includegraphics[width=0.2641\textwidth]{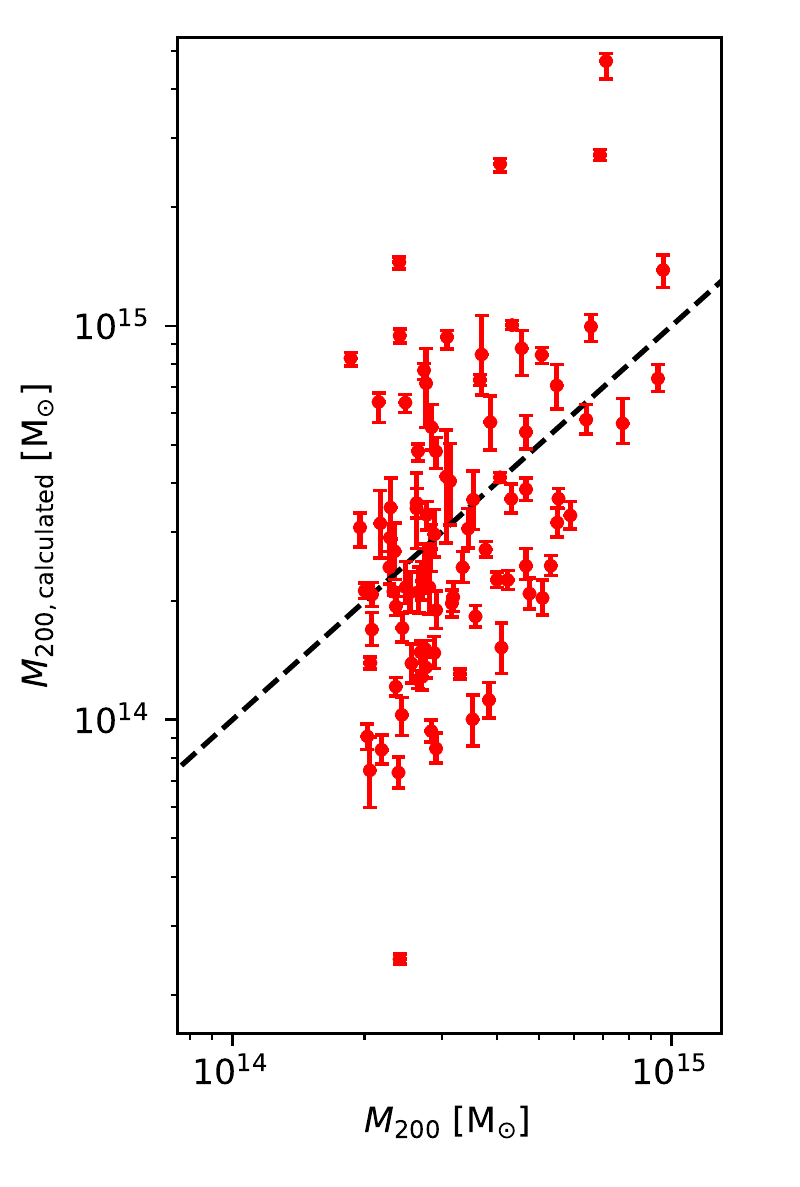}
    \includegraphics[width=0.2135\textwidth]{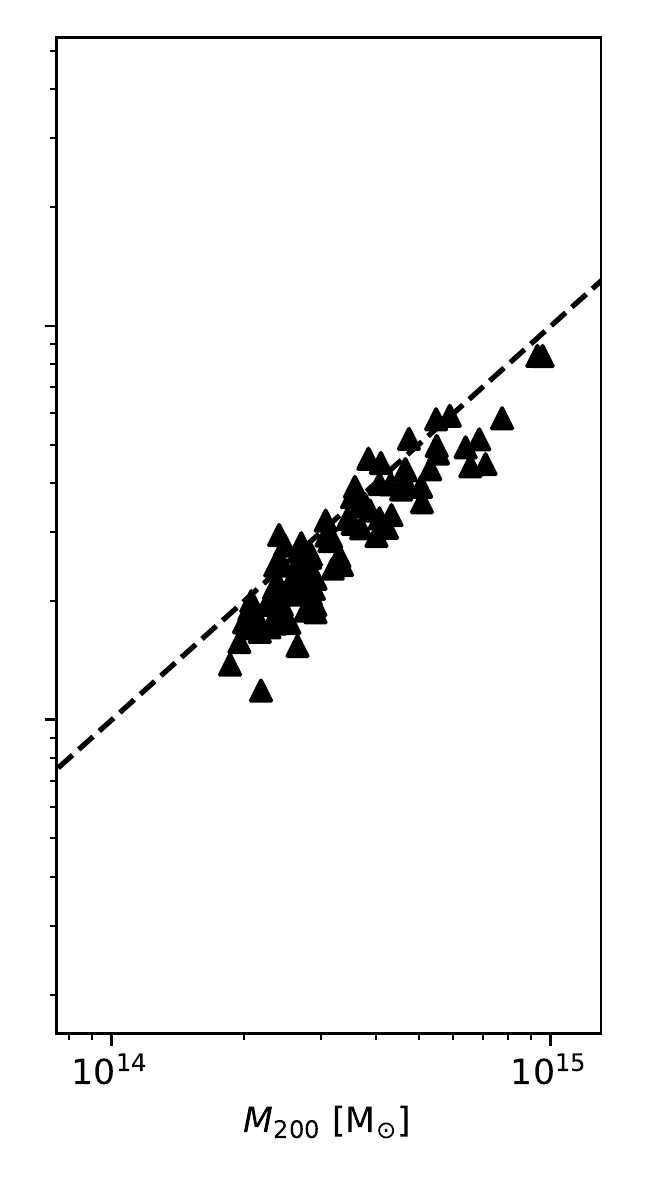}
    \caption{Idealized observations: calculated mass values $M_{\mathrm{500, calculated}}$ (upper row)/ $M_{\mathrm{200, calculated}}$ (lower row) as a function of true mass values according to the simulations $M_{\mathrm{500}}$/ $M_{\mathrm{200}}$, respectively. The two left panels show the results for the fits with MBProj2; the right panels show the results for the hydrostatic mass according to theory.}
    \label{fig:M_calc}
\end{figure}

In the next step, we compute values for $r_{\mathrm{500}}$, $r_{\mathrm{200}}$ and the cumulative total masses at these radii independently of the information from the simulations. In order to do this, we use the corresponding mass profiles and the critical density of the universe at the corresponding redshift and calculate the radii, where the average overdensity of the cluster is 500 and 200 times the critical density, respectively (see Fig. \ref{fig:combi_plot_140_384}). We use the same approach for the profiles fitted with MBProj2 and for the hydrostatic mass profiles according to theory, while for the former, we have to resort to a linear extrapolation in logarithmic space for some clusters. We call the calculated values for the radii $r_{\mathrm{x, calculated}}$. We also determine the corresponding masses at these radii $M_{\mathrm{x, calculated}}$ (Fig. \ref{fig:M_calc}). One can see that for most of the clusters, the calculated masses based on the fit with MBProj2 are smaller than the true masses. Furthermore, they show a larger scattering amplitude.
The uncertainty bars of $M_{\mathrm{x, calculated}}$ are determined according to the lower and upper limits of the fitted cumulative total mass profiles: considering the lower limit of the mass profile determines the lower uncertainty of each data point and vice versa. If $M_{\mathrm{x, calculated}}$ is calculated involving the extrapolation of the mass profile, the uncertainty of the mass profile in the outermost fitted bin is decisive.

\begin{figure}
    \centering
    \includegraphics[width=0.23\textwidth]{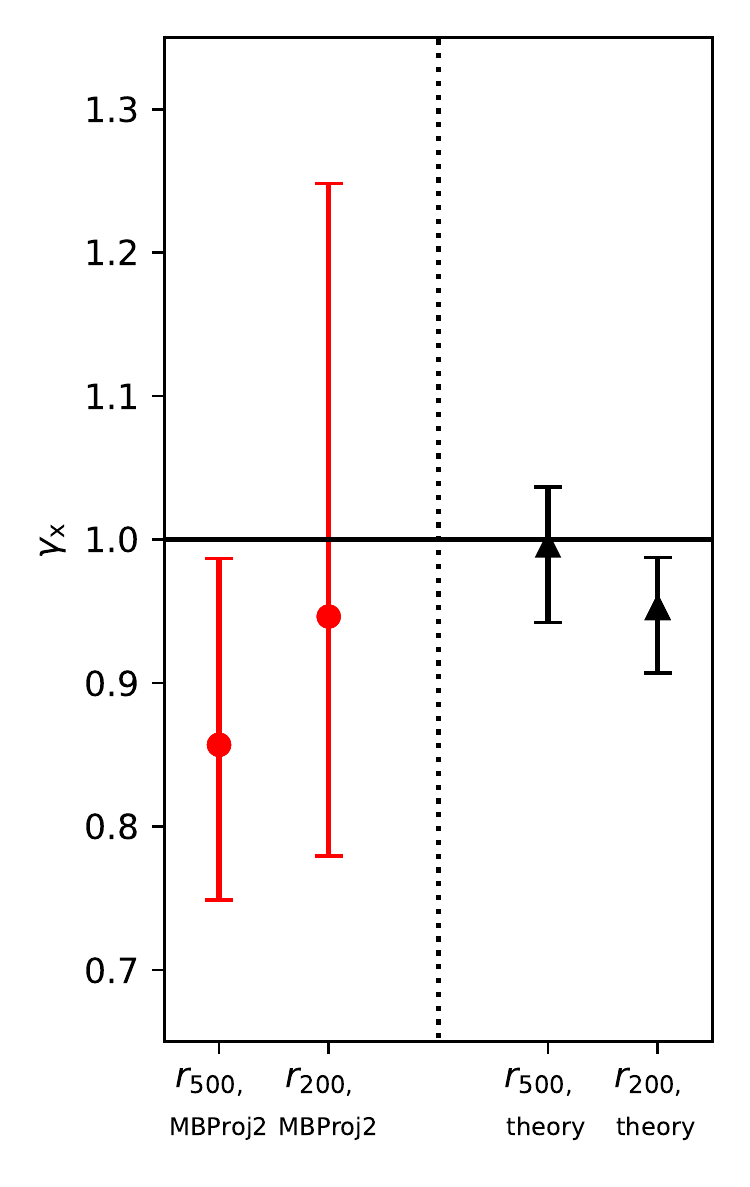}
    \hspace{0.01cm}
    \includegraphics[width=0.2343\textwidth]{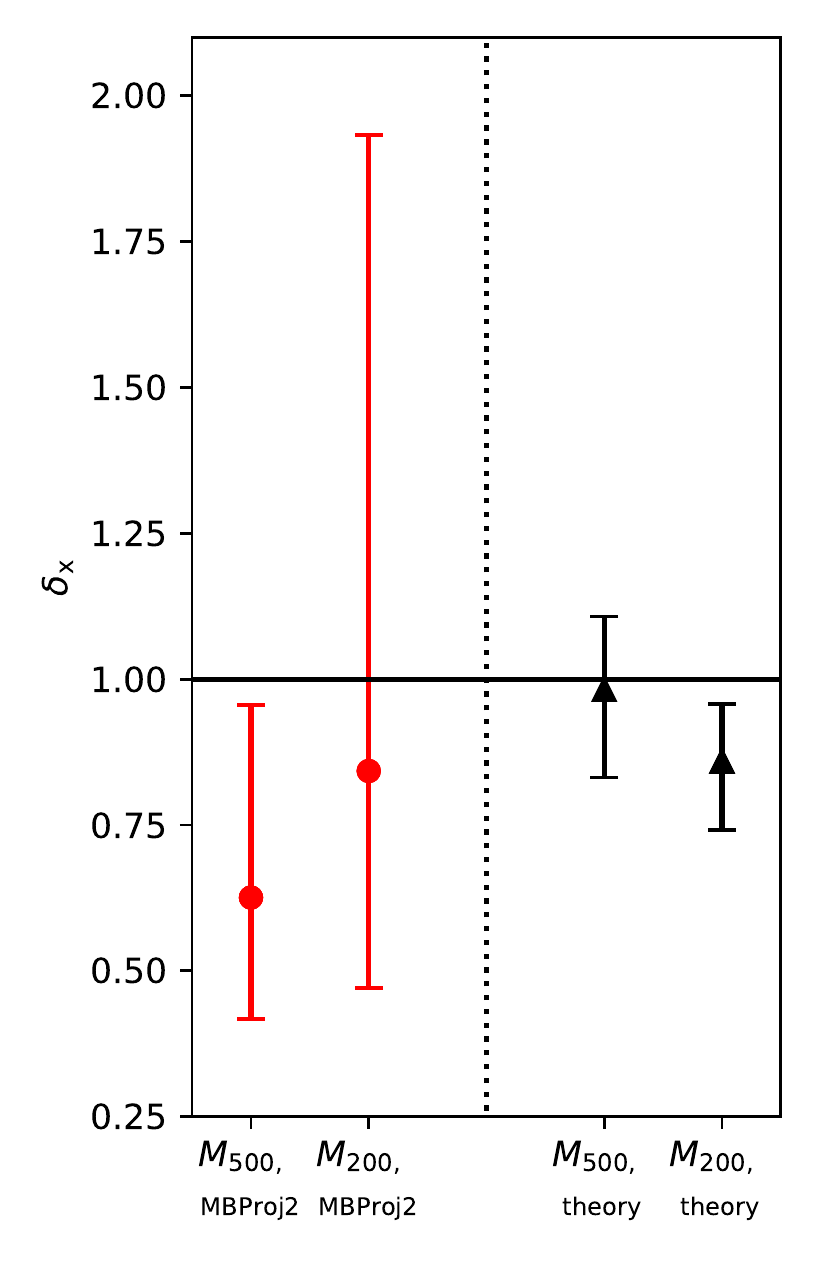}
    \caption{Idealized observations: medians and 1$\sigma$ intervals of $\gamma_{\mathrm{x}}$ and $\delta_{\mathrm{x}}$ (Eqs. \ref{equation_gamma_delta_definition}). The red bars show the results for the fits with MBProj2; the black bars show the results for the hydrostatic profiles according to theory.}
    \label{fig:rM_calc_bars}
\end{figure}

To compare the results of our calculation to the true values according to the simulations, we define two parameters:
\begin{align}
    \gamma_{\mathrm{x}} = \frac{r_{\mathrm{x, calculated}}}{r_{\mathrm{x}}} \hspace{0.75cm}\text{and}\hspace{0.75cm}
    \delta_{\mathrm{x}} = \frac{M_{\mathrm{x, calculated}}}{M_{\mathrm{x}}},
    \label{equation_gamma_delta_definition}
\end{align}
which we evaluate for $x=500$ and $x=200$. Note that $r_{\mathrm{x}}$ and $M_{\mathrm{x}}$ denote the true values according to the simulations.
The median values and 1$\sigma$ intervals of $\gamma_{\mathrm{x}}$ and $\delta_{\mathrm{x}}$ for our sample of clusters are shown in Fig. \ref{fig:rM_calc_bars}. Similarly to Fig. \ref{fig:zeta_bars} the calculation of the 1$\sigma$ intervals in Fig. \ref{fig:rM_calc_bars} does not consider the uncertainties of the values for the individual clusters. The medians and 1$\sigma$ intervals behave similarly to the corresponding values of $\zeta_{\mathrm{x}}$ (see Fig. \ref{fig:zeta_bars}).
Therefore, the fact that the median values of the results of the fits with MBProj2 are smaller than one is most likely a consequence of the hydrostatic mass bias. Moreover, the 1$\sigma$ intervals for $\delta_{\mathrm{x}}$ are much larger than for $\gamma_{\mathrm{x}}$.

\subsubsection{Influence of the shift of the center}
\label{section_Influence_of_the_shift_of_the_center}

\begin{figure}
    \centering
    \includegraphics[width=0.248\textwidth]{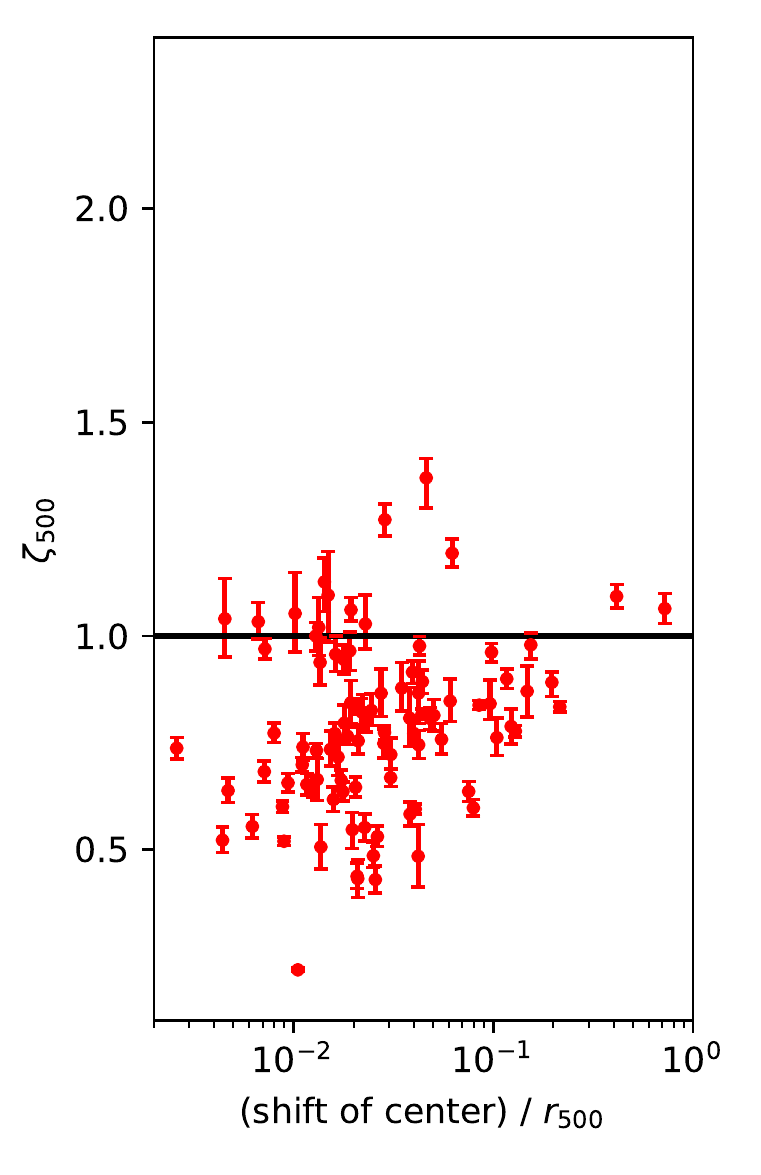}
    \includegraphics[width=0.21535\textwidth]{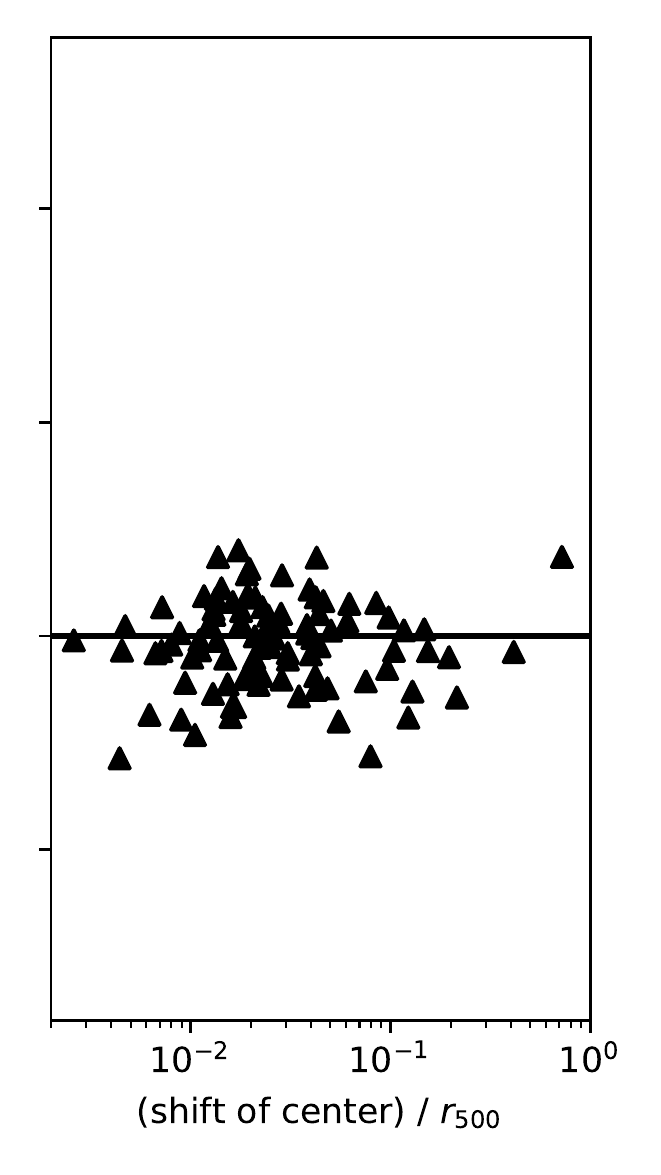}
    \includegraphics[width=0.248\textwidth]{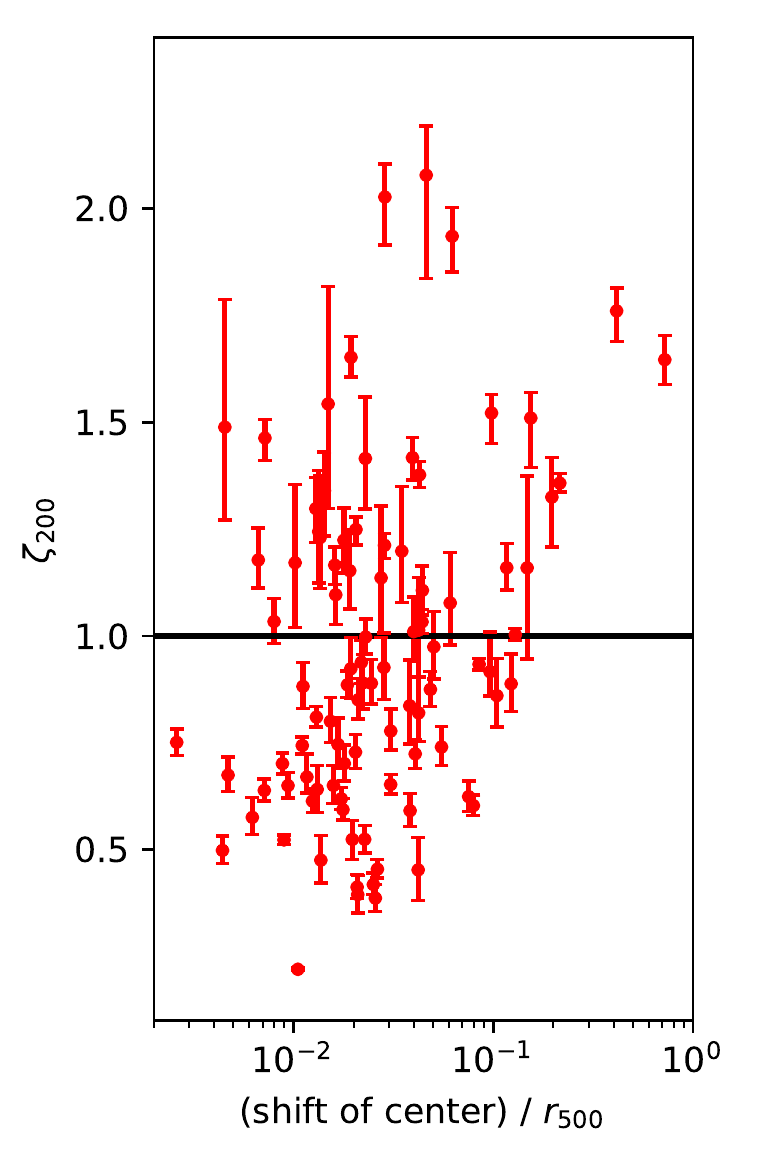}
    \includegraphics[width=0.21535\textwidth]{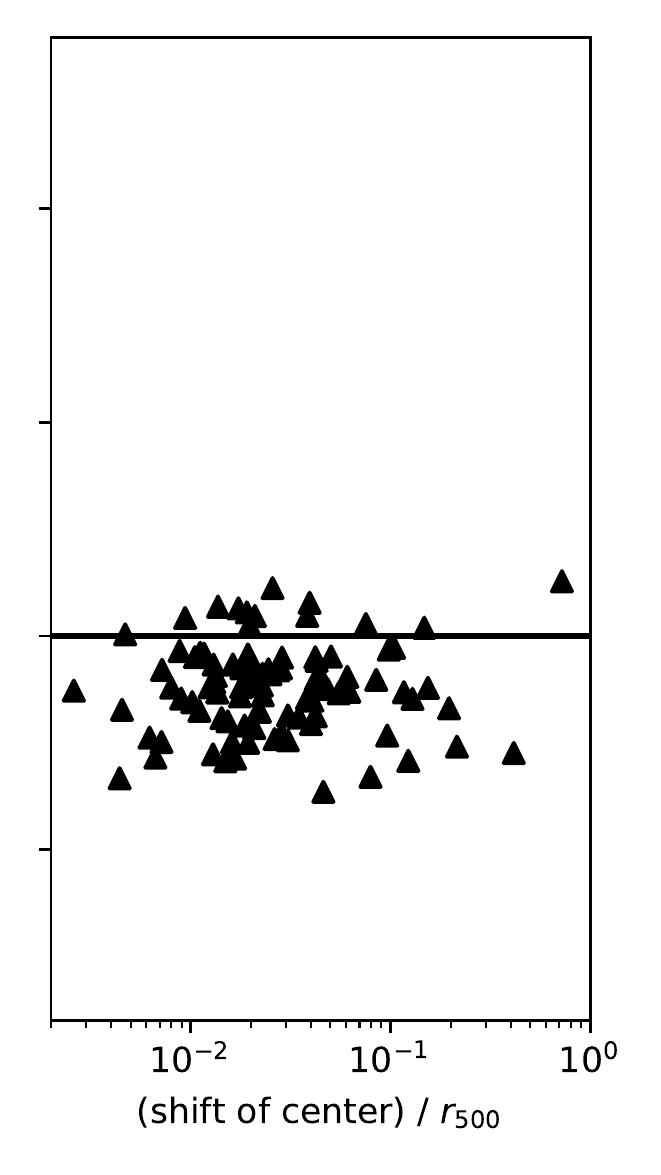}
    \caption{Idealized observations: $\zeta$ at the true $r_{\mathrm{500}}$ (upper row)/ $r_{\mathrm{200}}$ (lower row) as a function of the normalized shift of the center (see Sect. \ref{section_Comparing_profiles_from_simulations_and_observations}). The two left panels show the results for the fits with MBProj2; the right panels show the results for the hydrostatic mass according to theory.}
    \label{fig:shift}
\end{figure}

We have already mentioned in Sect. \ref{section_Comparing_profiles_from_simulations_and_observations} that the shift of the center position might introduce discrepancies when comparing the profiles fitted with MBProj2 and the true profiles. To analyze this effect, we again show the fraction $\zeta_{\mathrm{x}}$, but this time as a function of the shift of the center divided by the true $r_{\mathrm{500}}$ (Fig. \ref{fig:shift}).
Apart from the methodological influence, the size of the shift could also be seen as an information about the properties of the cluster. Specifically, it is reasonable to assume that a larger shift often refers to a higher disturbance of the individual cluster \citep[compare to e.g.][]{Mohr_1993, Poole_2006, Rasia_2013, Weissmann_2013, Chon_2017}. 
As the hydrostatic profiles according to theory are based on the true profiles, they also assume the true center position. This allows conclusions about the correlation between disturbance of the cluster and the validity of hydrostatic equilibrium.
If the shift of the center is large compared to the true $r_{\mathrm{500}}$, one could expect the corresponding values of $\zeta$ to be far from one because the cluster might be comparatively disturbed then. In addition, this could produce systematic differences between the fitted and true profiles. Interestingly, a strong correlation between the normalized shift of the center and $\zeta$ is not discernible. This applies to both the results of the fit with MBProj2 and the hydrostatic profiles according to theory. Consequently, we cannot verify that the shift of the center taken as a measure of disturbance of a cluster is a substantial indicator for the validity of hydrostatic equilibrium.
However, the position of the true center is a three-dimensional quantity, whereas the image-derived one is two-dimensional. This could lead to projection effects smoothing the correlation.

\subsubsection{Evolution with redshift}

\begin{figure}
    \centering
    \subfigure[snap 140 (4 clusters)]{
    \includegraphics[width=0.247\textwidth]{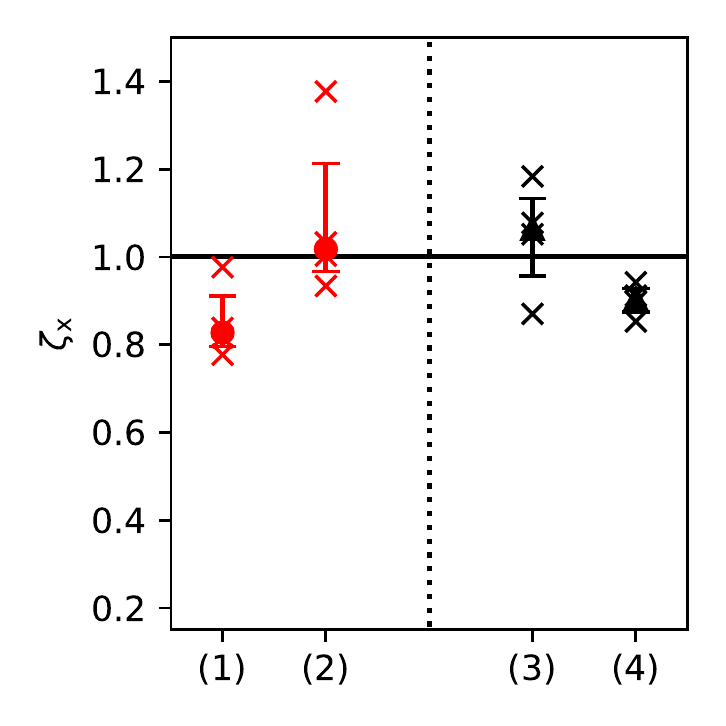}}
    \subfigure[snap 136 (6 clusters)]{
    \includegraphics[width=0.205\textwidth]{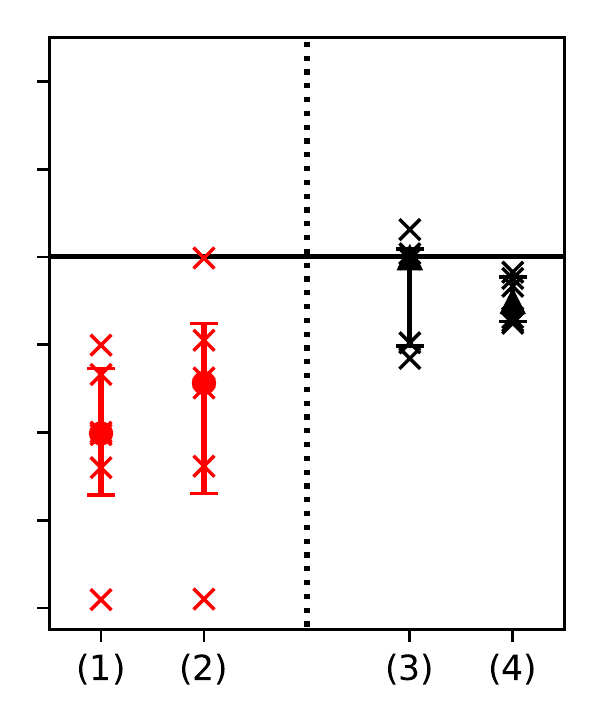}}
    \subfigure[snap 132 (17 clusters)]{
    \includegraphics[width=0.247\textwidth]{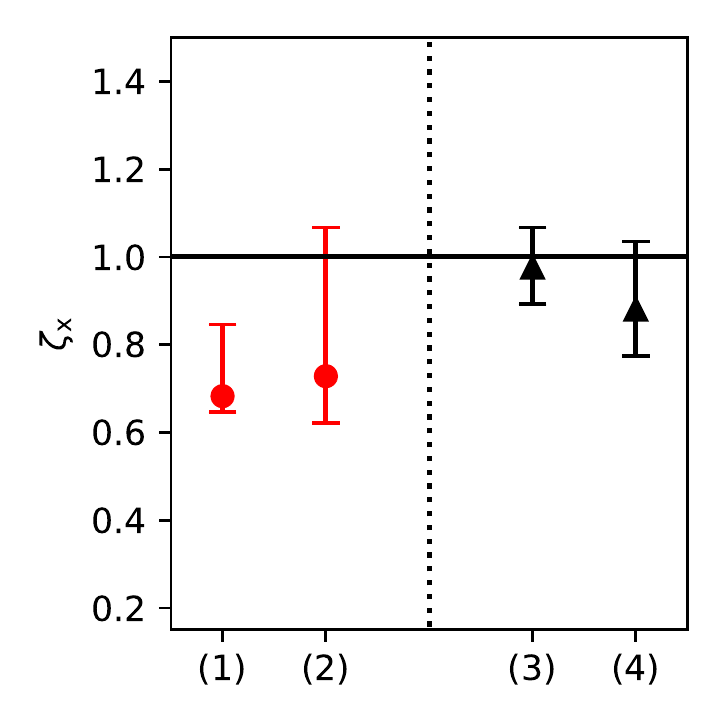}}
    \subfigure[snap 128 (27 clusters)]{
    \includegraphics[width=0.205\textwidth]{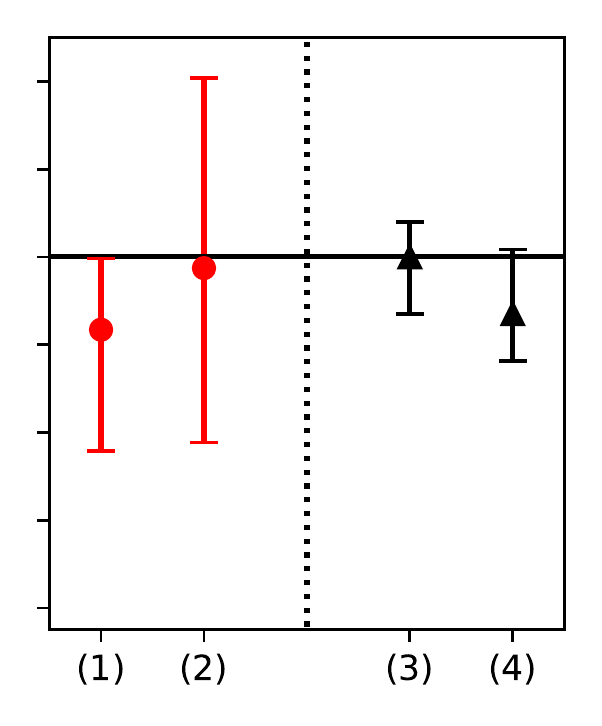}}
    \subfigure[snap 124 (39 clusters)]{
    \includegraphics[width=0.247\textwidth]{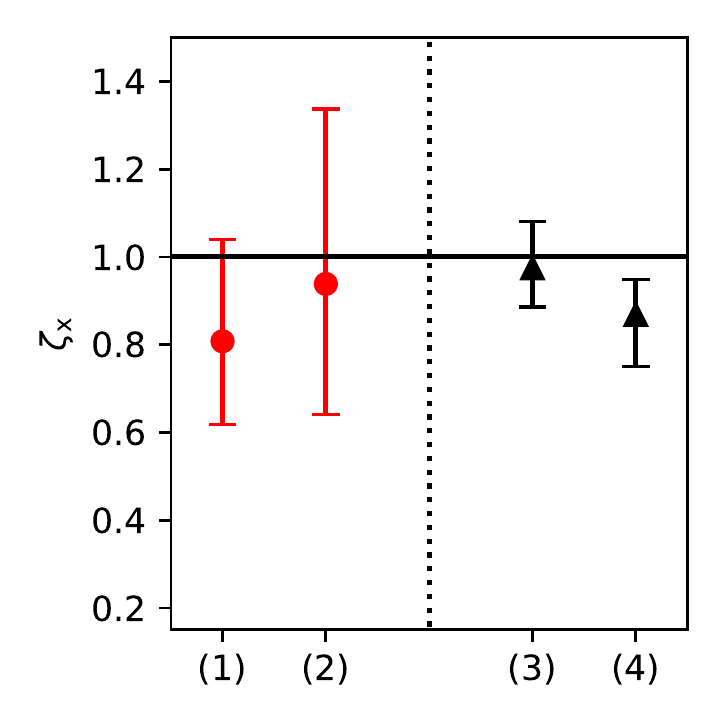}}
    \caption{Idealized observations: medians and 1$\sigma$ intervals of $\zeta_{\mathrm{x}}$ for each snap (comparable to Fig. \ref{fig:zeta_bars}). The results are shown at \newline(1) $r_{\mathrm{500}}$ for the fits with MBProj2,\newline(2) $r_{\mathrm{200}}$ for the fits with MBProj2, \newline(3) $r_{\mathrm{500}}$ for the hydrostatic profiles according to theory and \newline (4) $r_{\mathrm{200}}$ for the hydrostatic profiles according to theory.\newline The caption of each plot shows the number of the snap and how many clusters it contains. Each cluster in a snap has the same redshift (see Table \ref{tab:snaps}). Because of the low number of clusters in snap 140 and 136, the corresponding plots also show the values for the individual clusters as crosses.}
    \label{fig:part_plots}
\end{figure}

To investigate the influence of the redshift on the results, we analyze the fraction $\zeta$ for the clusters in the different snaps separately (Fig. \ref{fig:part_plots}). The redshift of all clusters in the same snap is identical (see Sect. \ref{section_Magneticum_Pathfinder_simulations}). The results of the fits are susceptible to statistical fluctuations due to low numbers of sources, leading to too small median values for snap 136. Nevertheless, for the fits with MBProj2, an increase of the median value from $\zeta_{\mathrm{500}}$ to $\zeta_{\mathrm{200}}$ is discernible for every redshift. Therefore the too high steepness of the cumulative MBProj2 total mass profiles is apparent for all tested redshifts (see Sect. \ref{section_Steepness_of_the_mass_profiles}).
In contrast, the results of the hydrostatic profiles according to theory have the opposite trend, which suggests the influence of the hydrostatic mass bias for clusters with all tested redshifts. We do not find a clear trend of the medians with redshift.

\subsection{Clusters from Magneticum Pathfinder simulations in realistic observations}
\label{section_Clusters_from_Magneticum_Pathfinder_simulations_in_realistic_observations}

\begin{figure}
    \centering
    \subfigure[idealized observations]{\includegraphics[width=0.2423\textwidth]{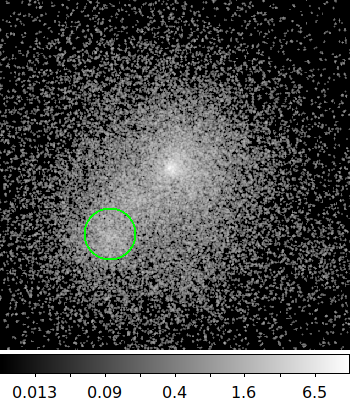}}
    \subfigure[realistic observations]{\includegraphics[width=0.2423\textwidth]{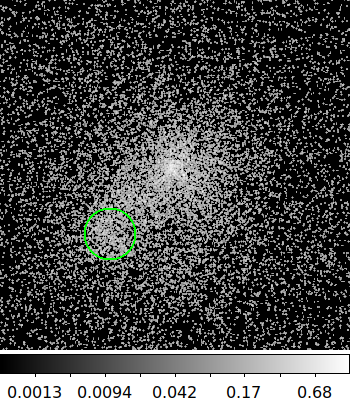}}
    \caption{Cluster 384 in snap 140 from the Magneticum Pathfinder simulations. The images use a logarithmic scale, covers energies from \SI{0.6}{keV} to \SI{1.1}{keV} and are smoothed with a gaussian. The numbers on the scales of the colorbars show the numbers of counts per pixel. They differ for idealized and realistic observations because of the relative exposure times. The green circles have a radius of about 3.1 arcmin and indicate the masked region.}
    \label{fig:images}
\end{figure}

In this section, we analyze the clusters of our sample from the Magneticum Pathfinder simulations under realistic observational conditions. The X-ray data are generated in the eROSITA Final Equatorial Depth Survey (eFEDS) area \citep{Brunner_2021}. eFEDS was observed in the performance verification phase of eROSITA. The clusters are produced to have the same true center position as for the idealized observations in Sect. \ref{section_Clusters_from_Magneticum_Pathfinder_simulations_in_idealized_observations}. We assume the same center position as for the idealized observations and apply the same masking (both determined using idealized images; see Sects. \ref{section_Estimating_the_position_of_the_center_of_the_cluster} and \ref{section_Masking}). This would of course not be possible with real data, but ensures the results in Sect. \ref{section_Clusters_from_Magneticum_Pathfinder_simulations_in_idealized_observations} can be compared.
For the realistic observations, we use the standard eROSITA PSF \citep{Predehl_2021, Merloni_2012, Dauser_2019}. Furthermore, we use exposure maps with realistic exposure times. The exposure maps also take into account the eFEDS attitude data to model the pointing of the telescope with time. The version of the eROSITA instrument definition used is 1.8.2. Our background model makes use of the eFEDS survey data to make a particle background model and we incorporate an X-ray foreground model from \cite{Sanders_2021}. When fitting the synthetic data, we assume the survey average PSF. We do not include point sources in the synthetic data.
We do not use the energy band with the highest energy for the realistic observations (see definition of energy band in Sect. \ref{section_Fitting_process}).
Fig. \ref{fig:images} shows the same cluster under idealized and realistic observational conditions for comparison in the energy band \SI{0.6}{keV} to \SI{1.1}{keV}. The cluster is taken from snap 140, the snap with the lowest redshift.

We assume a modified $\beta$-model with one $\beta$-component to model the electron gas density with MBProj2. This model seems to be more appropriate for realistic data quality than the interpolation model used in Sect. \ref{section_Clusters_from_Magneticum_Pathfinder_simulations_in_idealized_observations}. For the dark matter mass, we again use an NFW model.
Background surface brightness profiles we created from cluster-free simulations using the same background model.
When fitting the data an extra parameter, which introduces a scaling between the model and fitted backgrounds levels, was included in the fits, with a gaussian prior.
Equivalent to Sect. \ref{section_Clusters_from_Magneticum_Pathfinder_simulations_in_idealized_observations}, we assume that $N_\mathrm{H} = \SI{0.01e22}{cm^{-2}}$.

\begin{figure}
    \centering
    \includegraphics[width=0.485\textwidth]{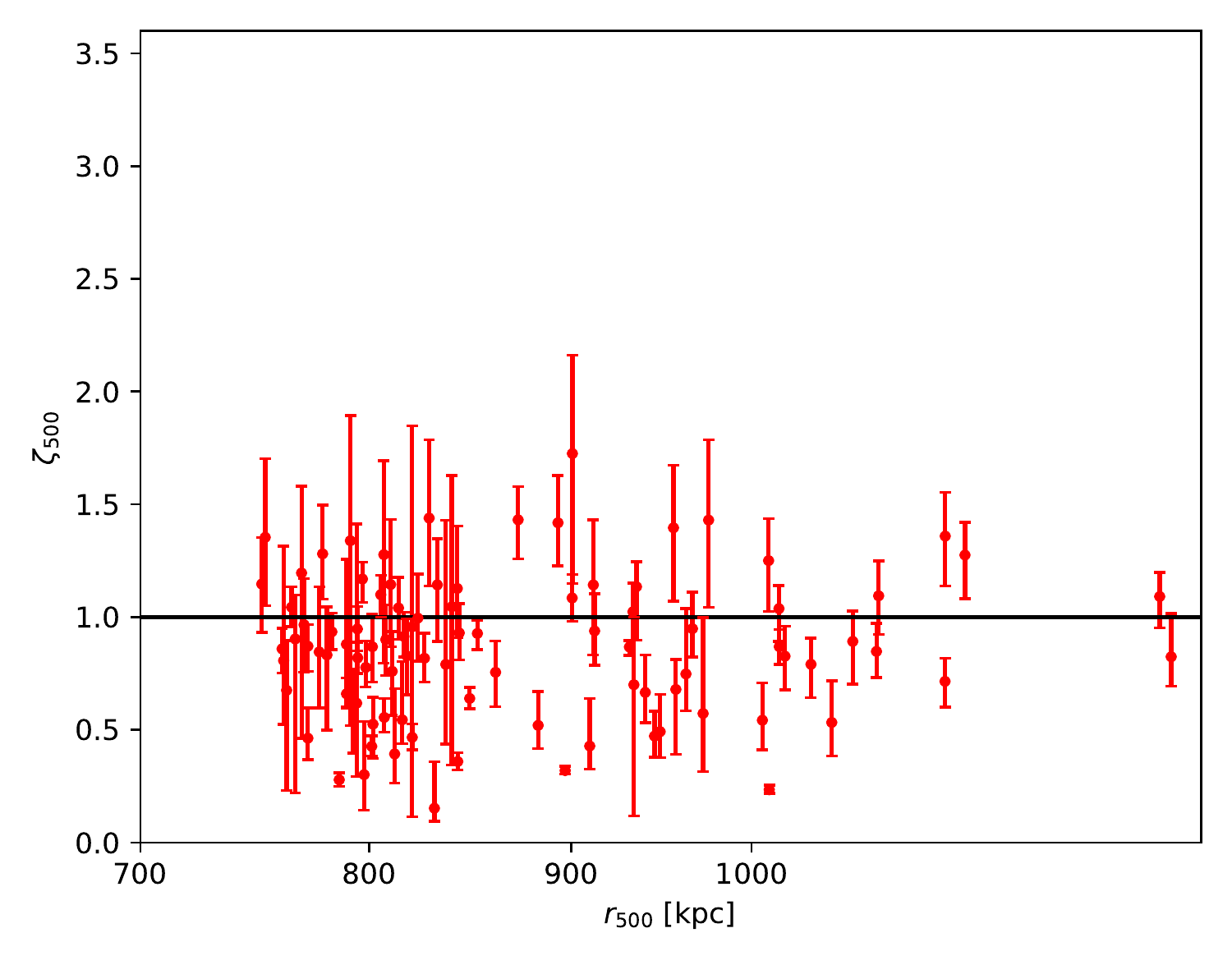}
    \includegraphics[width=0.485\textwidth]{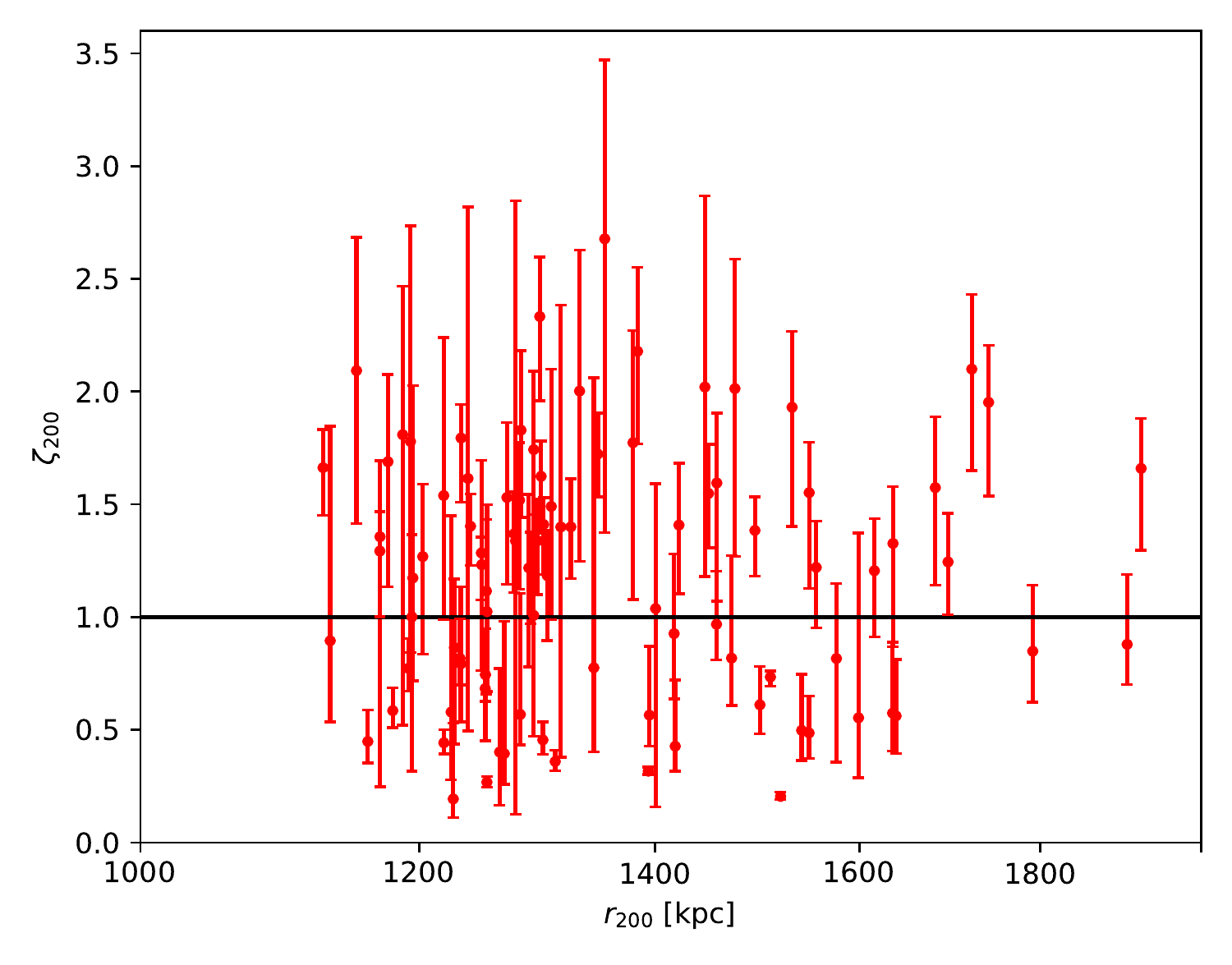}
    \caption{Realistic observations: $\zeta$ at the true $r_{\mathrm{500}}$ (upper panel)/ $r_{\mathrm{200}}$ (lower panel), according to the fits with MBProj2.}
    \label{fig:zeta_realistic}
\end{figure}

\begin{figure}
    \centering
    \includegraphics[width=0.485\textwidth]{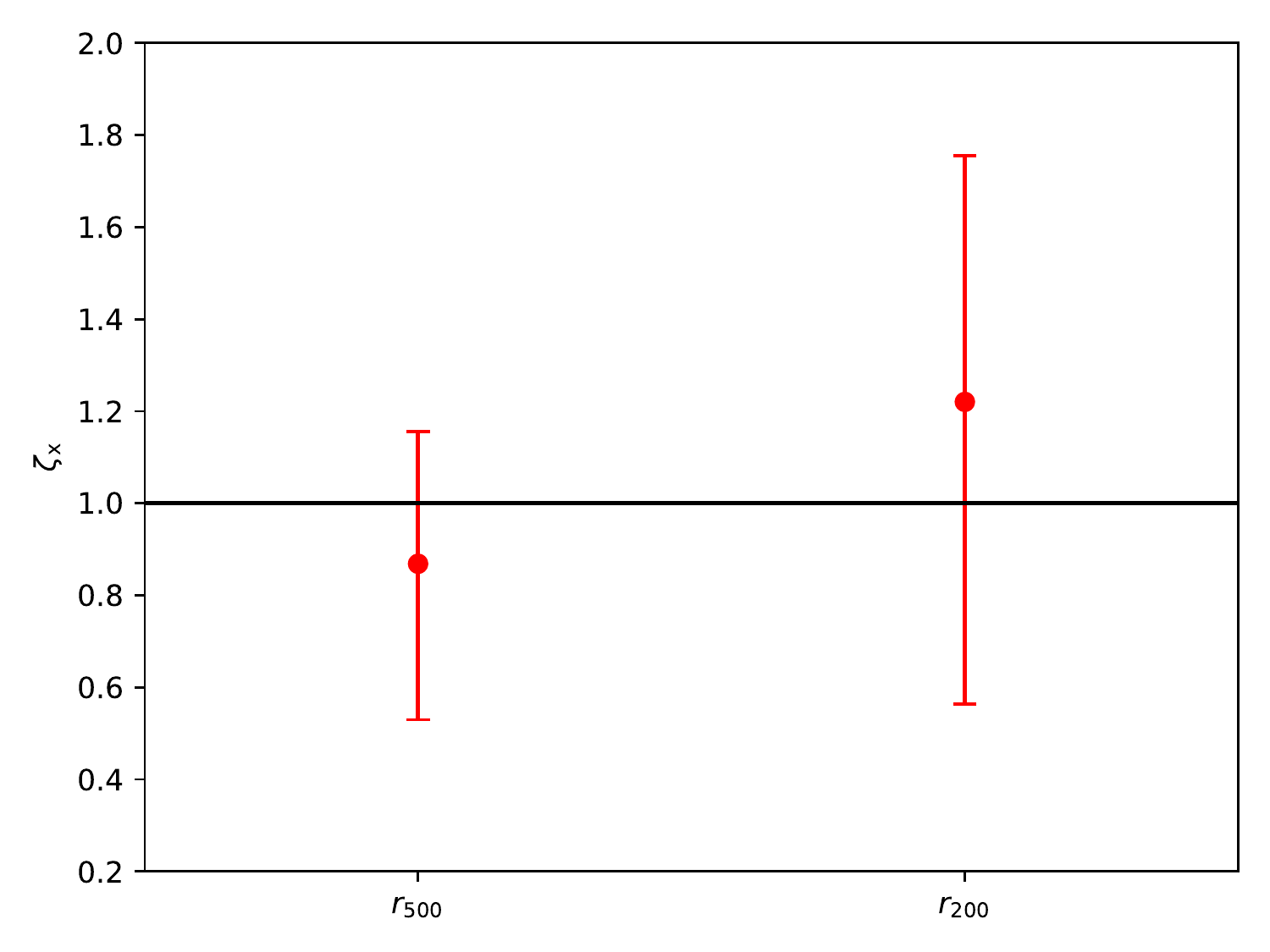}
    \caption{Realistic observations: medians and 1$\sigma$ intervals of $\zeta_{\mathrm{x}}$, according to the fits with MBProj2.}
    \label{fig:zeta_bars_realistic}
\end{figure}

We analyze the data in the same way like in Sect. \ref{section_Clusters_from_Magneticum_Pathfinder_simulations_in_idealized_observations}. Fig. \ref{fig:zeta_realistic} shows the fraction $\zeta$ (Eq. \ref{equation_zeta}) for the clusters from the Magneticum Pathfinder simulations under realistic observational conditions. Fig. \ref{fig:zeta_bars_realistic} shows the corresponding medians and 1$\sigma$ intervals. One can see that the uncertainty bars for the particular clusters are considerably larger than it is the case for the idealized observations. This is in agreement with our expectations regarding the effects of the lower data quality. Further, the 1$\sigma$ intervals of the total sample are larger.
Moreover, the medians lie higher than for the idealized observations, particularly for $\zeta_{\mathrm{200}}$.

\begin{figure}
    \centering
    \includegraphics[width=0.485\textwidth]{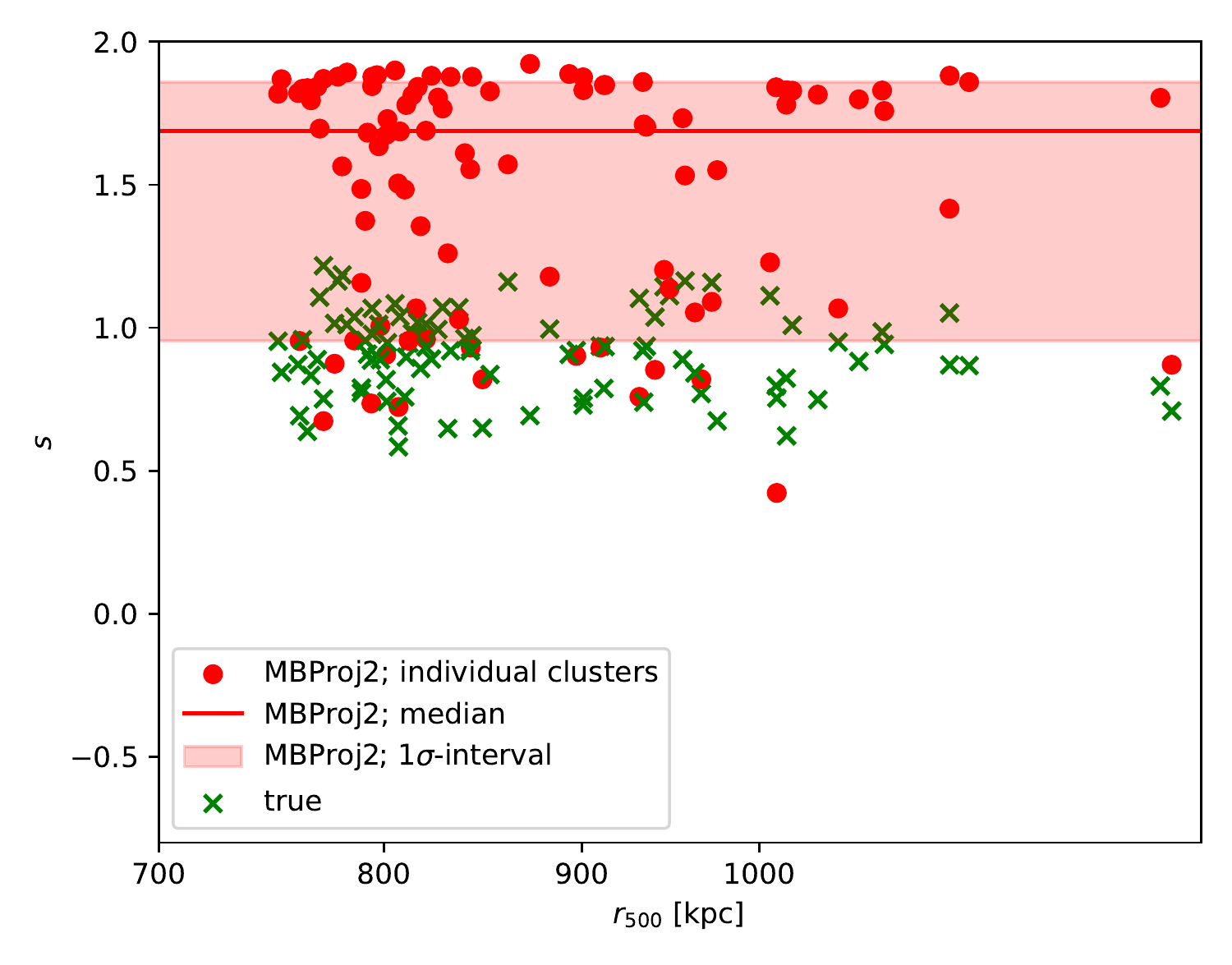}
    \caption{Realistic observations: steepness parameter $s$ for the results of the fits with MBProj2 and the true values for comparison. The horizontal red line shows the median for the results of the fits; the red region shows the corresponding 1$\sigma$-interval. This plot can be compared to Fig. \ref{fig:s_500}.}
    \label{fig:s_500_realistic}
\end{figure}

Next, we analyze the steepness value $s$ (Eq. \ref{equation_s}) of the cumulative total mass profiles. Fig. \ref{fig:s_500_realistic} shows the results for the particular clusters on the left and the corresponding median and 1$\sigma$ interval on the right. The median of $s$ is clearly higher than for the idealized observations. Apparently the lower data quality magnifies the problem of the too high steepness, probably also as we are forced to use the modified $\beta$-model with one $\beta$-component instead of the interpolation model for the density.
In contrast to the median, the highest values of $s$ are very similar for idealized and realistic observational conditions.

\begin{figure}
    \centering
    \includegraphics[width=0.485\textwidth]{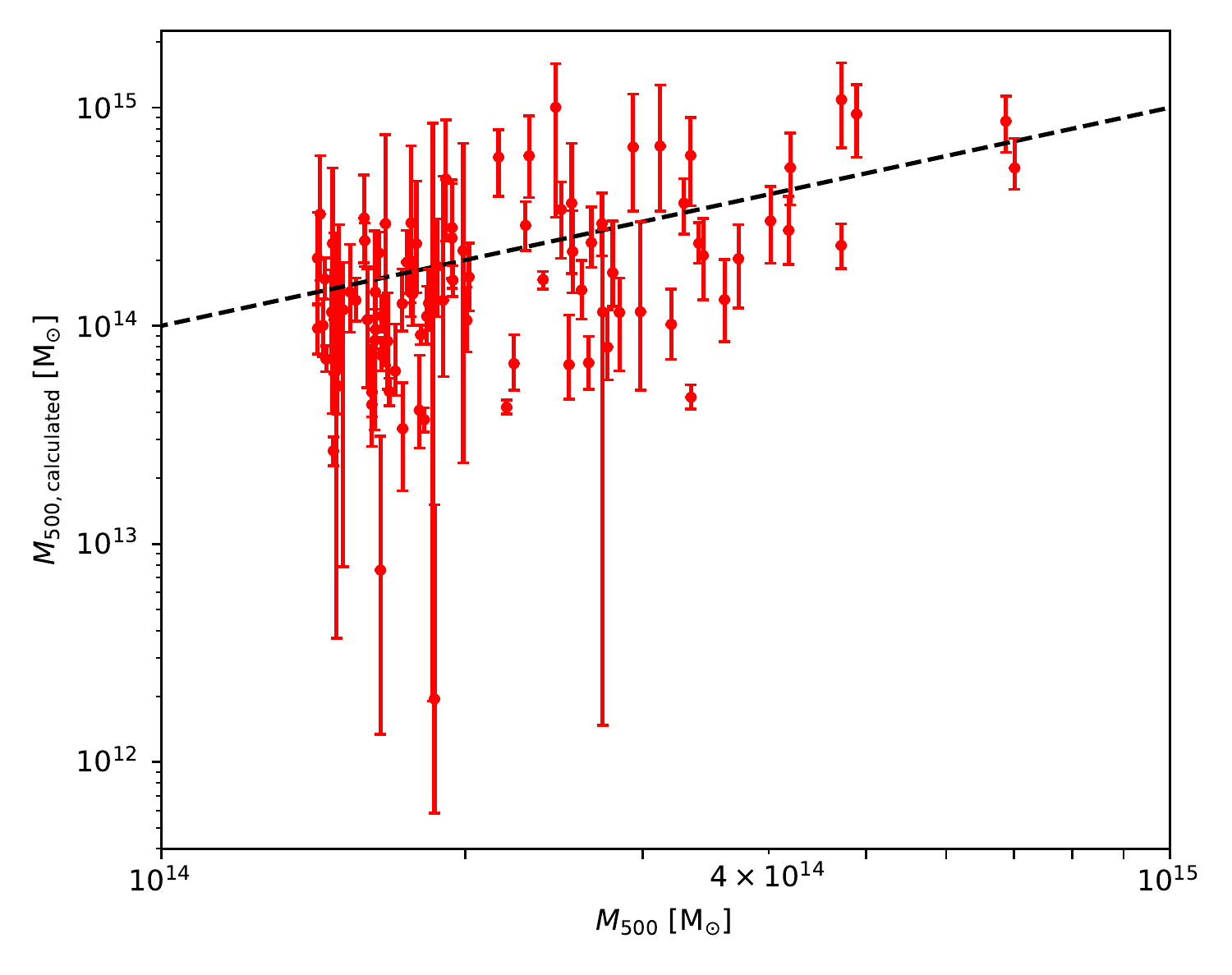}
    \includegraphics[width=0.485\textwidth]{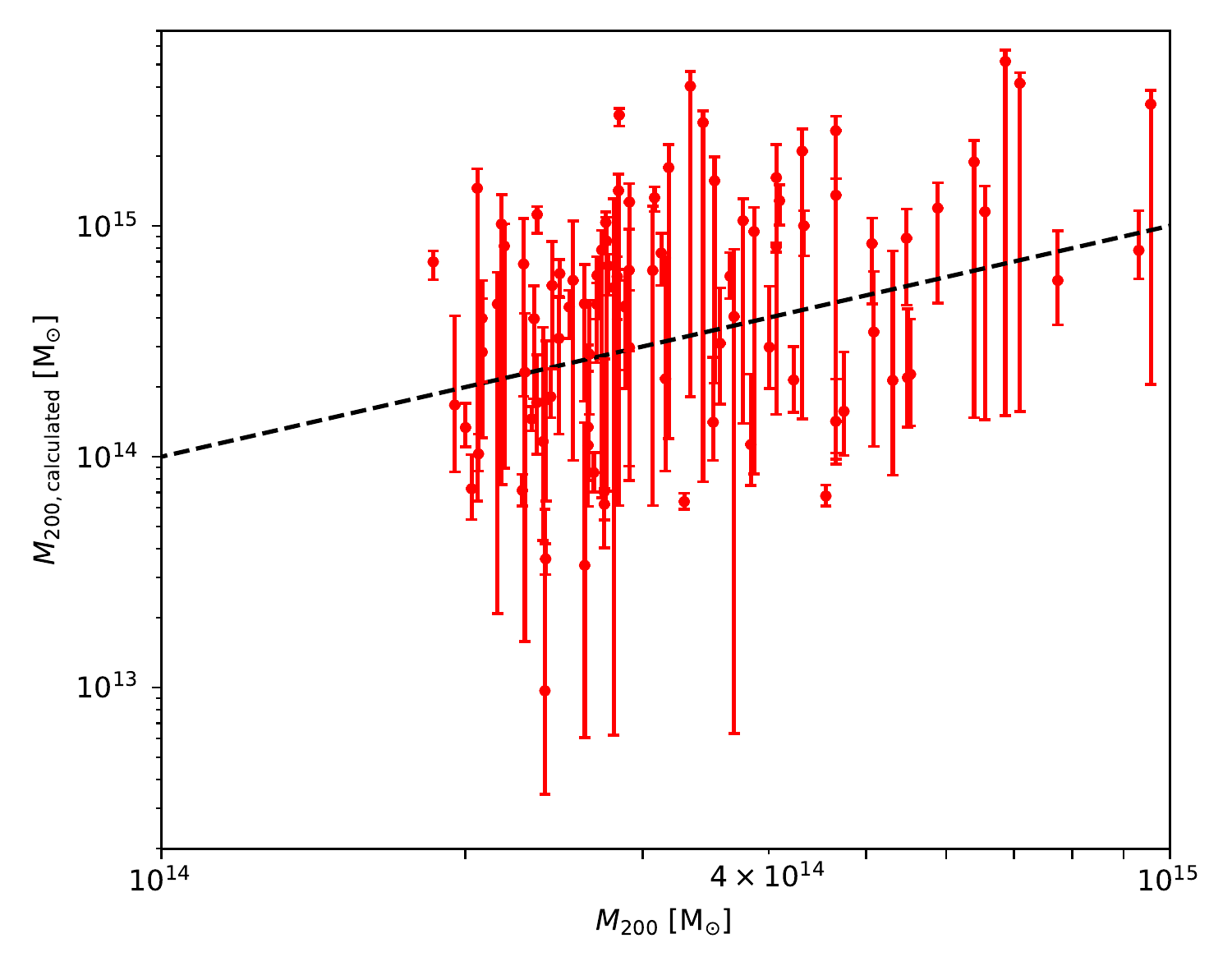}
    \caption{Realistic observations: calculated mass values $M_{\mathrm{500, calculated}}$ (upper panel)/ $M_{\mathrm{200, calculated}}$ (lower panel), based on the fits with MBProj2, as a function of true mass values according to the simulations $M_{\mathrm{500}}$/ $M_{\mathrm{200}}$, respectively.}
    \label{fig:M_calc_realistic}
\end{figure}

\begin{figure}
    \centering
    \includegraphics[width=0.23\textwidth]{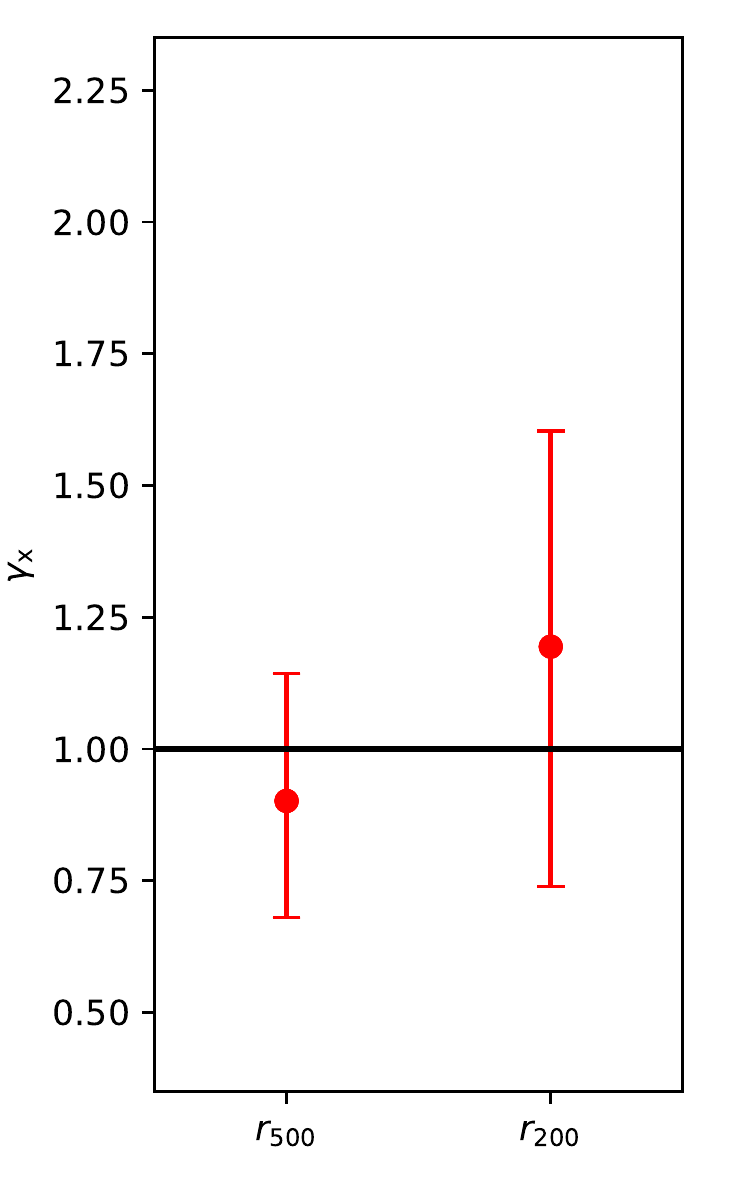}
    \hspace{0.01cm}
    \includegraphics[width=0.2343\textwidth]{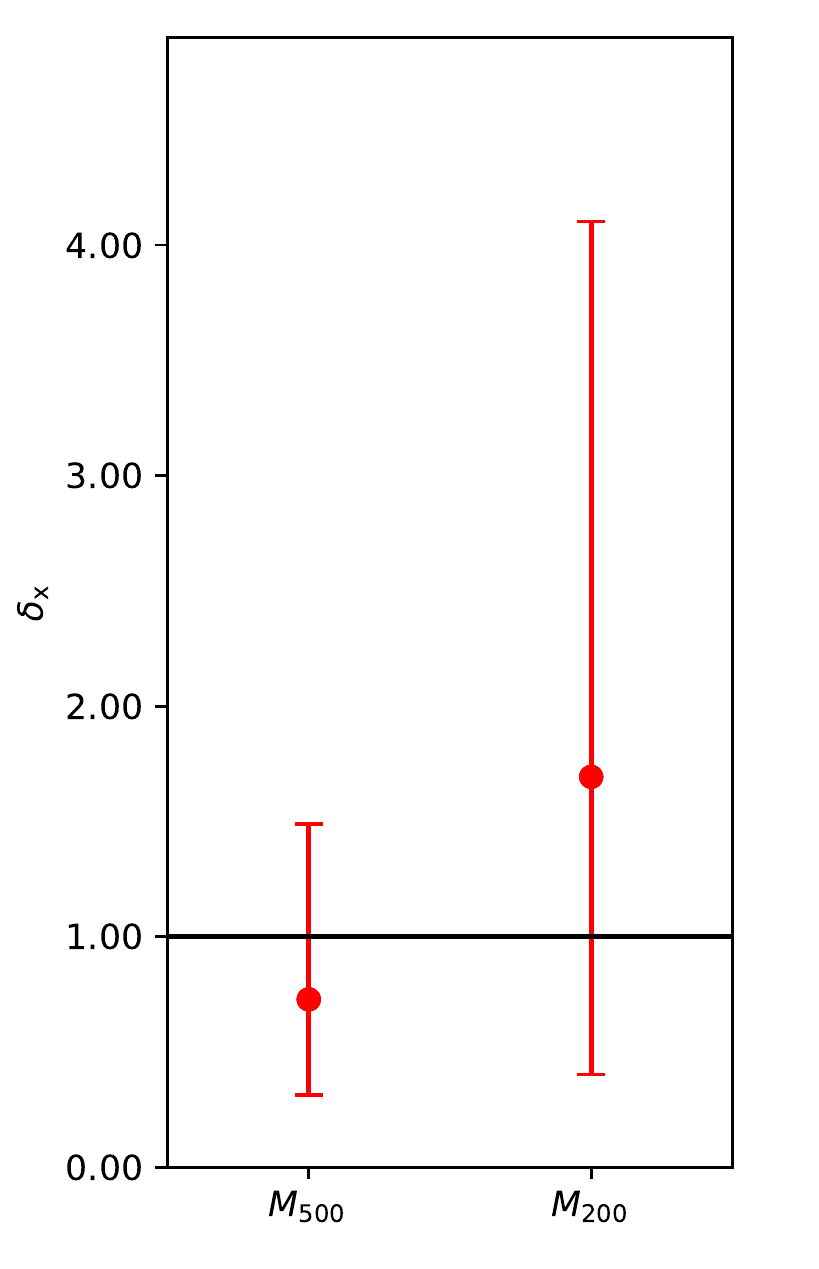}
    \caption{Realistic observations: medians and 1$\sigma$ intervals of $\gamma_{\mathrm{x}}$ and $\delta_{\mathrm{x}}$ (Eqs. \ref{equation_gamma_delta_definition}), based on the fits with MBProj2.}
    \label{fig:rM_calc_bars_realistic}
\end{figure}

Like for the idealized observations, we compute values for $r_{\mathrm{x, calculated}}$ and $M_{\mathrm{x, calculated}}$ based on the cumulative total mass profiles also for the realistic observations (see Sect. \ref{section_Calculation_of_rx_and_Mx_based_on_the_mass_profiles}). The calculated mass values for the individual clusters are shown in Fig. \ref{fig:M_calc_realistic}. Fig. \ref{fig:rM_calc_bars_realistic} shows the medians and 1$\sigma$ intervals of $\gamma$ and $\delta$ (Eqs. \ref{equation_gamma_delta_definition}) for the realistic observations. The medians of $\gamma_{\mathrm{200}}$ and $\delta_{\mathrm{200}}$ are remarkably higher than for the idealized observations; the 1$\sigma$ intervals are considerably larger in all cases.
The extremely large 1$\sigma$ interval of $\delta_{200}$ is probably also due to the fact that all values, where the corresponding $r_{\mathrm{200, calculated}}$ is substantially larger than the true $r_{\mathrm{200}}$, are based on a linear extrapolation of the fitted cumulative total mass profile (compare to Sect. \ref{section_Calculation_of_rx_and_Mx_based_on_the_mass_profiles}). The too high steepness of the mass profile then leads to a too high steepness of the extrapolation over a large radius range. This causes too large values of $r_{\mathrm{200, calculated}}$ and has even stronger implications for $M_{\mathrm{200, calculated}}$.

\begin{figure}
    \centering
    \subfigure[snap 140]{
    \includegraphics[width=0.1475\textwidth]{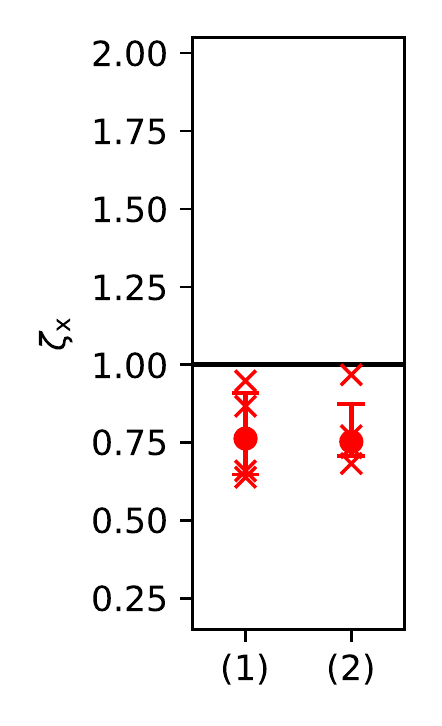}}
    \subfigure[snap 136]{
    \includegraphics[width=0.1002\textwidth]{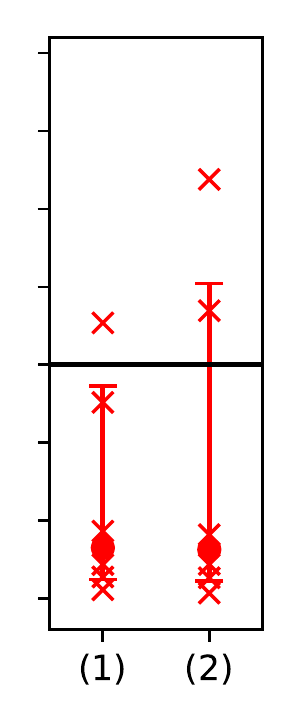}}
    \subfigure[snap 132]{
    \includegraphics[width=0.1002\textwidth]{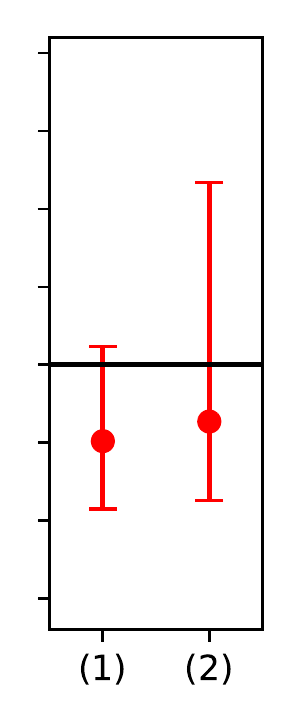}}
    \subfigure[snap 128]{
    \includegraphics[width=0.1475\textwidth]{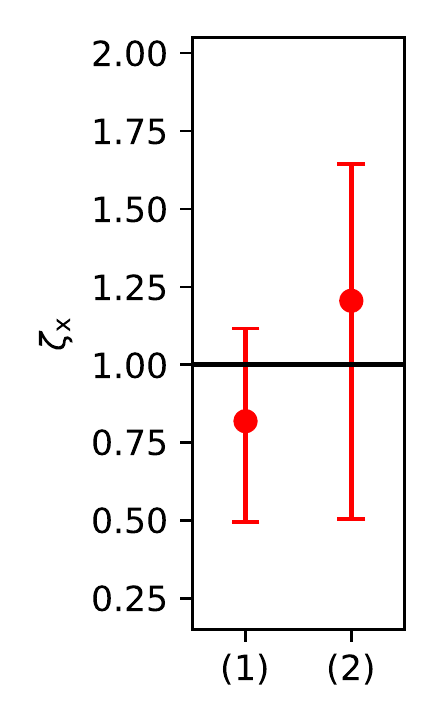}}
    \subfigure[snap 124]{
    \includegraphics[width=0.1002\textwidth]{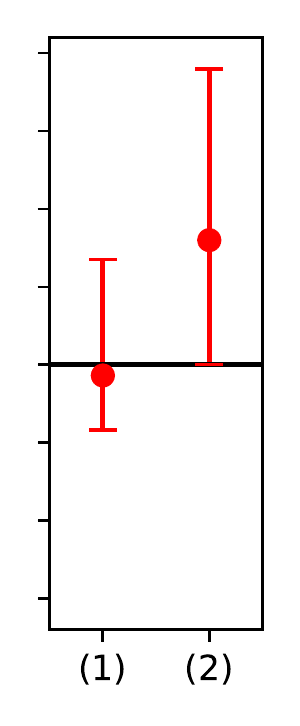}}
    \caption{Realistic observations: medians and 1$\sigma$ intervals of $\zeta_{\mathrm{x}}$ for each snap (comparable to Fig. \ref{fig:zeta_bars_realistic}). The results are shown at \newline (1) $r_{\mathrm{500}}$ for the fits with MBProj2 and \newline (2) $r_{\mathrm{200}}$ for the fits with MBProj2. \newline Each cluster in a snap has the same redshift (see Table \ref{tab:snaps}). Because of the low number of clusters in snap 140 and 136, the corresponding plots also show the values for the individual clusters as crosses.}
    \label{fig:part_plots_realistic}
\end{figure}

It is insightful to analyze the fraction $\zeta$ independently for each redshift again for the realistic observations (Fig. \ref{fig:part_plots_realistic}). We find that the trend towards systematically too high medians of $\zeta_{\mathrm{200}}$ is only discernible for the clusters in the snaps 128 and 124. For lower redshifts, the results are close to our expectations, neglecting some outliers in snap 136. Also the medians for $\zeta_{\mathrm{500}}$ for higher redshifts seem to be plausible. In contrast, the reaction of the fit to the realistic data quality causes a too high steepness especially for higher redshifts and therefore leads to unreasonable high values of $\zeta_{\mathrm{200}}$. Although the problem of too high steepness also exists for the idealized observations, the medians of $\zeta_{\mathrm{200}}$ show reasonable values there also for high redshifts.

\section{Discussion}

Our results for the fits with MBProj2 are partially based on the knowledge from the simulations, like the true values for $r_{\mathrm{500}}$ and $r_{\mathrm{200}}$. Specifically, we use a radius close to the true $r_{\mathrm{200}}$ as the maximum radius for the fit (see Sect. \ref{section_Fitting_process}). To avoid the influences of this choice, we perform the analysis of $\zeta$ at the true $r_{\mathrm{500}}$ and $r_{\mathrm{200}}$ for the clusters from the Magneticum Pathfinder simulations using a value for the maximum radius that is estimated by eye to roughly match $r_{\mathrm{200}}$. This estimate is based on the image of the cluster and naturally deviates from the true value for some clusters. All these estimated radii are larger than the true $r_{\mathrm{500}}$, but some of them are smaller than the true $r_{\mathrm{200}}$. In this case, we determine $\zeta$ at the true $r_{\mathrm{200}}$ using a linear extrapolation in logarithmic space. Like in the previous section, we determine the corresponding values for $\zeta$ at the true $r_{\mathrm{500}}$ and $r_{\mathrm{200}}$. For idealized observational conditions, we find that the corresponding median of $\zeta_{\mathrm{500}}$ deviates from the value in the previous section (see Fig. \ref{fig:zeta_bars}) by less than 0.5\%; for the median of $\zeta_{\mathrm{200}}$ we find a deviation of about 5\%. For realistic observational conditions, the corresponding deviations are about 1\% at $r_{\mathrm{500}}$ and about 4\% at $r_{\mathrm{200}}$. Also the 1$\sigma$ intervals are mostly similar for both choices of maximum radii for the fit.

\cite{Barnes_2021} also derive hydrostatic masses of galaxy clusters based on simulations, including clusters in a similar mass range. Further, they also investigate the fraction of hydrostatic mass to true mass. They find median values and 1$\sigma$ intervals in the same order of magnitude as our results, assuming idealized observational conditions.
They find the hydrostatic mass bias to be mass-dependent. In our results a clear dependence of the fraction $\zeta$ on the radius of the cluster is not discernible. Nevertheless, our results for the hydrostatic mass bias obtained with the fit with MBProj2 seem to be affected by the problem of too high steepness.

We recognize that it is difficult to predict if results for a cluster will be close to the true values by looking at the image of the cluster. As a dependence of $\zeta$ on the relative shift of the center is not clearly visible, we suspect that also other aspects have an influence on the results, like for example the degree of symmetry of the cluster \citep[see e.g.][]{Piffaretti_2003, Morandi_2010, Eckert_2015}.
The validity of hydrostatic equilibrium in the clusters could also be affected by mergers \citep[see e.g.][]{Nelson_2012, Reiprich_2004, Puchwein_2007}.
To gain more knowledge about individual clusters and the validity of the assumption of hydrostatic equilibrium, hydrostatic masses can be compared to results of other measurements \citep{Ettori19}, like X-ray based scaling relations \citep{Vikhlinin_2009}, galaxy dynamics \citep{Zhang_2017, Rines_2016}, weak lensing \citep{Zhang_2010} and the Sunyaev–Zeldovich effect \citep{Kay_2012}. \cite{Ettori19} identified the masses of several different measurements to be mostly congruent with the results based on the assumption of hydrostatic equilibrium.

\section{Conclusions}

We investigate hydrostatic equilibrium in clusters of galaxies using a sample of 93 clusters from the Magneticum Pathfinder simulations as well as several clusters with idealized properties.
We fit hydrostatic models to the X-ray images of the clusters using the fitting code MBProj2, while testing different models for the electron density in the ICM and for the dark matter mass.
For comparison, we compute hydrostatic mass profiles using the true profiles for gas electron density and pressure known from the simulations.
We draw the following conclusions:

\begin{enumerate}
    \item Assuming isothermal clusters with different temperatures under idealized observational conditions, MBProj2 is able to fit flat temperature profiles that deviate from the true values by a few percent at the maximum; for most cases, the deviations are at about 1\% or smaller.
    \item Under idealized observational conditions, the cumulative total mass profile of a cluster fulfilling perfect hydrostatic equilibrium can be reproduced such that the deviations of the cumulative total mass at the true $r_{\mathrm{500}}$ and $r_{\mathrm{200}}$ are below 7\%; this holds for all tested models.
    \item Considering the clusters from the Magneticum Pathfinder simulations under idealized observational conditions and assuming a modified $\beta$-model with one $\beta$-component for the gas electron density and an NFW model for the dark matter mass, the fit with MBProj2 leads to medians for the fraction $\zeta$ (Eq. \ref{equation_zeta}) that are between 0.77 and 0.9, which roughly fits to our expectations of $b \sim 0.1-0.2$ as found from simulations (see Sect. \ref{section_Introduction}). The 1$\sigma$ interval for $\zeta_{\mathrm{500}}$ has an adequate size, while the results for $\zeta_{\mathrm{200}}$ exhibit a noticeably larger scattering amplitude. Furthermore, we recognize an on average too large steepness of the fitted cumulative mass profiles.
    \item The hydrostatic mass profiles based on theory show results that match our assumptions for almost all clusters: The medians of $\zeta_{\mathrm{500}}$ and $\zeta_{\mathrm{200}}$ are congruent with the expectations and the corresponding 1$\sigma$ intervals are considerably smaller than for the fit with MBProj2.
    \item The shift of the center position, possibly used as a proxy for the dynamic state of a cluster, does not make a substantial difference in the accuracy of hydrostatic masses in our case.
    \item Considering realistic eROSITA data quality, we find medians and 1$\sigma$ intervals for $\zeta_{\mathrm{500}}$ which are consistent with our expectations; for $\zeta_{\mathrm{200}}$ we find too large medians for clusters with higher redshifts.
    \item The values for $r_{\mathrm{500, calculated}}$, $r_{\mathrm{200, calculated}}$, $M_{\mathrm{500, calculated}}$ and $M_{\mathrm{200, calculated}}$ - calculated based on the cumulative mass profiles and the critical density of the universe at the corresponding redshift - show reasonable medians for idealized observational conditions; however, they also seem to be affected by the problem of the too high steepness of the fitted cumulative total mass profiles. For realistic observational conditions, the magnified steepness problem leads to too high medians of $\gamma_{\mathrm{200}}$ and $\delta_{\mathrm{200}}$ (see Eqs. \ref{equation_gamma_delta_definition}); these two quantities further show unreasonable large 1$\sigma$ intervals, probably also driven by the extrapolation of the profiles.
    \item Hydrostatic masses obtained with MBProj2 are an interesting and helpful method, also in the context of the eROSITA survey.
\end{enumerate}

A future application of mass measurements of galaxy clusters with MBProj2 can be cosmological studies, like already mentioned in Sect. \ref{section_Introduction}.
For this, the code will be applied to real data from eROSITA. The exact behavior of the fit as well as further issues with real data can be the subject of future work. Furthermore, it can be a future prospect to compare the results from real data to the results obtained with Chandra and XMM-Newton and to those from weak lensing. In summary, there will be a number of interesting future projects that are likely to yield insightful results.


\begin{acknowledgements}
      The Magneticum Pathfinder simulations have been performed at the Leibniz-Rechenzentrum with CPU time assigned to the projects pr86re and pr83li.
      The synthetic data shown here were processed using the eSASS software system developed by the German eROSITA consortium.
      KD acknowledges support by the COMPLEX project from the European Research Council (ERC) under the European Union’s Horizon 2020 research and innovation program grant agreement ERC-2019-AdG 882679 and support by the Deutsche Forschungsgemeinschaft (DFG, German Research Foundation) under Germany’s Excellence Strategy - EXC-2094 - 390783311.
      VB acknowledges support by the DFG project nr. 415510302. EB acknowledge financial support from the European Research Council (ERC) Consolidator Grant under the European Union's Horizon 2020 research and innovation programme (grant agreement CoG DarkQuest No 101002585).
\end{acknowledgements}

\tiny{\noindent Parts of the content are identical or similar to the master's thesis \qq{Mass Profiles of Galaxy Clusters in the eROSITA survey} by Dominik Scheck, submitted on October 19, 2021 at the Department of Physics of the Technical University Munich.}

%
%

\bibliographystyle{aa}
\bibliography{aanda}

\begin{thebibliography}{78}
\expandafter\ifx\csname natexlab\endcsname\relax\def\natexlab#1{#1}\fi

\bibitem[{{Anders} \& {Grevesse}(1989)}]{Anders_1989}
{Anders}, E. \& {Grevesse}, N. 1989, \gca, 53, 197

\bibitem[{{Arnaud}(1996)}]{Arnaud_1996}
{Arnaud}, K.~A. 1996, in Astronomical Society of the Pacific Conference Series,
  Vol. 101, Astronomical Data Analysis Software and Systems V, ed. G.~H.
  {Jacoby} \& J.~{Barnes}, 17

\bibitem[{{Balucinska-Church} \& {McCammon}(1992)}]{Balucinska-Church_1992}
{Balucinska-Church}, M. \& {McCammon}, D. 1992, \apj, 400, 699

\bibitem[{{Barnes} {et~al.}(2021){Barnes}, {Vogelsberger}, {Pearce}, {Pop},
  {Kannan}, {Cao}, {Kay}, \& {Hernquist}}]{Barnes_2021}
{Barnes}, D.~J., {Vogelsberger}, M., {Pearce}, F.~A., {et~al.} 2021, \mnras,
  506, 2533

\bibitem[{{Biffi} {et~al.}(2016){Biffi}, {Borgani}, {Murante}, {Rasia},
  {Planelles}, {Granato}, {Ragone-Figueroa}, {Beck}, {Gaspari}, \&
  {Dolag}}]{Biffi16}
{Biffi}, V., {Borgani}, S., {Murante}, G., {et~al.} 2016, \apj, 827, 112

\bibitem[{{Biffi} {et~al.}(2013){Biffi}, {Dolag}, \&
  {B{\"o}hringer}}]{Biffi_2013}
{Biffi}, V., {Dolag}, K., \& {B{\"o}hringer}, H. 2013, \mnras, 428, 1395

\bibitem[{{Biffi} {et~al.}(2012){Biffi}, {Dolag}, {B{\"o}hringer}, \&
  {Lemson}}]{Biffi_2012}
{Biffi}, V., {Dolag}, K., {B{\"o}hringer}, H., \& {Lemson}, G. 2012, \mnras,
  420, 3545

\bibitem[{{Biffi} {et~al.}(2018){Biffi}, {Dolag}, \& {Merloni}}]{Biffi_2018}
{Biffi}, V., {Dolag}, K., \& {Merloni}, A. 2018, \mnras, 481, 2213

\bibitem[{{Biffi} {et~al.}(2022){Biffi}, {Dolag}, {Reiprich}, {Veronica},
  {Ramos-Ceja}, {Bulbul}, {Ota}, \& {Ghirardini}}]{Biffi_2021}
{Biffi}, V., {Dolag}, K., {Reiprich}, T.~H., {et~al.} 2022, \aap, 661, A17

\bibitem[{{Boehringer} \& {Werner}(2009)}]{Boehringer_2009}
{Boehringer}, H. \& {Werner}, N. 2009, arXiv e-prints, arXiv:0907.4277

\bibitem[{{Brunner} {et~al.}(2018){Brunner}, {Boller}, {Coutinho}, {Dauser},
  {Dennerl}, {Dwelly}, {Freyberg}, {F{\"u}rmetz}, {Georgakakis}, {Grossberger},
  {Kreykenbohm}, {Lamer}, {Meidinger}, {M{\"u}ller}, {Predehl}, {Robrade},
  {Sanders}, \& {Wilms}}]{Brunner_2018}
{Brunner}, H., {Boller}, T., {Coutinho}, D., {et~al.} 2018, in Society of
  Photo-Optical Instrumentation Engineers (SPIE) Conference Series, Vol. 10699,
  Space Telescopes and Instrumentation 2018: Ultraviolet to Gamma Ray, ed.
  J.-W.~A. {den Herder}, S.~{Nikzad}, \& K.~{Nakazawa}, 106995G

\bibitem[{{Brunner} {et~al.}(2022){Brunner}, {Liu}, {Lamer}, {Georgakakis},
  {Merloni}, {Brusa}, {Bulbul}, {Dennerl}, {Friedrich}, {Liu}, {Maitra},
  {Nandra}, {Ramos-Ceja}, {Sanders}, {Stewart}, {Boller}, {Buchner}, {Clerc},
  {Comparat}, {Dwelly}, {Eckert}, {Finoguenov}, {Freyberg}, {Ghirardini},
  {Gueguen}, {Haberl}, {Kreykenbohm}, {Krumpe}, {Osterhage}, {Pacaud},
  {Predehl}, {Reiprich}, {Robrade}, {Salvato}, {Santangelo}, {Schrabback},
  {Schwope}, \& {Wilms}}]{Brunner_2021}
{Brunner}, H., {Liu}, T., {Lamer}, G., {et~al.} 2022, \aap, 661, A1

\bibitem[{{Bryan} \& {Norman}(1998)}]{Bryan_1998}
{Bryan}, G.~L. \& {Norman}, M.~L. 1998, \apj, 495, 80

\bibitem[{{Bulbul} {et~al.}(2022){Bulbul}, {Liu}, {Pasini}, {Comparat},
  {Hoang}, {Klein}, {Ghirardini}, {Salvato}, {Merloni}, {Seppi}, {Wolf},
  {Anderson}, {Bahar}, {Brusa}, {Br{\"u}ggen}, {Buchner}, {Dwelly},
  {Ibarra-Medel}, {Ider Chitham}, {Liu}, {Nandra}, {Ramos-Ceja}, {Sanders}, \&
  {Shen}}]{Bulbul_2022}
{Bulbul}, E., {Liu}, A., {Pasini}, T., {et~al.} 2022, \aap, 661, A10

\bibitem[{{Cavaliere} \& {Fusco-Femiano}(1978)}]{Cavaliere_1978}
{Cavaliere}, A. \& {Fusco-Femiano}, R. 1978, \aap, 70, 677

\bibitem[{{Chiu} {et~al.}(2016){Chiu}, {Mohr}, {McDonald}, {Bocquet}, {Ashby},
  {Bayliss}, {Benson}, {Bleem}, {Brodwin}, {Desai}, {Dietrich}, {Forman},
  {Gangkofner}, {Gonzalez}, {Hennig}, {Liu}, {Reichardt}, {Saro}, {Stalder},
  {Stanford}, {Song}, {Schrabback}, {{\v{S}}uhada}, {Strazzullo}, \&
  {Zenteno}}]{Chiu_2016}
{Chiu}, I., {Mohr}, J., {McDonald}, M., {et~al.} 2016, \mnras, 455, 258

\bibitem[{{Chiu} {et~al.}(2022){Chiu}, {Ghirardini}, {Liu}, {Grandis},
  {Bulbul}, {Bahar}, {Comparat}, {Bocquet}, {Clerc}, {Klein}, {Liu}, {Li},
  {Miyatake}, {Mohr}, {More}, {Oguri}, {Okabe}, {Pacaud}, {Ramos-Ceja},
  {Reiprich}, {Schrabback}, \& {Umetsu}}]{Chiu21}
{Chiu}, I.~N., {Ghirardini}, V., {Liu}, A., {et~al.} 2022, \aap, 661, A11

\bibitem[{{Chon} \& {B{\"o}hringer}(2017)}]{Chon_2017}
{Chon}, G. \& {B{\"o}hringer}, H. 2017, \aap, 606, L4

\bibitem[{{Clerc} {et~al.}(2018){Clerc}, {Ramos-Ceja}, {Ridl}, {Lamer},
  {Brunner}, {Hofmann}, {Comparat}, {Pacaud}, {K{\"a}fer}, {Reiprich},
  {Merloni}, {Schmid}, {Brand}, {Wilms}, {Friedrich}, {Finoguenov}, {Dauser},
  \& {Kreykenbohm}}]{Clerc_2018}
{Clerc}, N., {Ramos-Ceja}, M.~E., {Ridl}, J., {et~al.} 2018, \aap, 617, A92

\bibitem[{{Dauser} {et~al.}(2019){Dauser}, {Falkner}, {Lorenz}, {Kirsch},
  {Peille}, {Cucchetti}, {Schmid}, {Brand}, {Oertel}, {Smith}, \&
  {Wilms}}]{Dauser_2019}
{Dauser}, T., {Falkner}, S., {Lorenz}, M., {et~al.} 2019, \aap, 630, A66

\bibitem[{{Eckert} {et~al.}(2015){Eckert}, {Roncarelli}, {Ettori}, {Molendi},
  {Vazza}, {Gastaldello}, \& {Rossetti}}]{Eckert_2015}
{Eckert}, D., {Roncarelli}, M., {Ettori}, S., {et~al.} 2015, \mnras, 447, 2198

\bibitem[{{Ettori} {et~al.}(2013){Ettori}, {Donnarumma}, {Pointecouteau},
  {Reiprich}, {Giodini}, {Lovisari}, \& {Schmidt}}]{Ettori_2013}
{Ettori}, S., {Donnarumma}, A., {Pointecouteau}, E., {et~al.} 2013, \ssr, 177,
  119

\bibitem[{{Ettori} {et~al.}(2019){Ettori}, {Ghirardini}, {Eckert},
  {Pointecouteau}, {Gastaldello}, {Sereno}, {Gaspari}, {Ghizzardi},
  {Roncarelli}, \& {Rossetti}}]{Ettori19}
{Ettori}, S., {Ghirardini}, V., {Eckert}, D., {et~al.} 2019, \aap, 621, A39

\bibitem[{{Foreman-Mackey} {et~al.}(2013){Foreman-Mackey}, {Hogg}, {Lang}, \&
  {Goodman}}]{Foreman-Mackey_2013}
{Foreman-Mackey}, D., {Hogg}, D.~W., {Lang}, D., \& {Goodman}, J. 2013, \pasp,
  125, 306

\bibitem[{{George} {et~al.}(2007){George}, {Arnaud}, {Pence}, {Ruamsuwan}, \&
  {Corcoran}}]{George_2007}
{George}, I.~M., {Arnaud}, K.~A., {Pence}, B., {Ruamsuwan}, L., \& {Corcoran},
  M.~F. 2007, The Calibration Requirements for Spectral Analysis (Definition of
  RMF and ARF file formats), Technical report, Code 662, NASA/GSFC, Greenbelt,
  MD20771

\bibitem[{{Ghirardini} {et~al.}(2018){Ghirardini}, {Ettori}, {Eckert},
  {Molendi}, {Gastaldello}, {Pointecouteau}, {Hurier}, \&
  {Bourdin}}]{Ghirardini18}
{Ghirardini}, V., {Ettori}, S., {Eckert}, D., {et~al.} 2018, \aap, 614, A7

\bibitem[{{Goodman} \& {Weare}(2010)}]{Goodman_2010}
{Goodman}, J. \& {Weare}, J. 2010, Communications in Applied Mathematics and
  Computational Science, 5, 65

\bibitem[{{Grandis} {et~al.}(2021){Grandis}, {Bocquet}, {Mohr}, {Klein}, \&
  {Dolag}}]{Grandis21}
{Grandis}, S., {Bocquet}, S., {Mohr}, J.~J., {Klein}, M., \& {Dolag}, K. 2021,
  \mnras, 507, 5671

\bibitem[{{Hoekstra} {et~al.}(2015){Hoekstra}, {Herbonnet}, {Muzzin}, {Babul},
  {Mahdavi}, {Viola}, \& {Cacciato}}]{Hoekstra_2015}
{Hoekstra}, H., {Herbonnet}, R., {Muzzin}, A., {et~al.} 2015, \mnras, 449, 685

\bibitem[{{Kay} {et~al.}(2012){Kay}, {Peel}, {Short}, {Thomas}, {Young},
  {Battye}, {Liddle}, \& {Pearce}}]{Kay_2012}
{Kay}, S.~T., {Peel}, M.~W., {Short}, C.~J., {et~al.} 2012, \mnras, 422, 1999

\bibitem[{{Komatsu} {et~al.}(2011){Komatsu}, {Smith}, {Dunkley}, {Bennett},
  {Gold}, {Hinshaw}, {Jarosik}, {Larson}, {Nolta}, {Page}, {Spergel},
  {Halpern}, {Hill}, {Kogut}, {Limon}, {Meyer}, {Odegard}, {Tucker}, {Weiland},
  {Wollack}, \& {Wright}}]{Komatsu_2011}
{Komatsu}, E., {Smith}, K.~M., {Dunkley}, J., {et~al.} 2011, \apjs, 192, 18

\bibitem[{{Lau} {et~al.}(2009){Lau}, {Kravtsov}, \& {Nagai}}]{Lau_2009}
{Lau}, E.~T., {Kravtsov}, A.~V., \& {Nagai}, D. 2009, \apj, 705, 1129

\bibitem[{{Lau} {et~al.}(2013){Lau}, {Nagai}, \& {Nelson}}]{Lau13}
{Lau}, E.~T., {Nagai}, D., \& {Nelson}, K. 2013, \apj, 777, 151

\bibitem[{{Liu} {et~al.}(2022){Liu}, {Bulbul}, {Ghirardini}, {Liu}, {Klein},
  {Clerc}, {{\"O}zsoy}, {Ramos-Ceja}, {Pacaud}, {Comparat}, {Okabe}, {Bahar},
  {Biffi}, {Brunner}, {Br{\"u}ggen}, {Buchner}, {Ider Chitham}, {Chiu},
  {Dolag}, {Gatuzz}, {Gonzalez}, {Hoang}, {Lamer}, {Merloni}, {Nandra},
  {Oguri}, {Ota}, {Predehl}, {Reiprich}, {Salvato}, {Schrabback}, {Sanders},
  {Seppi}, \& {Thibaud}}]{Liu_2022}
{Liu}, A., {Bulbul}, E., {Ghirardini}, V., {et~al.} 2022, \aap, 661, A2

\bibitem[{{Martino} {et~al.}(2014){Martino}, {Mazzotta}, {Bourdin}, {Smith},
  {Bartalucci}, {Marrone}, {Finoguenov}, \& {Okabe}}]{Martino14}
{Martino}, R., {Mazzotta}, P., {Bourdin}, H., {et~al.} 2014, \mnras, 443, 2342

\bibitem[{{Mazzotta} {et~al.}(2004){Mazzotta}, {Rasia}, {Moscardini}, \&
  {Tormen}}]{Mazzotta_2004}
{Mazzotta}, P., {Rasia}, E., {Moscardini}, L., \& {Tormen}, G. 2004, \mnras,
  354, 10

\bibitem[{{Merloni} {et~al.}(2012){Merloni}, {Predehl}, {Becker},
  {B{\"o}hringer}, {Boller}, {Brunner}, {Brusa}, {Dennerl}, {Freyberg},
  {Friedrich}, {Georgakakis}, {Haberl}, {Hasinger}, {Meidinger}, {Mohr},
  {Nandra}, {Rau}, {Reiprich}, {Robrade}, {Salvato}, {Santangelo}, {Sasaki},
  {Schwope}, {Wilms}, \& {German eROSITA Consortium}}]{Merloni_2012}
{Merloni}, A., {Predehl}, P., {Becker}, W., {et~al.} 2012, arXiv e-prints,
  arXiv:1209.3114

\bibitem[{{Mohr} {et~al.}(1993){Mohr}, {Fabricant}, \& {Geller}}]{Mohr_1993}
{Mohr}, J.~J., {Fabricant}, D.~G., \& {Geller}, M.~J. 1993, \apj, 413, 492

\bibitem[{{Morandi} {et~al.}(2010){Morandi}, {Pedersen}, \&
  {Limousin}}]{Morandi_2010}
{Morandi}, A., {Pedersen}, K., \& {Limousin}, M. 2010, \apj, 713, 491

\bibitem[{{Morrison} \& {McCammon}(1983)}]{Morrison_1983}
{Morrison}, R. \& {McCammon}, D. 1983, \apj, 270, 119

\bibitem[{{Navarro} {et~al.}(1996){Navarro}, {Frenk}, \&
  {White}}]{Navarro_1996}
{Navarro}, J.~F., {Frenk}, C.~S., \& {White}, S. D.~M. 1996, \apj, 462, 563

\bibitem[{{Nelson} {et~al.}(2012){Nelson}, {Rudd}, {Shaw}, \&
  {Nagai}}]{Nelson_2012}
{Nelson}, K., {Rudd}, D.~H., {Shaw}, L., \& {Nagai}, D. 2012, \apj, 751, 121

\bibitem[{{Piffaretti} {et~al.}(2003){Piffaretti}, {Jetzer}, \&
  {Schindler}}]{Piffaretti_2003}
{Piffaretti}, R., {Jetzer}, P., \& {Schindler}, S. 2003, \aap, 398, 41

\bibitem[{{Pillepich} {et~al.}(2012){Pillepich}, {Porciani}, \&
  {Reiprich}}]{Pillepich12}
{Pillepich}, A., {Porciani}, C., \& {Reiprich}, T.~H. 2012, \mnras, 422, 44

\bibitem[{{Pointecouteau} {et~al.}(2004){Pointecouteau}, {Arnaud}, {Kaastra},
  \& {de Plaa}}]{Pointecouteau_2004}
{Pointecouteau}, E., {Arnaud}, M., {Kaastra}, J., \& {de Plaa}, J. 2004, \aap,
  423, 33

\bibitem[{{Pointecouteau} {et~al.}(2005){Pointecouteau}, {Arnaud}, \&
  {Pratt}}]{Pointecouteau05}
{Pointecouteau}, E., {Arnaud}, M., \& {Pratt}, G.~W. 2005, \aap, 435, 1

\bibitem[{{Poole} {et~al.}(2006){Poole}, {Fardal}, {Babul}, {McCarthy},
  {Quinn}, \& {Wadsley}}]{Poole_2006}
{Poole}, G.~B., {Fardal}, M.~A., {Babul}, A., {et~al.} 2006, \mnras, 373, 881

\bibitem[{{Pratt} {et~al.}(2019){Pratt}, {Arnaud}, {Biviano}, {Eckert},
  {Ettori}, {Nagai}, {Okabe}, \& {Reiprich}}]{Pratt_2019}
{Pratt}, G.~W., {Arnaud}, M., {Biviano}, A., {et~al.} 2019, \ssr, 215, 25

\bibitem[{{Predehl} {et~al.}(2021){Predehl}, {Andritschke}, {Arefiev},
  {Babyshkin}, {Batanov}, {Becker}, {B{\"o}hringer}, {Bogomolov}, {Boller},
  {Borm}, {Bornemann}, {Br{\"a}uninger}, {Br{\"u}ggen}, {Brunner}, {Brusa},
  {Bulbul}, {Buntov}, {Burwitz}, {Burkert}, {Clerc}, {Churazov}, {Coutinho},
  {Dauser}, {Dennerl}, {Doroshenko}, {Eder}, {Emberger}, {Eraerds},
  {Finoguenov}, {Freyberg}, {Friedrich}, {Friedrich}, {F{\"u}rmetz},
  {Georgakakis}, {Gilfanov}, {Granato}, {Grossberger}, {Gueguen}, {Gureev},
  {Haberl}, {H{\"a}lker}, {Hartner}, {Hasinger}, {Huber}, {Ji}, {Kienlin},
  {Kink}, {Korotkov}, {Kreykenbohm}, {Lamer}, {Lomakin}, {Lapshov}, {Liu},
  {Maitra}, {Meidinger}, {Menz}, {Merloni}, {Mernik}, {Mican}, {Mohr},
  {M{\"u}ller}, {Nandra}, {Nazarov}, {Pacaud}, {Pavlinsky}, {Perinati},
  {Pfeffermann}, {Pietschner}, {Ramos-Ceja}, {Rau}, {Reiffers}, {Reiprich},
  {Robrade}, {Salvato}, {Sanders}, {Santangelo}, {Sasaki}, {Scheuerle},
  {Schmid}, {Schmitt}, {Schwope}, {Shirshakov}, {Steinmetz}, {Stewart},
  {Str{\"u}der}, {Sunyaev}, {Tenzer}, {Tiedemann}, {Tr{\"u}mper}, {Voron},
  {Weber}, {Wilms}, \& {Yaroshenko}}]{Predehl_2021}
{Predehl}, P., {Andritschke}, R., {Arefiev}, V., {et~al.} 2021, \aap, 647, A1

\bibitem[{{Puchwein} \& {Bartelmann}(2007)}]{Puchwein_2007}
{Puchwein}, E. \& {Bartelmann}, M. 2007, \aap, 474, 745

\bibitem[{{Rasia} {et~al.}(2013){Rasia}, {Meneghetti}, \&
  {Ettori}}]{Rasia_2013}
{Rasia}, E., {Meneghetti}, M., \& {Ettori}, S. 2013, The Astronomical Review,
  8, 40

\bibitem[{{Reiprich} {et~al.}(2004){Reiprich}, {Sarazin}, {Kempner}, \&
  {Tittley}}]{Reiprich_2004}
{Reiprich}, T.~H., {Sarazin}, C.~L., {Kempner}, J.~C., \& {Tittley}, E. 2004,
  \apj, 608, 179

\bibitem[{{Retana-Montenegro} {et~al.}(2012){Retana-Montenegro}, {van Hese},
  {Gentile}, {Baes}, \& {Frutos-Alfaro}}]{Retana-Montenegro_2012}
{Retana-Montenegro}, E., {van Hese}, E., {Gentile}, G., {Baes}, M., \&
  {Frutos-Alfaro}, F. 2012, \aap, 540, A70

\bibitem[{{Rines} {et~al.}(2016){Rines}, {Geller}, {Diaferio}, \&
  {Hwang}}]{Rines_2016}
{Rines}, K.~J., {Geller}, M.~J., {Diaferio}, A., \& {Hwang}, H.~S. 2016, \apj,
  819, 63

\bibitem[{{Sanders} {et~al.}(2022){Sanders}, {Biffi}, {Br{\"u}ggen}, {Bulbul},
  {Dennerl}, {Dolag}, {Erben}, {Freyberg}, {Gatuzz}, {Ghirardini}, {Hoang},
  {Klein}, {Liu}, {Merloni}, {Pacaud}, {Ramos-Ceja}, {Reiprich}, \&
  {ZuHone}}]{Sanders_2021}
{Sanders}, J.~S., {Biffi}, V., {Br{\"u}ggen}, M., {et~al.} 2022, \aap, 661, A36

\bibitem[{{Sanders} {et~al.}(2014){Sanders}, {Fabian}, {Hlavacek-Larrondo},
  {Russell}, {Taylor}, {Hofmann}, {Tremblay}, \& {Walker}}]{Sanders_2014}
{Sanders}, J.~S., {Fabian}, A.~C., {Hlavacek-Larrondo}, J., {et~al.} 2014,
  \mnras, 444, 1497

\bibitem[{{Sanders} {et~al.}(2018){Sanders}, {Fabian}, {Russell}, \&
  {Walker}}]{Sanders_2018}
{Sanders}, J.~S., {Fabian}, A.~C., {Russell}, H.~R., \& {Walker}, S.~A. 2018,
  \mnras, 474, 1065

\bibitem[{{Schellenberger} {et~al.}(2015){Schellenberger}, {Reiprich},
  {Lovisari}, {Nevalainen}, \& {David}}]{Schellenberger15}
{Schellenberger}, G., {Reiprich}, T.~H., {Lovisari}, L., {Nevalainen}, J., \&
  {David}, L. 2015, \aap, 575, A30

\bibitem[{{Schmidt} \& {Allen}(2007)}]{Schmidt_2007}
{Schmidt}, R.~W. \& {Allen}, S.~W. 2007, \mnras, 379, 209

\bibitem[{{Schoenberg}(1964)}]{Schoenberg_1964}
{Schoenberg}, I.~J. 1964, Proceedings of the National Academy of Science, 51,
  24

\bibitem[{{Simionescu} {et~al.}(2017){Simionescu}, {Werner}, {Mantz}, {Allen},
  \& {Urban}}]{Simionescu_2017}
{Simionescu}, A., {Werner}, N., {Mantz}, A., {Allen}, S.~W., \& {Urban}, O.
  2017, \mnras, 469, 1476

\bibitem[{{Smith} {et~al.}(2001){Smith}, {Brickhouse}, {Liedahl}, \&
  {Raymond}}]{Smith_2001}
{Smith}, R.~K., {Brickhouse}, N.~S., {Liedahl}, D.~A., \& {Raymond}, J.~C.
  2001, \apjl, 556, L91

\bibitem[{{Springel} \& {Hernquist}(2003{\natexlab{a}})}]{Springel_2003_3}
{Springel}, V. \& {Hernquist}, L. 2003{\natexlab{a}}, in Astrophysical
  Supercomputing using Particle Simulations, ed. J.~{Makino} \& P.~{Hut}, Vol.
  208, 273

\bibitem[{{Springel} \& {Hernquist}(2003{\natexlab{b}})}]{Springel_2003_2}
{Springel}, V. \& {Hernquist}, L. 2003{\natexlab{b}}, \mnras, 339, 289

\bibitem[{{Springel} \& {Hernquist}(2003{\natexlab{c}})}]{Springel_2003}
{Springel}, V. \& {Hernquist}, L. 2003{\natexlab{c}}, \mnras, 339, 312

\bibitem[{{Tanabashi} {et~al.}(2018){Tanabashi}, {Hagiwara}, {Hikasa},
  {Nakamura}, {Sumino}, {Takahashi}, {Tanaka}, {Agashe}, {Aielli}, {Amsler},
  {Antonelli}, {Asner}, {Baer}, {Banerjee}, {Barnett}, {Basaglia}, {Bauer},
  {Beatty}, {Belousov}, {Beringer}, {Bethke}, {Bettini}, {Bichsel}, {Biebel},
  {Black}, {Blucher}, {Buchmuller}, {Burkert}, {Bychkov}, {Cahn}, {Carena},
  {Ceccucci}, {Cerri}, {Chakraborty}, {Chen}, {Chivukula}, {Cowan}, {Dahl},
  {D'Ambrosio}, {Damour}, {de Florian}, {de Gouv{\^e}a}, {DeGrand}, {de Jong},
  {Dissertori}, {Dobrescu}, {D'Onofrio}, {Doser}, {Drees}, {Dreiner}, {Dwyer},
  {Eerola}, {Eidelman}, {Ellis}, {Erler}, {Ezhela}, {Fetscher}, {Fields},
  {Firestone}, {Foster}, {Freitas}, {Gallagher}, {Garren}, {Gerber}, {Gerbier},
  {Gershon}, {Gershtein}, {Gherghetta}, {Godizov}, {Goodman}, {Grab},
  {Gritsan}, {Grojean}, {Groom}, {Gr{\"u}newald}, {Gurtu}, {Gutsche}, {Haber},
  {Hanhart}, {Hashimoto}, {Hayato}, {Hayes}, {Hebecker}, {Heinemeyer},
  {Heltsley}, {Hern{\'a}ndez-Rey}, {Hisano}, {H{\"o}cker}, {Holder},
  {Holtkamp}, {Hyodo}, {Irwin}, {Johnson}, {Kado}, {Karliner}, {Katz}, {Klein},
  {Klempt}, {Kowalewski}, {Krauss}, {Kreps}, {Krusche}, {Kuyanov}, {Kwon},
  {Lahav}, {Laiho}, {Lesgourgues}, {Liddle}, {Ligeti}, {Lin}, {Lippmann},
  {Liss}, {Littenberg}, {Lugovsky}, {Lugovsky}, {Lusiani}, {Makida}, {Maltoni},
  {Mannel}, {Manohar}, {Marciano}, {Martin}, {Masoni}, {Matthews},
  {Mei{\ss}ner}, {Milstead}, {Mitchell}, {M{\"o}nig}, {Molaro}, {Moortgat},
  {Moskovic}, {Murayama}, {Narain}, {Nason}, {Navas}, {Neubert}, {Nevski},
  {Nir}, {Olive}, {Pagan Griso}, {Parsons}, {Patrignani}, {Peacock},
  {Pennington}, {Petcov}, {Petrov}, {Pianori}, {Piepke}, {Pomarol}, {Quadt},
  {Rademacker}, {Raffelt}, {Ratcliff}, {Richardson}, {Ringwald}, {Roesler},
  {Rolli}, {Romaniouk}, {Rosenberg}, {Rosner}, {Rybka}, {Ryutin}, {Sachrajda},
  {Sakai}, {Salam}, {Sarkar}, {Sauli}, {Schneider}, {Scholberg}, {Schwartz},
  {Scott}, {Sharma}, {Sharpe}, {Shutt}, {Silari}, {Sj{\"o}strand}, {Skands},
  {Skwarnicki}, {Smith}, {Smoot}, {Spanier}, {Spieler}, {Spiering}, {Stahl},
  {Stone}, {Sumiyoshi}, {Syphers}, {Terashi}, {Terning}, {Thoma}, {Thorne},
  {Tiator}, {Titov}, {Tkachenko}, {T{\"o}rnqvist}, {Tovey}, {Valencia}, {Van de
  Water}, {Varelas}, {Venanzoni}, {Verde}, {Vincter}, {Vogel}, {Vogt},
  {Wakely}, {Walkowiak}, {Walter}, {Wands}, {Ward}, {Wascko}, {Weiglein},
  {Weinberg}, {Weinberg}, {White}, {Wiencke}, {Willocq}, {Wohl}, {Womersley},
  {Woody}, {Workman}, {Yao}, {Zeller}, {Zenin}, {Zhu}, {Zhu}, {Zimmermann},
  {Zyla}, {Anderson}, {Fuller}, {Lugovsky}, {Schaffner}, \& {Particle Data
  Group}}]{Tanabashi_2018}
{Tanabashi}, M., {Hagiwara}, K., {Hikasa}, K., {et~al.} 2018, \prd, 98, 030001

\bibitem[{{Tornatore} {et~al.}(2007){Tornatore}, {Borgani}, {Dolag}, \&
  {Matteucci}}]{Tornatore_2007}
{Tornatore}, L., {Borgani}, S., {Dolag}, K., \& {Matteucci}, F. 2007, \mnras,
  382, 1050

\bibitem[{{Tornatore} {et~al.}(2004){Tornatore}, {Borgani}, {Matteucci},
  {Recchi}, \& {Tozzi}}]{Tornatore_2004}
{Tornatore}, L., {Borgani}, S., {Matteucci}, F., {Recchi}, S., \& {Tozzi}, P.
  2004, \mnras, 349, L19

\bibitem[{{Vikhlinin}(2006)}]{Vikhlinin_2006_Predicting}
{Vikhlinin}, A. 2006, \apj, 640, 710

\bibitem[{{Vikhlinin} {et~al.}(2009){Vikhlinin}, {Burenin}, {Ebeling},
  {Forman}, {Hornstrup}, {Jones}, {Kravtsov}, {Murray}, {Nagai}, {Quintana}, \&
  {Voevodkin}}]{Vikhlinin_2009}
{Vikhlinin}, A., {Burenin}, R.~A., {Ebeling}, H., {et~al.} 2009, \apj, 692,
  1033

\bibitem[{{Vikhlinin} {et~al.}(1999){Vikhlinin}, {Forman}, \&
  {Jones}}]{Vikhlinin_1999}
{Vikhlinin}, A., {Forman}, W., \& {Jones}, C. 1999, \apj, 525, 47

\bibitem[{{Vikhlinin} {et~al.}(2006){Vikhlinin}, {Kravtsov}, {Forman}, {Jones},
  {Markevitch}, {Murray}, \& {Van Speybroeck}}]{Vikhlinin06}
{Vikhlinin}, A., {Kravtsov}, A., {Forman}, W., {et~al.} 2006, \apj, 640, 691

\bibitem[{{von der Linden} {et~al.}(2014){von der Linden}, {Allen},
  {Applegate}, {Kelly}, {Allen}, {Ebeling}, {Burchat}, {Burke}, {Donovan},
  {Morris}, {Blandford}, {Erben}, \& {Mantz}}]{von_der_Linden_2014}
{von der Linden}, A., {Allen}, M.~T., {Applegate}, D.~E., {et~al.} 2014,
  \mnras, 439, 2

\bibitem[{{Wei{\ss}mann} {et~al.}(2013){Wei{\ss}mann}, {B{\"o}hringer},
  {{\v{S}}uhada}, \& {Ameglio}}]{Weissmann_2013}
{Wei{\ss}mann}, A., {B{\"o}hringer}, H., {{\v{S}}uhada}, R., \& {Ameglio}, S.
  2013, \aap, 549, A19

\bibitem[{{Wyithe} {et~al.}(2001){Wyithe}, {Turner}, \&
  {Spergel}}]{Wyithe_2001}
{Wyithe}, J.~S.~B., {Turner}, E.~L., \& {Spergel}, D.~N. 2001, \apj, 555, 504

\bibitem[{{Zhang} {et~al.}(2010){Zhang}, {Okabe}, {Finoguenov}, {Smith},
  {Piffaretti}, {Valdarnini}, {Babul}, {Evrard}, {Mazzotta}, {Sanderson}, \&
  {Marrone}}]{Zhang_2010}
{Zhang}, Y.-Y., {Okabe}, N., {Finoguenov}, A., {et~al.} 2010, \apj, 711, 1033

\bibitem[{{Zhang} {et~al.}(2017){Zhang}, {Reiprich}, {Schneider}, {Clerc},
  {Merloni}, {Schwope}, {Borm}, {Andernach}, {Caretta}, \& {Wu}}]{Zhang_2017}
{Zhang}, Y.-Y., {Reiprich}, T.~H., {Schneider}, P., {et~al.} 2017, \aap, 599,
  A138

\bibitem[{{Zhao}(1996)}]{Zhao_1996}
{Zhao}, H. 1996, \mnras, 278, 488

\end{thebibliography}

\end{document}